\documentclass[]{jfm}

\usepackage{graphicx}
\usepackage{epstopdf,epsfig}
\usepackage{newtxtext}
\usepackage{newtxmath}
\usepackage{natbib}
\usepackage{hyperref}
\hypersetup{
    colorlinks = true,
    urlcolor   = blue,
    citecolor  = black,
}

\newcommand{\RomanNumeralCaps}[1]
\linenumbers

\usepackage{latexsym}
\usepackage{amsmath}
\usepackage{amssymb}
\usepackage{tabularx}
\usepackage{multirow}
\usepackage{subfigure}
\usepackage{calligra}
\usepackage{calrsfs}
\usepackage{mathrsfs}
\usepackage{extarrows}
\usepackage{booktabs}
\usepackage{bm}
\usepackage{color}
\usepackage{xcolor}

\title{Noise-expansion cascade:  an origin of randomness of turbulence}

\author{
Shijun Liao\aff{1,2}
\corresp{\email{sjliao@sjtu.edu.cn}}
\and 
Shijie Qin\aff{2}
}

\affiliation{
\aff{1} State Key Laboratory of Ocean Engineering, Shanghai 200240, China
\aff{2} School of Ocean and Civil Engineering, Shanghai Jiao Tong University, Shanghai 200240, China
}

\begin{document}
\begin{sloppypar}
\maketitle

\begin{abstract}
Randomness is one of the most important characteristics of turbulence, but its origin remains an open question. By means of a ``thought experiment'' via several  clean  numerical experiments based on the Navier-Stokes equations for two-dimensional turbulent Kolmogorov flow, we reveal a new phenomenon, which we call the ``noise-expansion cascade'' whereby all micro-level noises/disturbances at different orders of magnitudes in the initial condition of Navier-Stokes equations enlarge consistently, say, one by one like an inverse cascade, to macro-level. More importantly, each noise/disturbance input may greatly change the macro-level characteristics and statistics of the resulting turbulence, clearly indicating that micro-level noise/disturbance might have great influence on macro-level characteristics and statistics of turbulence.    Besides, the noise-expansion cascade closely connects randomness of micro-level noise/disturbance and macro-level disorder of turbulence, thus revealing an origin of randomness of turbulence. This also highly suggests that unavoidable thermal fluctuations must be considered when simulating turbulence, even if such fluctuations are several orders of magnitudes smaller than other external environmental disturbances. Hopefully,  the ``noise-expansion cascade'' as a fundamental property of the NS equations could greatly deepen our understandings about turbulence, and besides is helpful for attacking the fourth  millennium problem posed by Clay Mathematics Institute in 2000.  
\end{abstract}

\hspace{-0.4cm}{\bf Keyword} turbulence, cascade, noise, randomness, clean numerical simulation

\section{Motivation}

``Turbulence is the last great unsolved problem of classical physics'', as pointed out by the Nobel Prize winner Richard Feynman. Especially, randomness is one of the most important characteristics of turbulence \citep{Davidson2004}, but its origin remains an {\em open} question until now, to the best of our knowledge.  
 
Today it is widely accepted by the scientific community that turbulent flows can be described mathematically by the Navier-Stokes (NS) equations.
The NS equations are so important and fundamental that their solution becomes the fourth millennium problem posed by \citet{MillenniumProblem}. 
In a pioneering paper, \citet{Orszag1970} proposed the ``direct numerical simulation'' (DNS) which numerically solved the NS equations without any turbulence mode. 
DNS has since become a useful tool in fundamental research of turbulence \citep{Rogallo1981NASA, She1990Nature, Nelkin1992Science, FeracoScience2024, Coleman2010DNS}, and each result given by DNS has been regarded as a ``clean'' benchmark solution, because it is widely believed that numerical noise of DNS would not grow to reach the macroscopic level due to fluid viscosity.
\citet{Coleman2010DNS} pointed out that DNS has ``the ability to perform fundamental studies of {\em clean} flows {\em unaffected} by numerical, modelling and measurement errors'' and ``the complete control of the initial and boundary conditions, and each term in the governing equations, also leads to profound advantages over laboratory and field studies.''  It should be emphasized here that  researchers traditionally focus on  ``the initial and boundary conditions'' and flow domain of turbulence but mostly neglect the influence of micro-level noises such as thermal fluctuation and environmental disturbances.   

However, turbulence governed by the Naver-Stokes equations should be chaotic, i.e. its spatiotemporal trajectories are very sensitive (i.e. unstable) to the initial conditions \citep{Deissler1986PoF, boffetta2017chaos, berera2018chaotic}. Recently, \citet{Vassilicos2023JFM} reported that the average uncertainty energy of three-dimensional Navier-Stokes turbulence grows exponentially.  The similar phenomenon, sometimes called ``inverse error cascade'',  was  also reported in some previous publications \citep{aurell1996growth, boffetta2001predictability, boffetta2010evidence, boffetta2012two, Lin2017SCPMA, JunZhang2024JFM}, which was due to the famous butterfly-effect of chaos.  Note that  the difference in initial condition of these previous publications is at the {\em same} order of magnitude.    
Different from these previous publications, 
 in this paper, we focus on the influence of several micro-level noises/disturbances at quite {\em different} orders of magnitudes.   Here, the noises/disturbances might be either physical environmental noises (such as those caused by thermal fluctuations) or artificial disturbances.      It should be emphasized that, as pointed out by  \citet{Coleman2010DNS},  ``the profound advantages over laboratory and field studies'' of numerical experiment is  the ``complete control of the initial and boundary conditions''.   
Therefore,  we can use such kind of freedom in choice of initial/boundary condition of the NS equations to do some ``thought experiments'' so as to deepen our understandings about turbulence.      Here,  
  let $f({\bf r}) + \Delta_{1}({\bf r}) +\Delta_{2}({\bf r})$ denote an initial condition of the NS equations, where ${\bf r}$ is a spatial vector, $f({\bf r})$ is a function at the macro-level, $\Delta_{1}({\bf r})$ is a small disturbance at a micro-level, for example at the order $10^{-20}$ of magnitude, and $\Delta_{2}({\bf r})$ is an even smaller disturbance, for example at the order $10^{-40}$ of magnitude, respectively.    
Traditionally, it is widely believed that the 2nd disturbance $\Delta_{2}({\bf r})$ could be negligible since it is 20 orders of magnitude smaller than the first disturbance $\Delta_{1}({\bf r})$. Is this traditional viewpoint really {\em correct} for turbulent flow governed by NS equations? This is a fundamental problem, which, to the best of our knowledge, is also an {\em open} question.  
 
In order to answer the two open questions mentioned above, it is necessary to develop a new kind of numerical algorithm, whose numerical noise must be much smaller than micro-level physical disturbances and artificial numerical noise throughout a finite but long enough interval of time. 
In 2009, a method called ``clean numerical simulation'' (CNS) was proposed by \citet{Liao2009} for solving problems involving chaos and turbulence, and since then its computational efficiency has been increased, step by step, by several orders of magnitude \citep{Liao2023book, Liao2014SCPMA, Lin2017SCPMA, Hu2020JCP, Qin2020CSF, Liao2022AAMM, Qin2023JFM, Qin2023AAMM, Qin2024PhysicaD}. Unlike DNS, the CNS algorithm uses {\em multiple-precision} \citep{oyanarte1990mp} with sufficient significant digits and thus can decrease {\em both} the truncation error and round-off error to any given tiny level.
Thus, numerical noise of CNS can be {\em rigorously} negligible throughout a time interval $t\in[0,T_{c}]$ that is long enough for calculating statistics \citep{Liao2023book}, where $T_{c}$ is called ``the critical predictable time''.   
The CNS result is therefore much more accurate than its DNS counterpart over a finite but long enough interval of time and so can be used as a clean benchmark solution to check, for the first time,  the validity of DNS.  
 
For example, it was found by \citet{Qin2022JFM} that the DNS result of a two-dimensional turbulent Rayleigh-B\'{e}nard (RB) convection, which is excited by the thermal fluctuation as an initial condition, quickly departs from the corresponding CNS benchmark solution: the result initially exhibits a non-shearing vortical/roll-like convection but then it quickly turns into a kind of zonal flow, while the CNS benchmark solution consistently retains the same non-shearing vortical/roll-like convection behaviour over a {\em finite} but long enough time interval $t\in[0,T_{c}]$, where $T_{c}$, called ``the critical predictable time'', is equal to 500 for the RB convection under consideration.
To further confirm this, a 2D turbulent Kolmogorov flow, excited by an initial condition with a kind of spatial symmetry, was solved by DNS and CNS, respectively \citep{Qin2024JOES}. It was found that the spatiotemporal trajectory of the CNS benchmark solution retains the same spatial symmetry as the initial condition throughout the whole interval of time [0,1000], however the spatiotemporal trajectory of the corresponding DNS result is the same at the beginning as the CNS benchmark result but quickly loses the spatial symmetry, clearly indicating that the spatio-temporal trajectory of the 2D turbulent Kolmogorov flow given by DNS is badly polluted by artificial numerical noise that quickly increases to the same order of magnitude as the exact solution of NS equations. It clearly illustrated that the 2D turbulent Kolmogorov flow is a chaotic system in that its spatiotemporal trajectory is rather sensitive to small disturbance caused by artificial numerical noise \citep{Qin2024JOES}. 
More importantly,  as illustrated by \citet{Qin2022JFM} and \citet{Qin2024JOES}, the DNS result {\em sometimes} may deviate greatly from the CNS benchmark solution {\em not only} in flow type and/or spatial symmetry of flow field {\em but also} even in statistics.  These two successful applications of CNS illustrated that CNS can indeed 
provide us with the capability to carry out {\em clean} numerical experiments that enable us to investigate accurately the evolution and propagation of micro-level noises/disturbances in the initial condition of the NS equations for turbulence. It should be emphasized that this object cannot be realized by DNS whose numerical noise quickly increases to the same order of magnitude as true solution, as illustrated by \citet{Qin2022JFM} and \citet{Qin2024JOES}.  

In addition,  CNS has been successfully used to attack some rather difficult problems in classical mechanics. For example, the number of periodic orbits of the famous three-body problem, which can be traced back to Newton in 1687, has been increased several orders of magnitude by means of CNS \citep{Li2017SCPMA, Li2018PASJ, Liao2022NA}. 
This is because CNS, unlike other traditional numerical methods, can correctly calculate the essentially chaotic trajectories of the three-body system \citep{NewScientist2017,NewScientist2018}.   The three-body problem highlights the need to determine the precise spatiotemporal trajectory of certain complicated dynamic systems.
 
In this paper, greatly inspired by spatial symmetry of the CNS benchmark solution  \citep{Qin2024JOES} of the 2D turbulent Kolmogorov flow, we properly conceive/design several clean numerical experiments based on CNS for a two-dimensional (2D) turbulent Kolmogorov flow governed by the NS equations with specially chosen initial conditions that contain terms of different orders of magnitudes and have different spatial symmetries, so as to accurately investigate these terms' propagations, evolutions, and macro-scale influences on the turbulence.
These clean numerical experiments provide us with rigorous evidence that all noises/disturbances at different orders of magnitudes in initial condition of the NS equations could enlarge, one by one like an inverse cascade, to a macro-level, and moreover each of them could greatly change the later turbulence characteristics. 
Based on this interesting phenomenon, we propose a new concept which we call the ``noise-expansion cascade''.
The noise-expansion cascade closely connects the randomness of the initial micro-level noise/disturbance to the later macro-level disorder of turbulence, and thus reveals an {\em origin} of randomness of turbulence.
Besides, according to the concept of noise-expansion cascade, unavoidable thermal fluctuations must be considered when simulating turbulence, even if such fluctuations are many orders of magnitudes smaller than other external disturbances.

The paper is structured as follows. \S~2 describes the design of the clean numerical experiments. \S~3 reports the detailed results obtained using the clean numerical experiments, notably the noise-expansion cascade phenomenon. \S~4 discusses the main findings of the work. 

\section{Clean numerical experiment}

Consider the 2D incompressible Kolmogorov flow \citep{obukhov1983kolmogorov, chandler2013invariant, wu2021quadratic} in a square domain $[0, L]^2$ (with a periodic boundary condition) under Kolmogorov forcing, which is stationary, monochromatic, and cosusoidally varying in space, with an integer $n_K$ describing the forcing scale and $\chi$ representing the corresponding forcing amplitude per unit mass of fluid, respectively. Using the length scale $L/2\pi$ and the time scale $\sqrt{L/2\pi\chi}$, the non-dimensional Navier-Stokes equation of this 2D Kolmogorov flow in the form of stream function reads 
\begin{equation}
 \frac{\partial}{\partial t}\Big(\nabla^{2}\psi\Big)+\frac{\partial(\psi,\nabla^{2}\psi)}{\partial(x,y)}-\frac{1}{Re}\nabla^{4}\psi+n_K\cos(n_Ky)=0,       \label{eq_psi}
\end{equation}
where 
\[
Re=\frac{\sqrt{\chi}}{\nu}\left(\frac{L}{2\pi}\right)^{\frac{3}{2}}   
\]
is the Reynolds number, $\nu$ denotes the kinematic viscosity, $\psi$ is the stream function, 
 $x,y\in[0,2\pi]$ are horizontal and vertical coordinates, $t$ denotes the time, $\nabla^{2}$ is the Laplace operator,  $\nabla^{4}=\nabla^{2}\nabla^{2}$, and 
\[
 \frac{\partial(a,b)}{\partial(x,y)}=\frac{\partial a}{\partial x}\frac{\partial b}{\partial y}-\frac{\partial b}{\partial x}\frac{\partial a}{\partial y}      
\]
is the Jacobi operator, respectively. Note that the stream function $\psi$ always satisfies the periodic boundary condition
\begin{equation}
\psi(x, y, t)=\psi(x+2\pi, y, t)=\psi(x, y+2\pi, t).       \label{boundary_condition}
\end{equation}
In order to have a relatively strong state of turbulent flow, we choose  $n_K=16$ and $Re=2000$ for all cases considered in this paper.  

Let us consider the following three different initial conditions
\begin{eqnarray}
 \psi(x,y,0) & = & -\frac{1}{2}\big[\cos(x+y) + \cos(x-y)\big],       \label{initial_condition} \\
 \psi(x,y,0) & = & -\frac{1}{2}\big[\cos(x+y)+\cos(x-y) \big] + \delta' \sin(x+y),       \label{initial_condition-1} \\
 \psi(x,y,0) & = & -\frac{1}{2}\big[\cos(x+y)+\cos(x-y) \big] + \delta' \sin(x+y) +\delta'' \sin(x+2y),       \label{initial_condition-2}
\end{eqnarray}
respectively, where $\delta' $ and $\delta''$ are constants, corresponding to a Kolmogorov flow with different spatial symmetry, as mentioned below.   

Note that the initial condition (\ref{initial_condition}) has the spatial symmetry 
\begin{equation}
\left\{
\begin{array}{ll}
\mbox{rotation}&: \;\; \psi(x,y,t)=\psi(2\pi-x,2\pi-y,t),\\
 \mbox{translation}&: \;\; \psi(x,y,t) =\psi(x+\pi,y+\pi,t),
 \end{array}  
 \right.     \label{symmetry_psi:A}
\end{equation}
at $t=0$. Here, we emphasize that \citet{Qin2024JOES} solved the 2D turbulent Kolmogorov flow governed by Eqs.~(\ref{eq_psi}) and (\ref{boundary_condition}) subject to the initial condition (\ref{initial_condition}) in the case of $n_K=4$  and $Re = 40$ by means of DNS and CNS. They found that the spatio-temporal trajectory given by DNS agrees well with the CNS benchmark solution from the beginning and retains the spatial symmetry (\ref{symmetry_psi:A}) until $t \approx 120$ when the DNS result completely loses spatial symmetry, unlike the CNS benchmark solution which retains the spatial symmetry (\ref{symmetry_psi:A}) throughout the {\em whole} time interval $t\in[0,1500]$, clearly indicating that the spatio-temporal trajectory given by DNS is badly polluted by artificial numerical noise when $t \geq 120$. 
So, it is impossible to obtain rigorous, accurate prediction of the evolution and propagation of micro-level disturbance by DNS. Therefore, we have to give up DNS in this paper, but use CNS instead.
Importantly, \citet{Qin2024JOES} revealed such an important fact that the 2D Kolmogorov turbulent flow given by CNS retains the {\em same} spatial symmetry as its initial condition: we will use this fact to do a thought experiment via several clean numerical experiments based on CNS, given that the same findings about spatial symmetry should apply qualitatively for the case $n_K=16$ and $Re=2000$ considered in this paper.     

The CNS algorithm is now briefly described. First, to decrease the spatial truncation error to a small enough level, we, as in DNS, discretize the spatial domain of the flow field by a uniform mesh $N^2 = 1024^2$, and adopt the Fourier pseudo-spectral method for spatial approximation with the $3/2$ rule for dealiasing. In this way, the corresponding spatial resolution is fine enough for the considered Kolmogorov flow: the grid spacing is less than the average Kolmogorov scale and enstrophy dissipative scale, as mentioned by \citet{pope2001turbulent} and \citet{Boffetta2012ARFM}.  
Besides, in order to decrease the temporal truncation error to a small enough level, we, {\em unlike} DNS, use the 140th-order (i.e. $M = 140$) Taylor expansion with a time step $\Delta t = 10^{-3}$. Furthermore, {\em different} from DNS, we use {\em multiple-precision} with 260 significant digits (i.e. $N_s = 260$) for all physical/numerical variables and parameters so as to decrease the round-off error to a small enough level. In addition, the self-adaptive CNS strategy \citep{Qin2023AAMM} and parallel computing are adopted to dramatically increase the computational efficiency of the CNS algorithm.
Especially, another CNS result is given by the same CNS algorithm but with even smaller numerical noise (i.e. using even larger $M$ and/or $N_{s}$ than those mentioned above), which confirms (by comparison) that the numerical noise of the former CNS result (say, given by $N=1024$, $M=140$ and $N_{s} = 260$) remains rigorously negligible throughout the whole time interval $t\in[0,300]$ so that it can be used as a clean benchmark solution.  For further details, please refer to \citet{Qin2024JOES} and \citet{Liao2023book}.  Note that  the related code of CNS and some movies can be downloaded via GitHub (\url{https://github.com/sjtu-liao/2D-Kolmogorov-turbulence}).

\section{Results of clean numerical experiments}

Our  clean numerical experiments based on CNS comprise two stages:
\begin{enumerate}
\item In the first stage, we set $\delta'=O(1)$ and $\delta''=O(1)$ in (\ref{initial_condition-1}) and (\ref{initial_condition-2}), and then confirm by means of CNS that the 2D turbulent Kolmogorov flow subject to the initial condition (\ref{initial_condition}) or (\ref{initial_condition-1}) always retains the {\em same} spatial symmetry as its corresponding initial condition, but the turbulent flow subject to the initial condition (\ref{initial_condition-2}) has {\em no} spatial symmetry at all. 
\item In the second stage, we set $\delta'=10^{-20}$ and $\delta''=10^{-40}$ in (\ref{initial_condition-1}) and (\ref{initial_condition-2}), and then carry out the corresponding clean numerical experiments by means of CNS.
We name the CNS results subject to the three different initial conditions (\ref{initial_condition}), (\ref{initial_condition-1}) and (\ref{initial_condition-2}) as follows: Flow CNS, Flow CNS$'$ and Flow CNS$''$, respectively. The so-called noise-expansion cascade phenomenon is revealed by comparing the evolutions of spatial symmetry of these three turbulent flows.
\end{enumerate}
Details of our clean numerical experiments based on CNS are described below.  
     
\subsection{Spatial symmetry under different initial conditions} 

Eqs.~(\ref{eq_psi}) and (\ref{boundary_condition}) subject to the initial condition (\ref{initial_condition}) in the case of $Re = 2000$ and $n_K=16$ are numerically solved by means of CNS. It is found that this CNS flow always retains the same spatial symmetry (\ref{symmetry_psi:A}) throughout the whole time interval $t\in[0,300]$ exactly as the initial condition (\ref{initial_condition}). 
This accords with the finding about the spatial symmetry of the 2D Kolmogorov turbulent flow obtained by \citet{Qin2024JOES} using CNS for the different case of $Re = 40$ and $n_K=4$. According to the governing equation (\ref{eq_psi}), the periodic boundary condition (\ref{boundary_condition}), the initial condition (\ref{initial_condition}) and the spatial symmetry (\ref{symmetry_psi:A}), the corresponding solution should be in series form as follows,
\begin{equation}
 \psi(x,y,t)=\sum_{m+n=2r} a_{m,n}(t) \; \cos(mx+ny)+ \sum_{m-n=2q} b_{m,n}(t) \; \cos(mx-ny),       \label{form}\nonumber
\end{equation}
where $a_{m,n}(t)$, $b_{m,n}(t)$ are unknown time-dependent coefficients and $m\geq0$, $n\geq0$, $r>0$, $q$ are integers. Thus, the vorticity $\omega=\nabla^{2}\psi$ of the flow field naturally retains the same spatial symmetry throughout the time interval  $t\in[0,300]$, i.e. 
\begin{equation}
\left\{
\begin{array}{ll}
\mbox{rotation}&: \;\; \omega(x,y,t)=\omega(2\pi-x,2\pi-y,t),\\
 \mbox{translation}&: \;\; \omega(x,y,t) =\omega(x+\pi,y+\pi,t).
 \end{array}  
 \right.      \label{symmetry-omega:A}
\end{equation}

Note that the initial condition (\ref{initial_condition-1}) in the case of $\delta' = 1$ has spatial symmetry in translation, say,
\begin{equation}
 \psi(x,y,t)=\psi(x+\pi,y+\pi,t), \;\;\;  \omega(x,y,t) = \omega(x+\pi,y+\pi,t)        \label{symmetry_psi:B}
\end{equation}
for $t=0$ here. Similarly, it is found that the corresponding CNS solution governed by Eqs.~(\ref{eq_psi}) and (\ref{boundary_condition}) subject to the initial condition (\ref{initial_condition-1}) with $\delta' = 1$ always retains the same spatial symmetry (\ref{symmetry_psi:B}) throughout the whole time interval $t\in[0,300]$. 
Besides, it is further found that the same spatial symmetry (\ref{symmetry_psi:B}) is obtained as long as $\delta'$ is a constant at a macro-level $O(1)$, such as $\delta' = 2, -3, \pi$ and so on. 
It should be emphasized that the initial condition~(\ref{initial_condition}) and the spatial symmetry (\ref{symmetry_psi:A}) involves {\em two} kinds of spatial symmetry, i.e. rotation and translation, but the initial condition (\ref{initial_condition-1}) in the case of $\delta' = O(1)$ and the spatial symmetry (\ref{symmetry_psi:B}) has only {\em one}, i.e. translation.  

Note that the initial condition (\ref{initial_condition-2}) in the case of $\delta' = 1$ and $\delta'' = 1$ has {\em no} spatial symmetry, due to its 3rd term $\sin(x+2y)$. In a similar way, it is found that the corresponding CNS solution indeed has {\em no} spatial symmetry throughout the whole time interval $t\in[0,300]$. The same conclusion about the spatial symmetry is obtained as long as $\delta'$ and $\delta''$ are constants at a macro-level $O(1)$, such as $\delta'$, $\delta''  = 2$, $-3$, $\pi$ and so on.

Using the afore-mentioned findings from clean numerical experiments based on CNS, we discover the so-called noise-expansion cascade phenomenon by undertaking the clean numerical experiments described below.

\subsection{Discovery of noise-expansion cascade}    \label{Influence}

In the second stage of our clean numerical experiments based on CNS, we {\em thereafter} choose $\delta' = 10^{-20}$ and $\delta'' =10^{-40}$ in the initial condition (\ref{initial_condition-1}) and (\ref{initial_condition-2}), corresponding to two micro-level disturbances $10^{-20}\sin(x+y)$ and $10^{-40}\sin(x+2y)$, where the second is 20 orders of magnitude smaller than the first. Thus, the initial conditions (\ref{initial_condition-1}) and (\ref{initial_condition-2}) become thereafter 
\begin{eqnarray}
 \psi(x,y,0) & = & -\frac{1}{2}\big[\cos(x+y)+\cos(x-y) \big] + 10^{-20} \sin(x+y),       \label{initial_condition-11} \\
 \psi(x,y,0) & = & -\frac{1}{2}\big[\cos(x+y)+\cos(x-y) \big] + 10^{-20} \sin(x+y) \nonumber\\
 && +10^{-40} \sin(x+2y),       \label{initial_condition-22}
\end{eqnarray}
respectively.  
In other words, Eqs.~(\ref{eq_psi}) and (\ref{boundary_condition}) in the case of $Re=2000$ and $n_{K}=16$ are solved by means of CNS in the time interval $t\in[0,300]$, subject to the initial condition 
(\ref{initial_condition}), (\ref{initial_condition-11}) or (\ref{initial_condition-22}), whose clean numerical simulations are called thereafter Flow CNS, Flow CNS$'$ and Flow CNS$''$, for the sake of simplicity. Note that, according to the three different initial conditions (\ref{initial_condition}), (\ref{initial_condition-11}) and (\ref{initial_condition-22}), Flow CNS$'$ is equal to Flow CNS plus $\delta_{1}(x,y,t)$, and Flow CNS$''$ is equal to Flow CNS$'$ plus $\delta_{2}(x,y,t)$, where $\delta_{1}(x,y,t)$ and $\delta_{2}(x,y,t)$ denote the spatio-temporal evolution of the first disturbance $10^{-20} \sin(x+y)$ and the second disturbance $10^{-40}\sin(x+2y)$ in the initial conditions (\ref{initial_condition-11}) and (\ref{initial_condition-22}), respectively.   
 
Due to the butterfly-effect of chaos, a micro-level disturbance of a chaotic system grows exponentially to the macro-level  \citep{Deissler1986PoF, aurell1996growth, boffetta2017chaos, berera2018chaotic,boffetta2001predictability, boffetta2010evidence, boffetta2012two, Vassilicos2023JFM, JunZhang2024JFM}.  
Logically, the smaller the disturbance, the longer the time it requires to reach the macro-level.
According to \citet{Qin2024JOES}, the two turbulent Kolmogorov flows under consideration are chaotic systems.
Therefore, $\delta_{2}(x,y,t)$, corresponding to the second disturbance $10^{-40}\sin(x+2y)$,  requires more time to reach the macro-level than $\delta_{1}(x,y,t)$, corresponding to the first disturbance $10^{-20}\sin(x+2y)$.  
According to our previous clean numerical experiments mentioned in \S3.1, when both $\delta_{1}(x,y,t)$ and $\delta_{2}(x,y,t)$ are negligible, Flow CNS$'$ and Flow CNS$''$ should have the same spatial symmetry (\ref{symmetry_psi:A}) as the initial condition (\ref{initial_condition}). However, when $\delta_{1}(x,y,t)$ corresponding to the first disturbance enlarges to a macro-level $O(1)$ but $\delta_{2}(x,y,t)$ is still at a micro-level and thus negligible, the corresponding Flow CNS$'$ and Flow CNS$''$ should have the same spatial symmetry (\ref{symmetry_psi:B}) as the initial condition (\ref{initial_condition-1}) when $\delta' = O(1)$. In addition, 
when both $\delta_{1}(x,y,t)$ and $\delta_{2}(x,y,t)$ enlarge to a macro-level $O(1)$, the corresponding Flow CNS$''$ should have {\em no} spatial symmetry at all, just like 2D turbulent Kolmogorov flow subject to the initial condition (\ref{initial_condition-2}) when $\delta' = O(1)$ and $\delta'' = O(1)$. Thus, by comparing the spatial symmetry of  Flow CNS, Flow CNS$'$ and Flow CNS$''$ given by our clean numerical experiments, we can find out when the evolution $\delta_1(x,y,t)$, corresponding to the first disturbance $10^{-20}\sin(x+y)$, and the evolution $\delta_2(x,y,t)$, corresponding to the second disturbance $10^{-40}\sin(x+2y)$, enlarge to macro-level $O(1)$.       

\begin{figure}
    \begin{center}
        \begin{tabular}{cc}
             \includegraphics[width=2.0in]{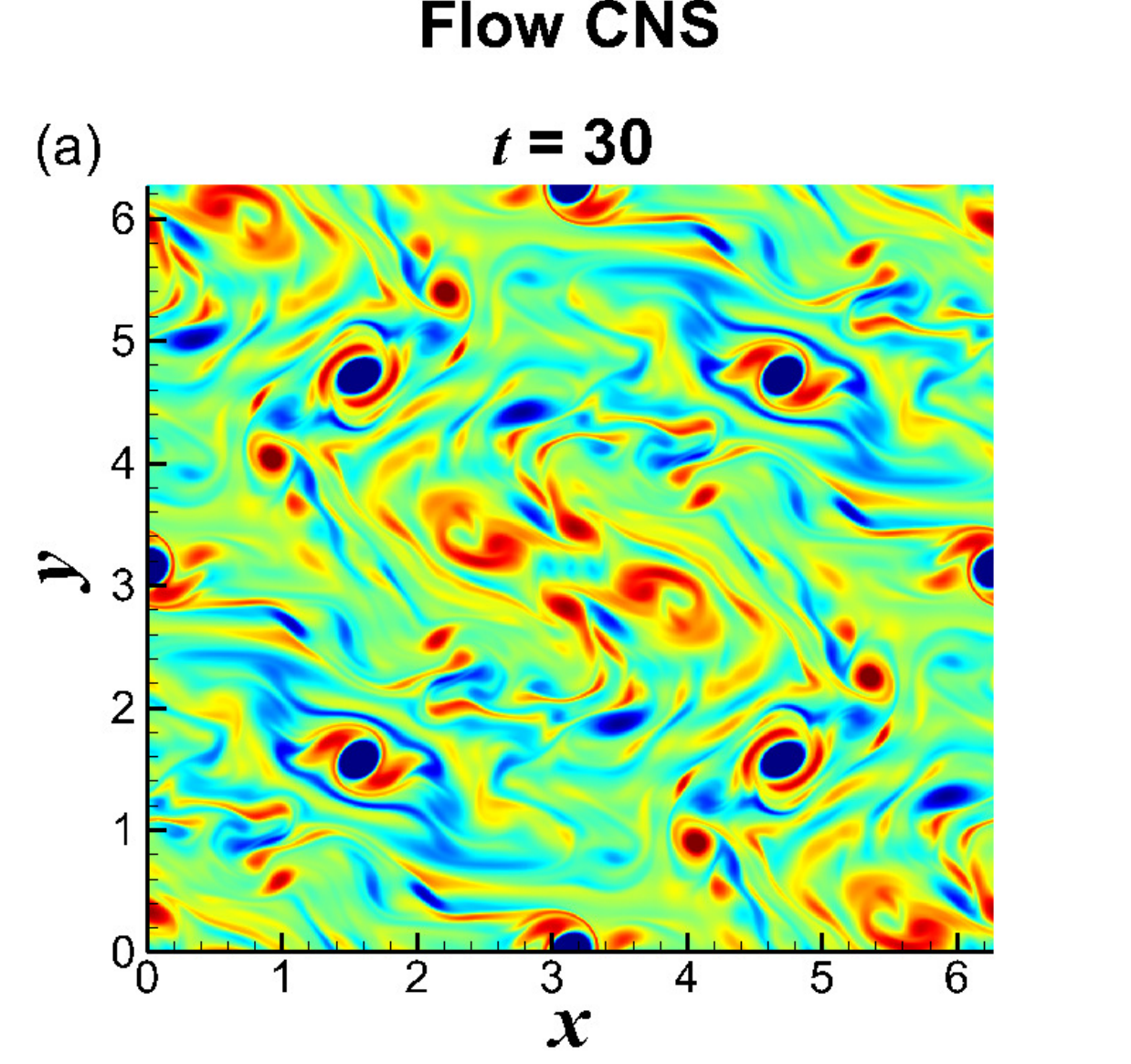}
             \includegraphics[width=2.0in]{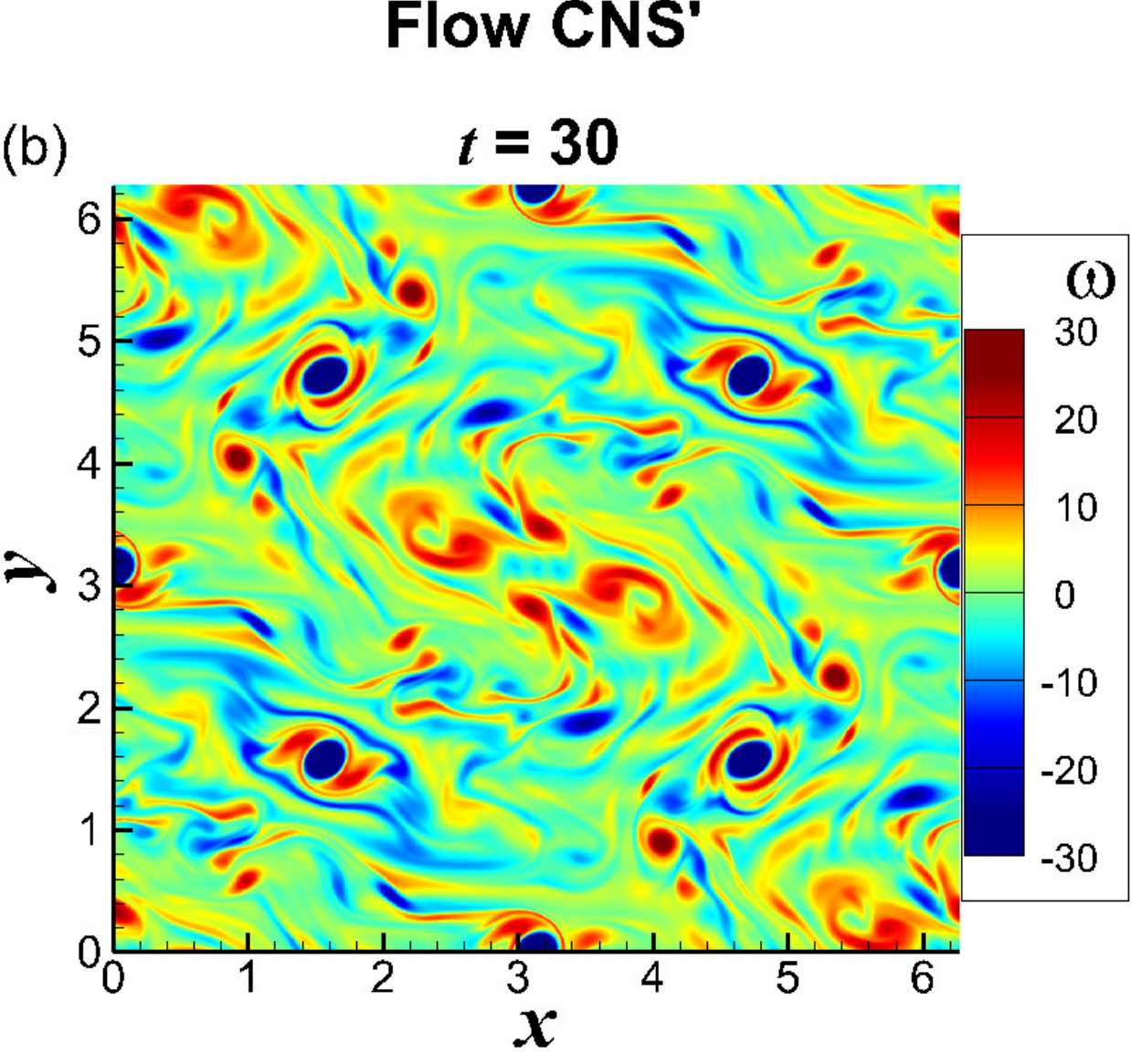}    \\
             \includegraphics[width=2.0in]{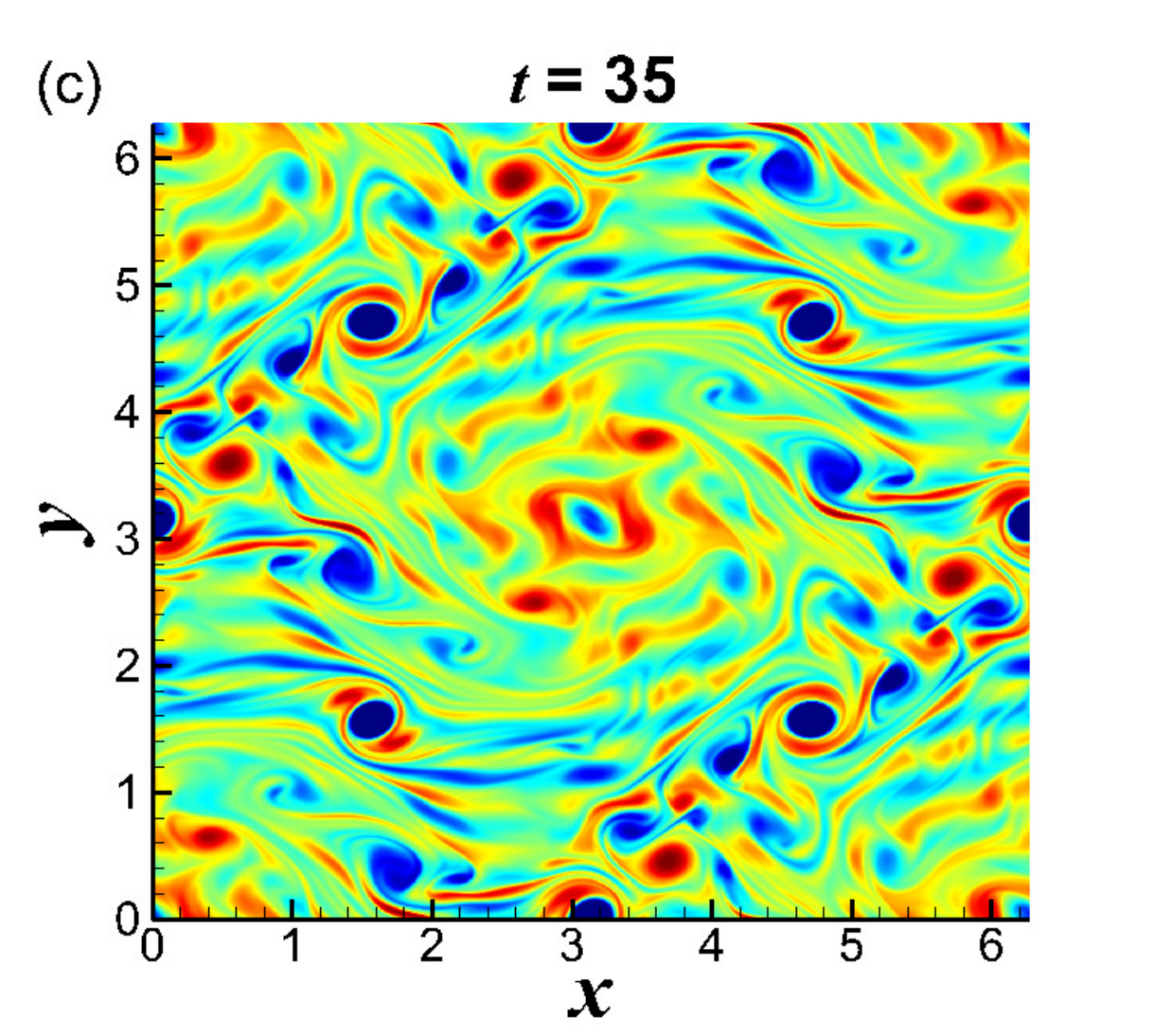}
             \includegraphics[width=2.0in]{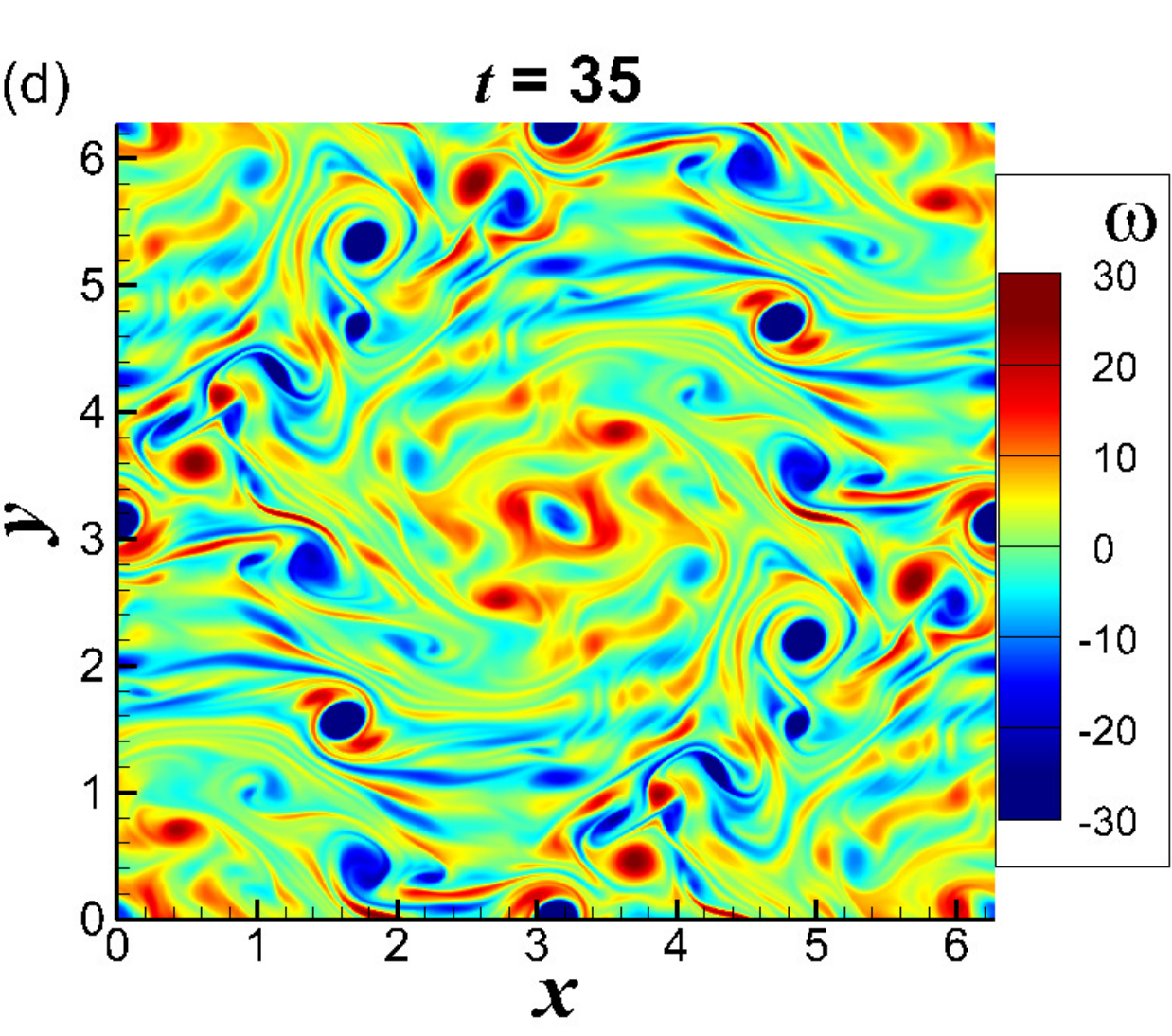}    \\
             \includegraphics[width=2.0in]{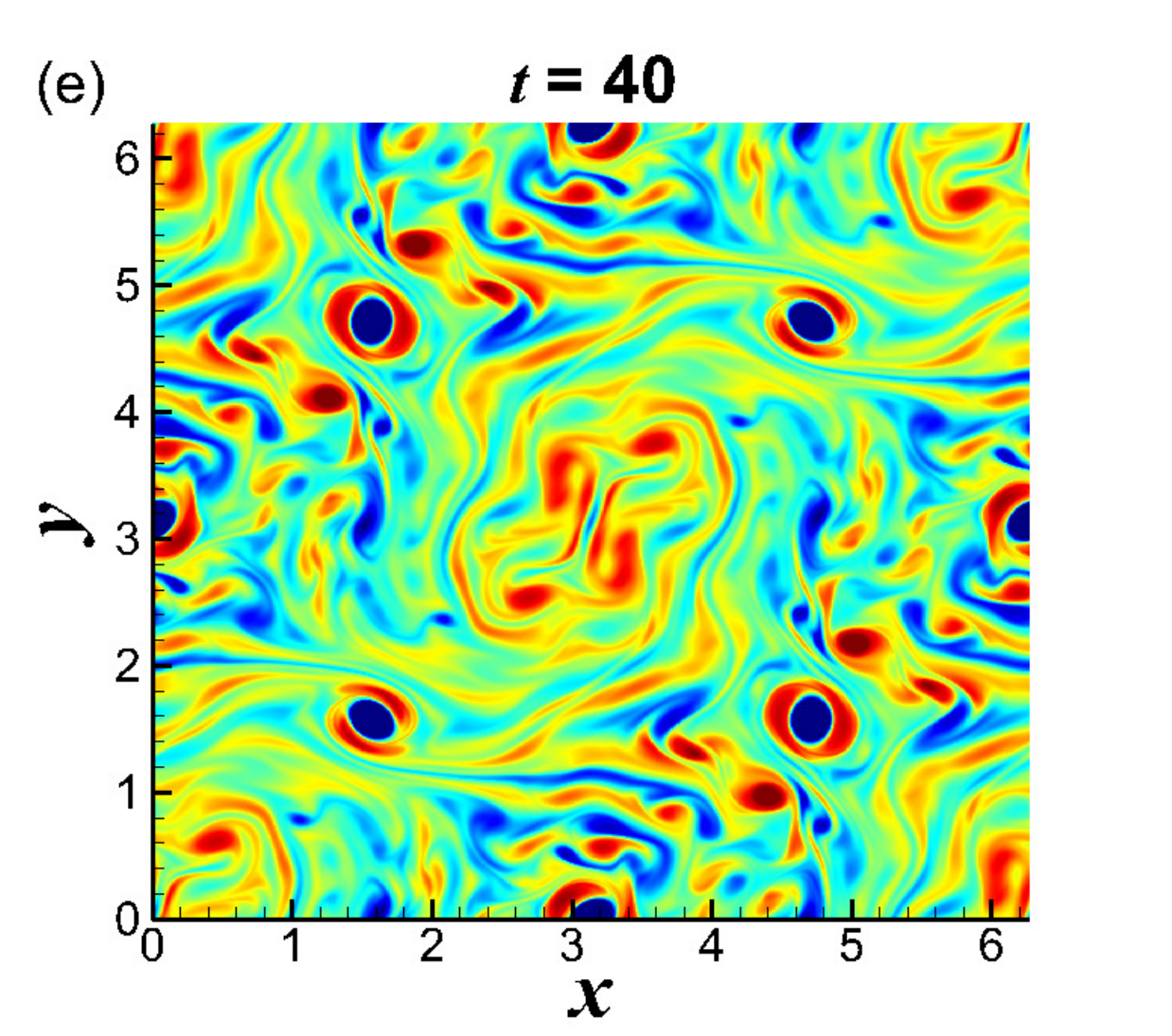}
             \includegraphics[width=2.0in]{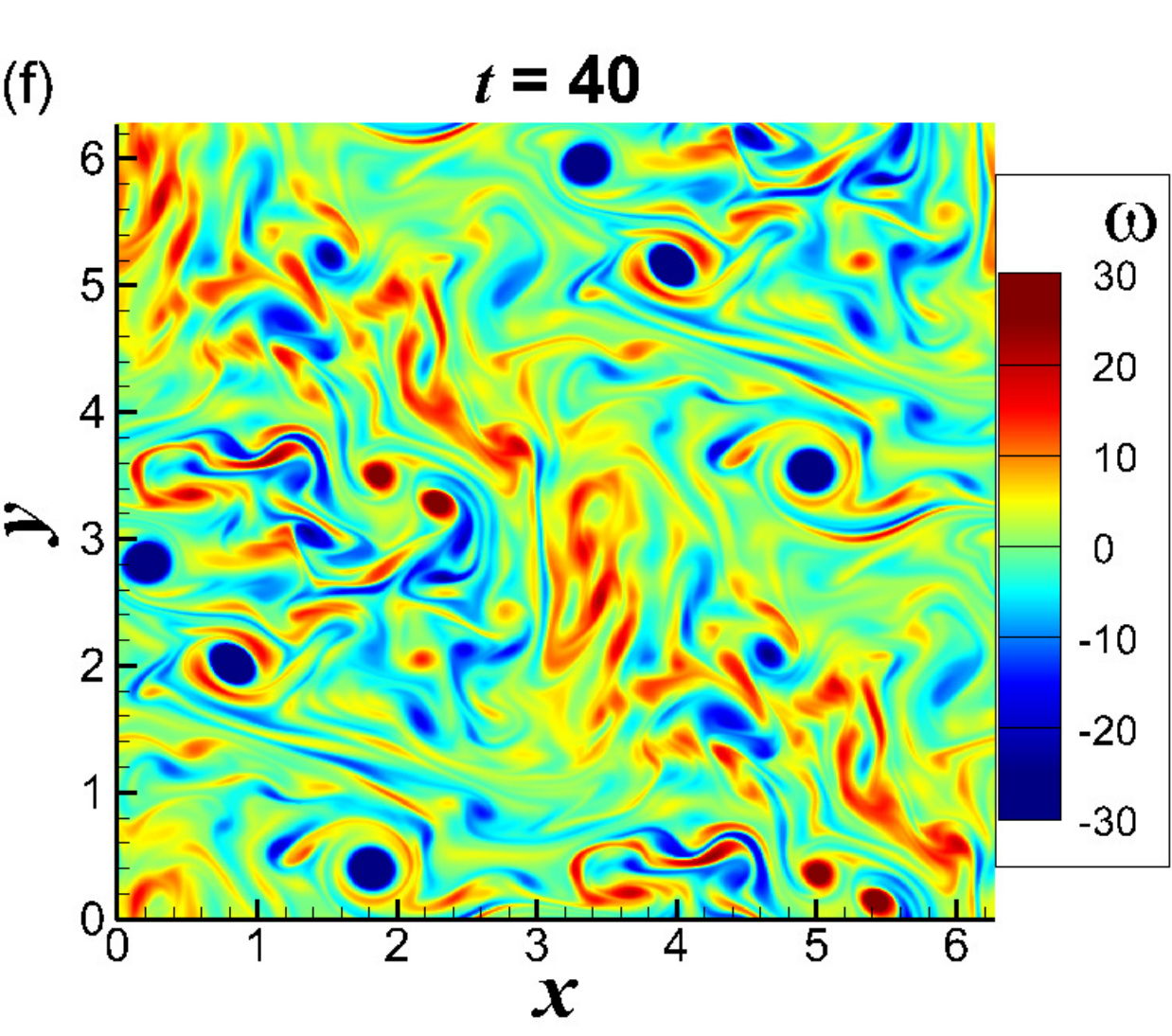}    \\
             \includegraphics[width=2.0in]{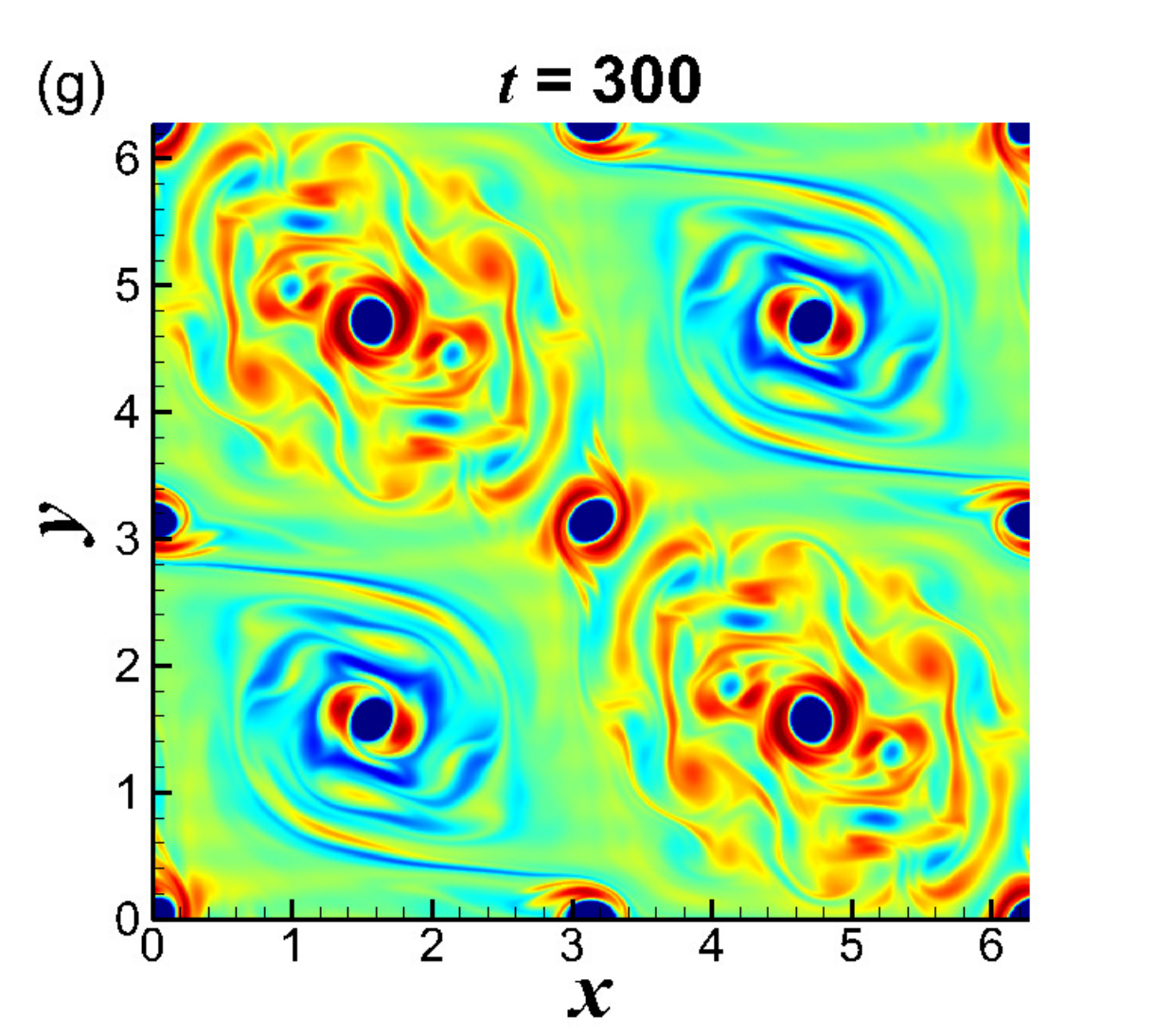}
             \includegraphics[width=2.0in]{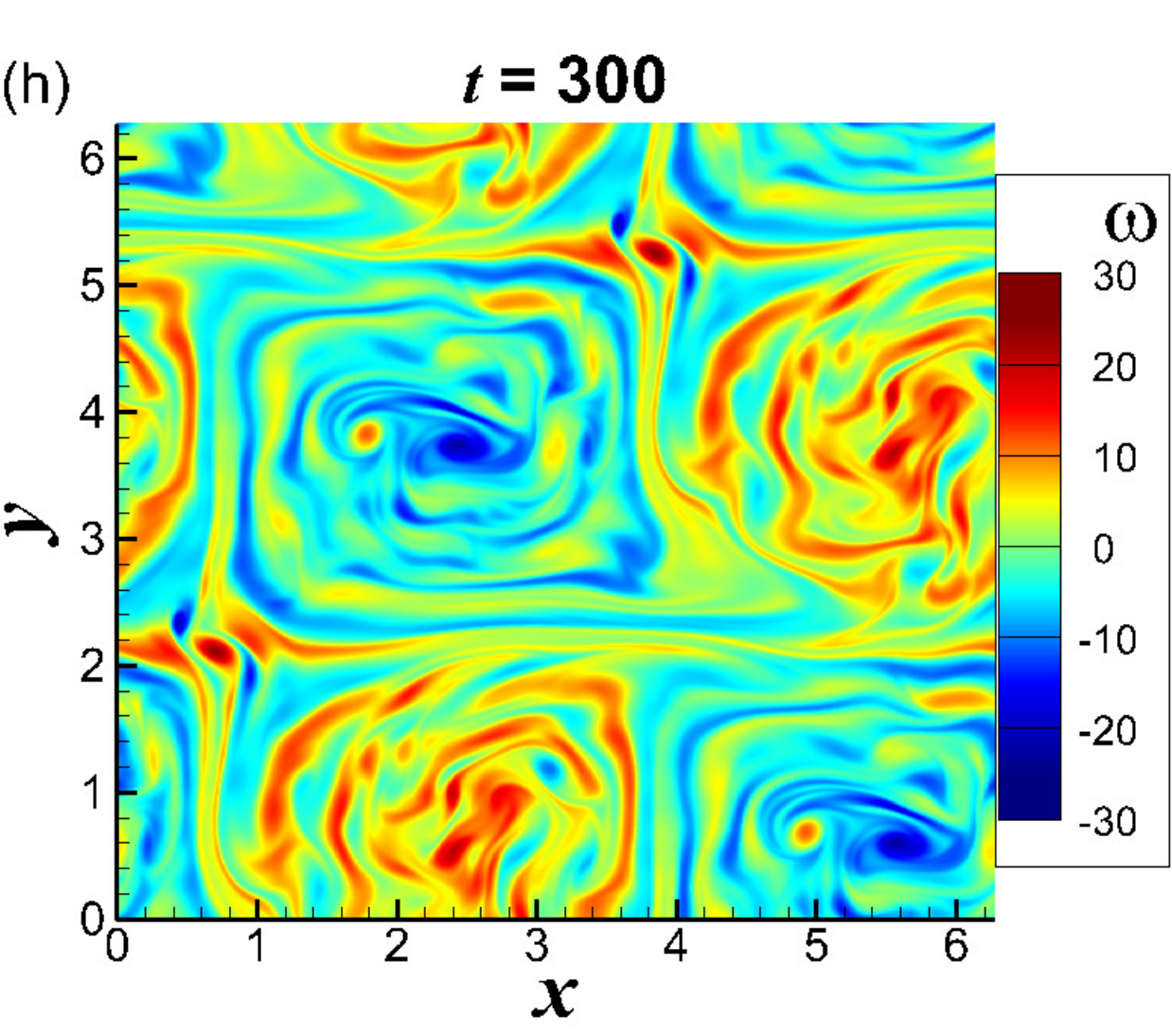}    \\
        \end{tabular}
    \caption{Vorticity fields $\omega(x,y)$ of the 2D turbulent Kolmogorov flow governed by (\ref{eq_psi}) and (\ref{boundary_condition}) for $n_K=16$ and $Re=2000$ given by CNS, subject to either the initial conditions (\ref{initial_condition})  (left, marked by Flow CNS) or (\ref{initial_condition-11}) (right, marked by Flow CNS$'$), at different times: \\(a)-(b) $t=30$, (c)-(d) $t=35$, (e)-(f) $t=40$, and (g)-(h) $t=300$. See the supplementary Movie~1 for the whole evolution process of vorticity field, which can be downloaded via GitHub (\url{https://github.com/sjtu-liao/2D-Kolmogorov-turbulence}).
}     \label{Vor_Evolutions-1}
    \end{center}
\end{figure}

\begin{figure}
    \begin{center}
        \begin{tabular}{cc}
             \includegraphics[width=2.0in]{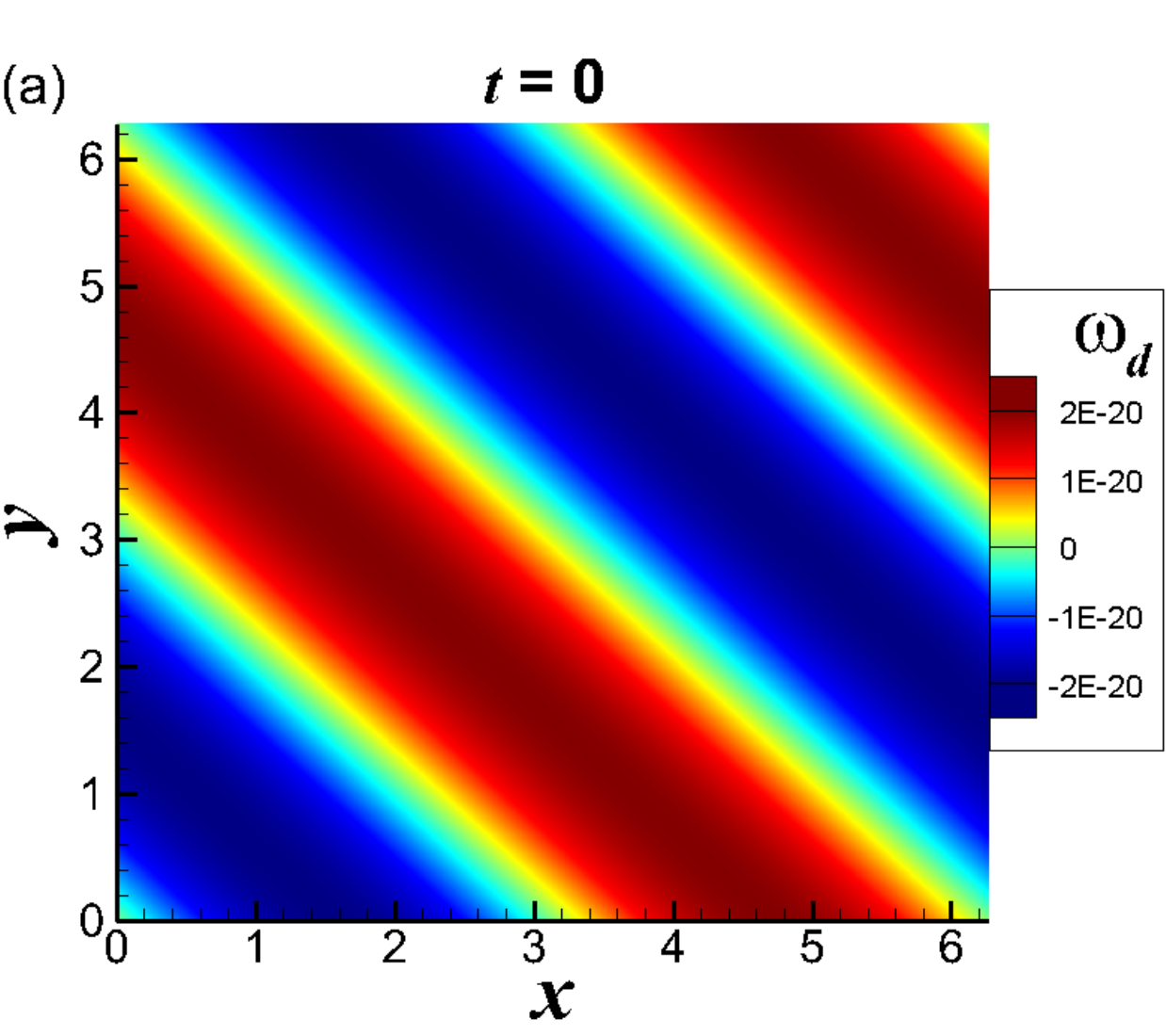}
             \includegraphics[width=2.0in]{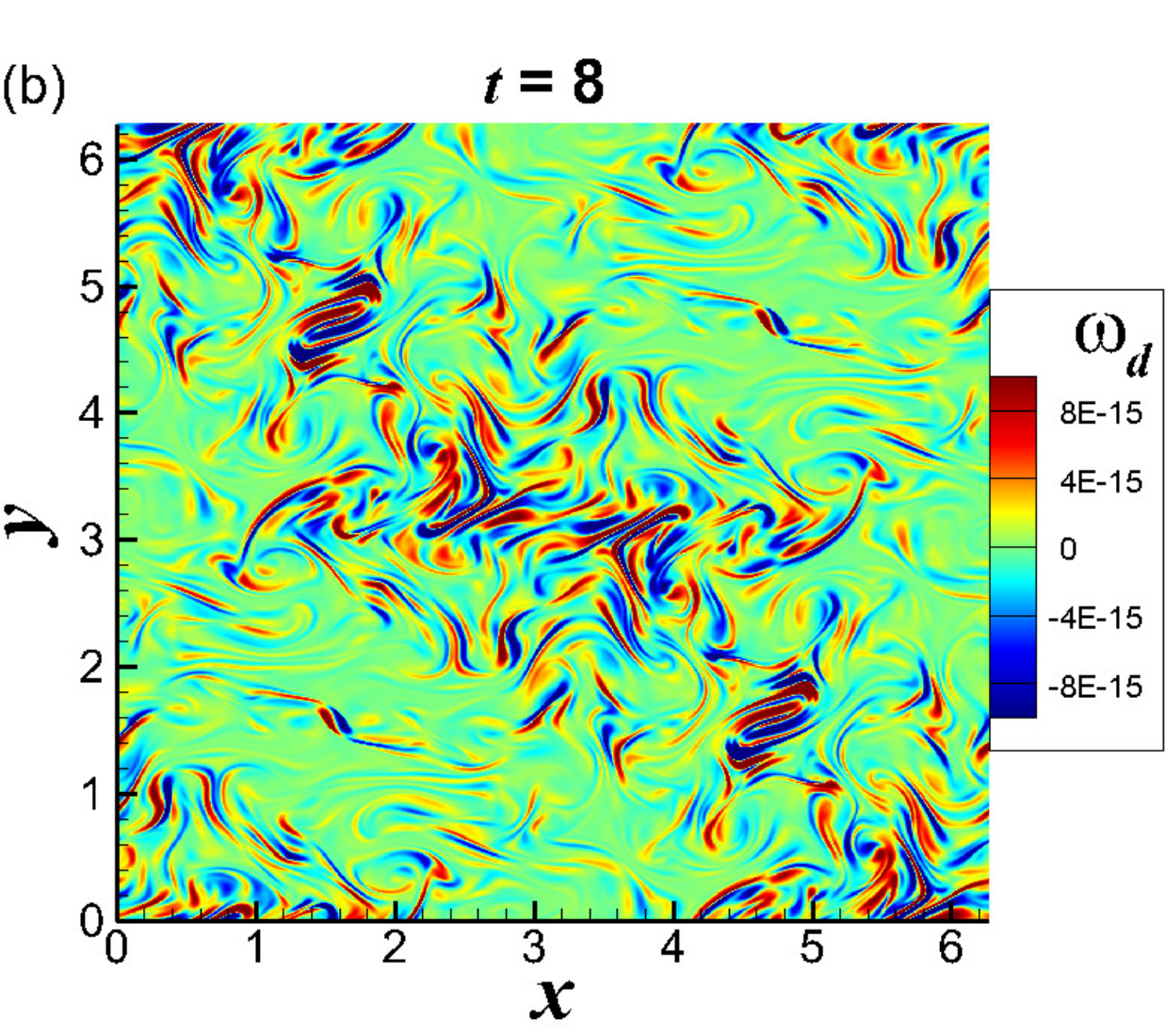}    \\
             \includegraphics[width=2.0in]{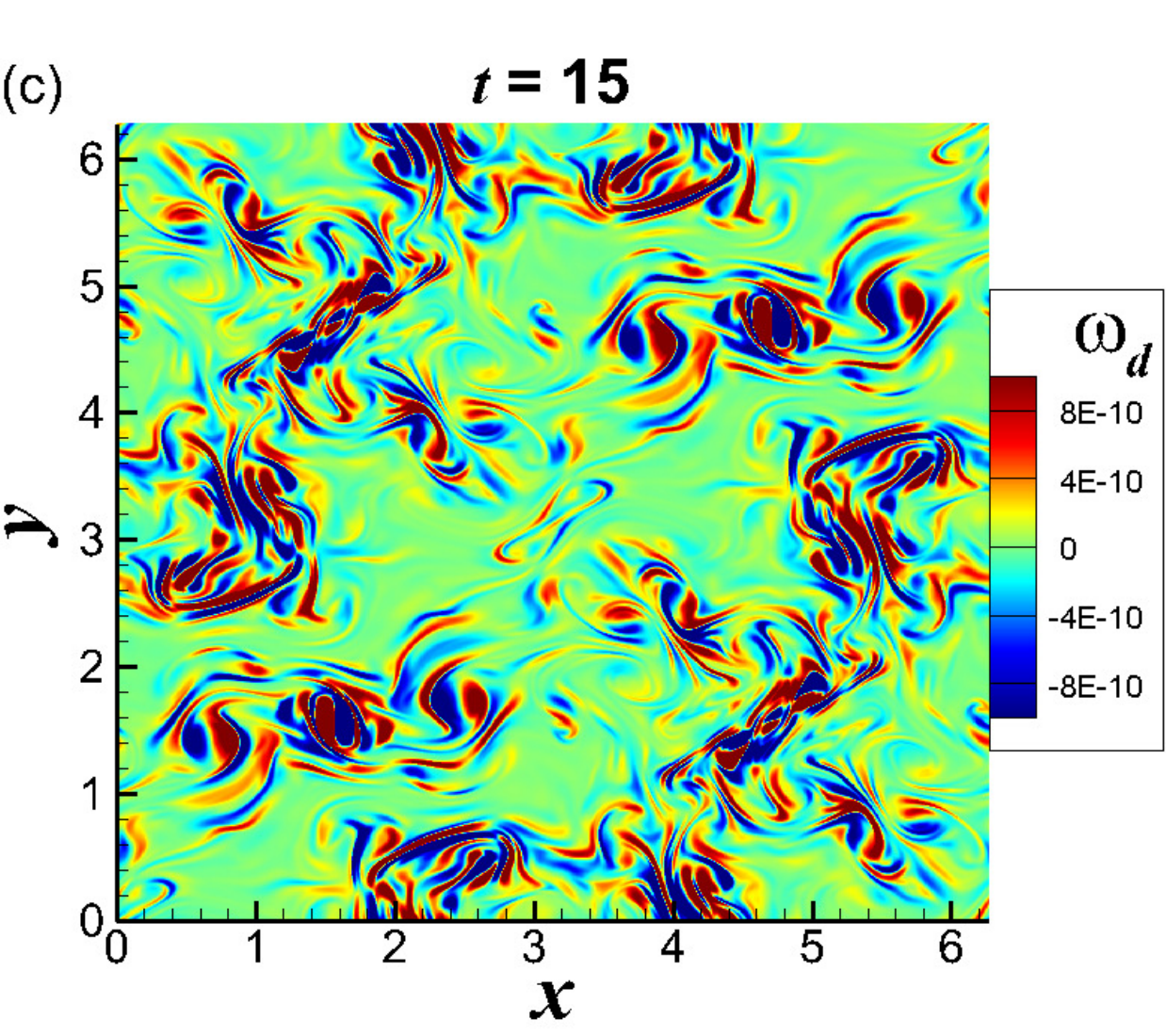}
             \includegraphics[width=2.0in]{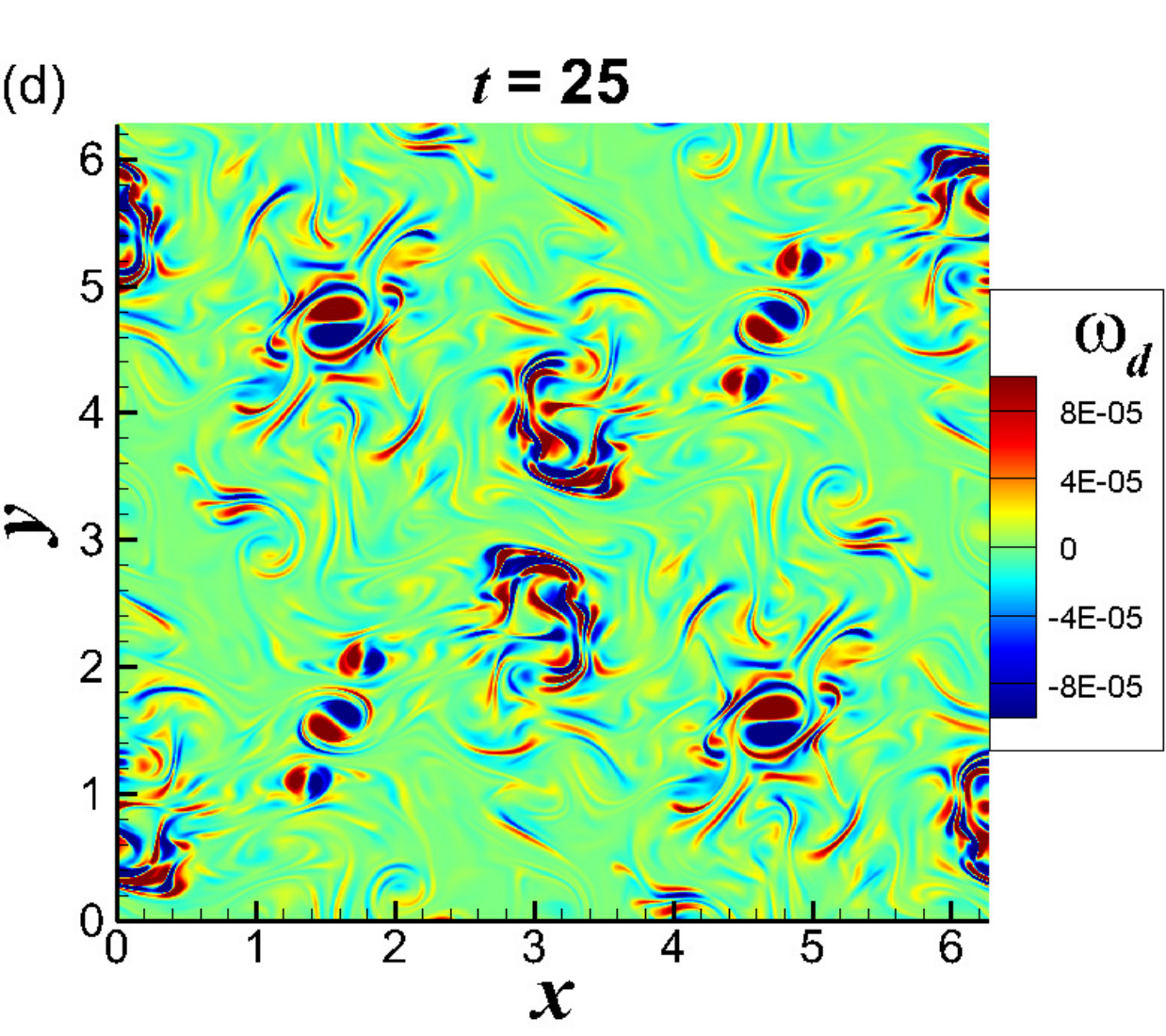}    \\
             \includegraphics[width=2.0in]{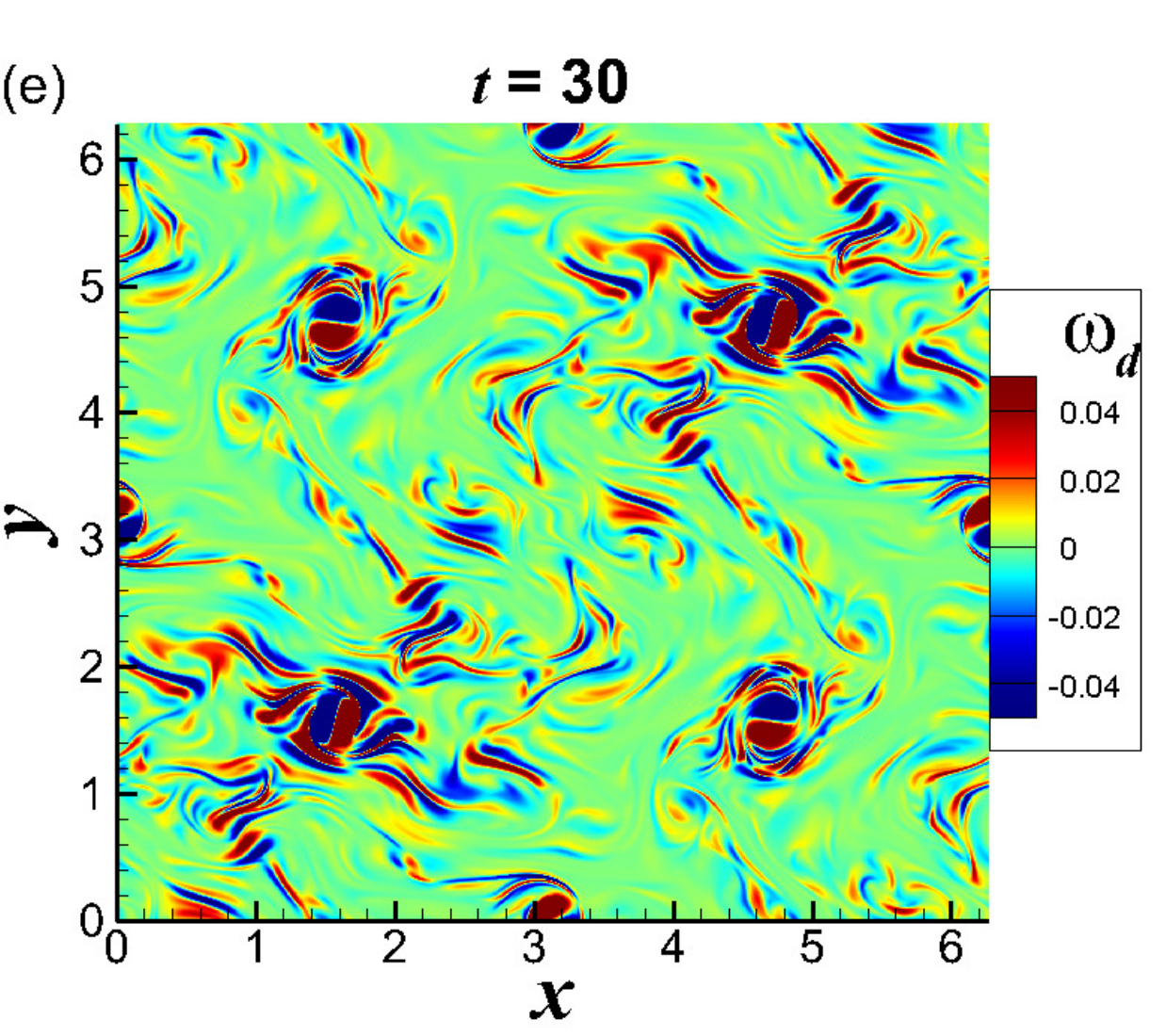}
             \includegraphics[width=2.0in]{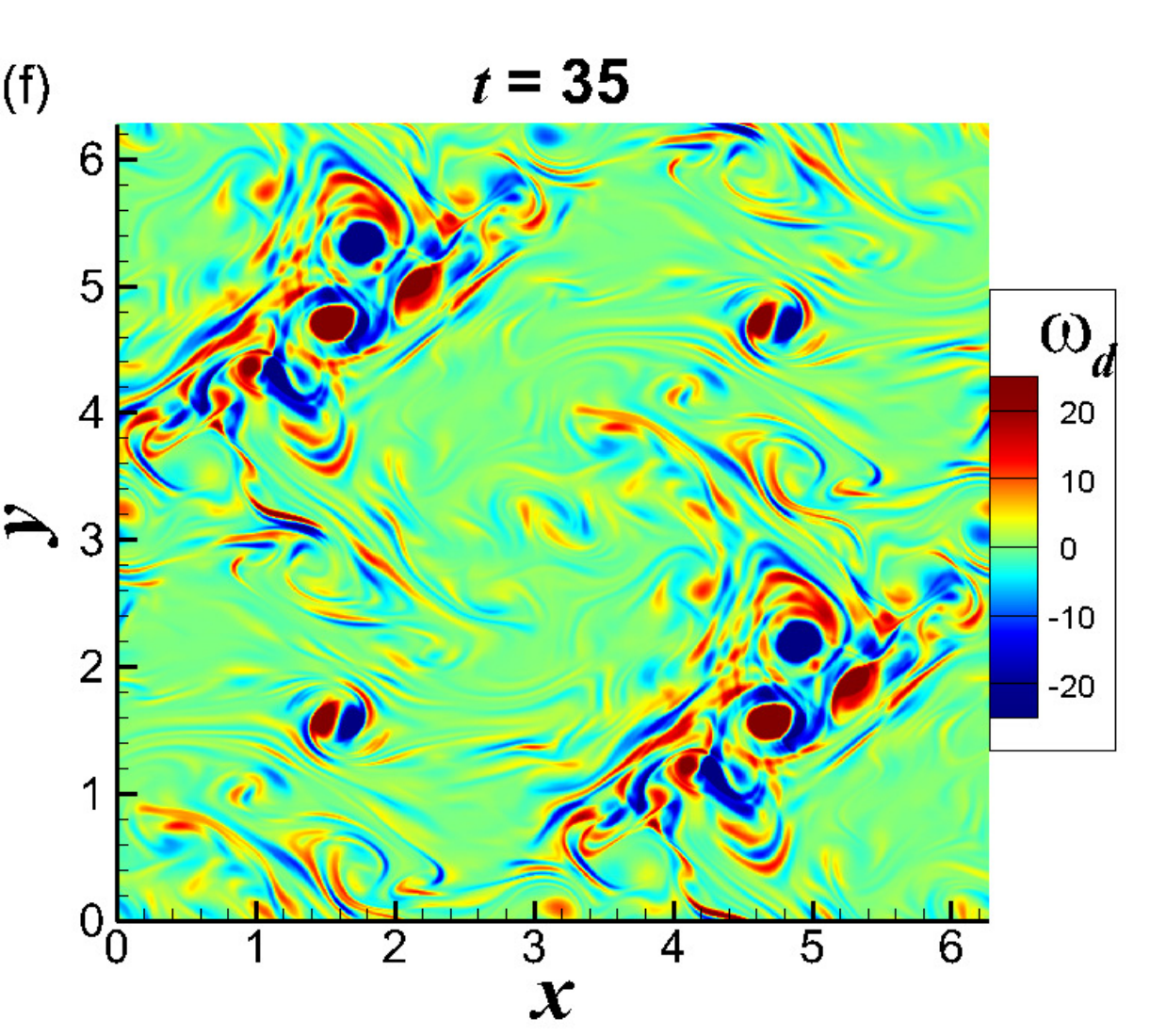}    \\
             \includegraphics[width=2.0in]{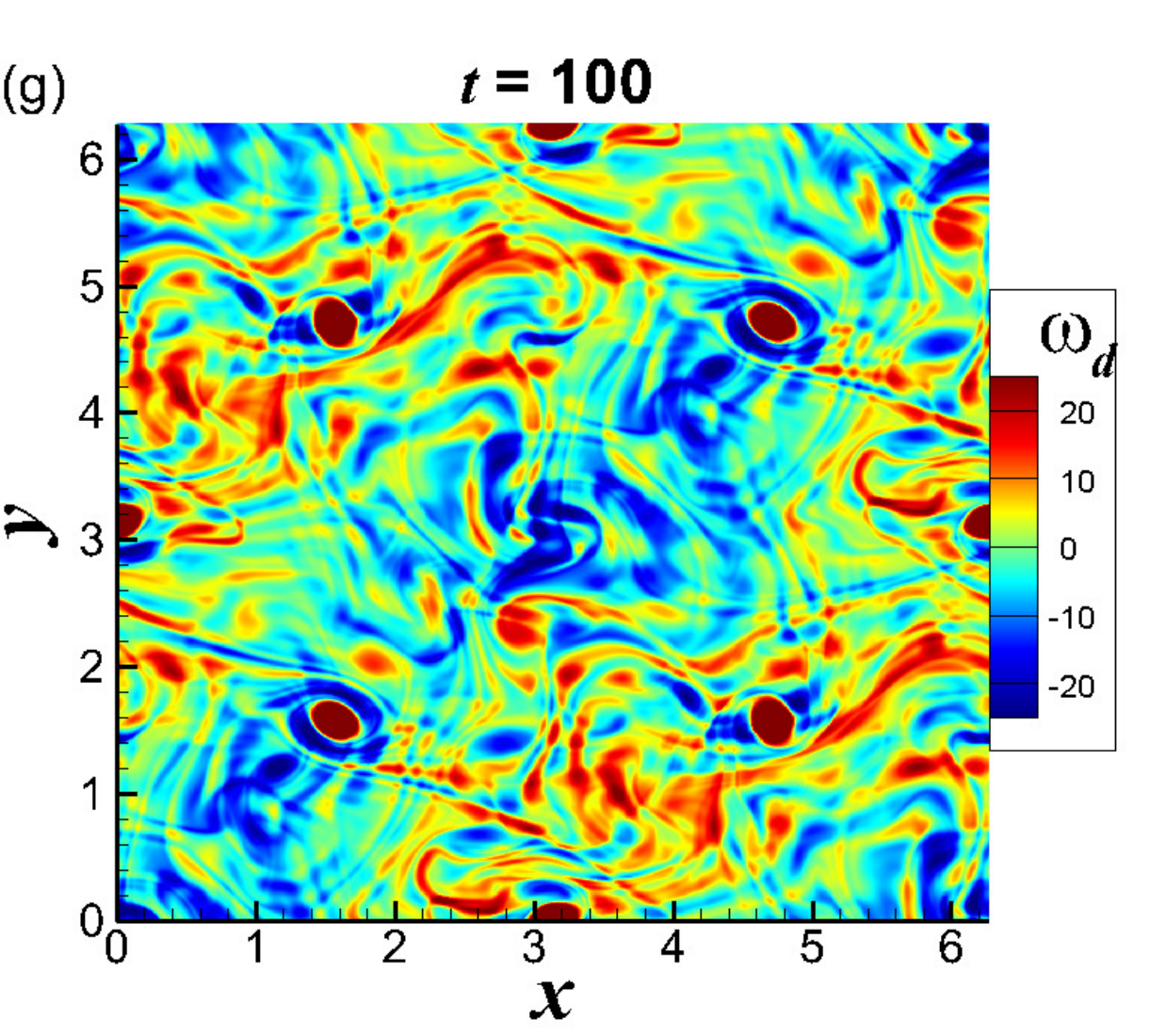}
             \includegraphics[width=2.0in]{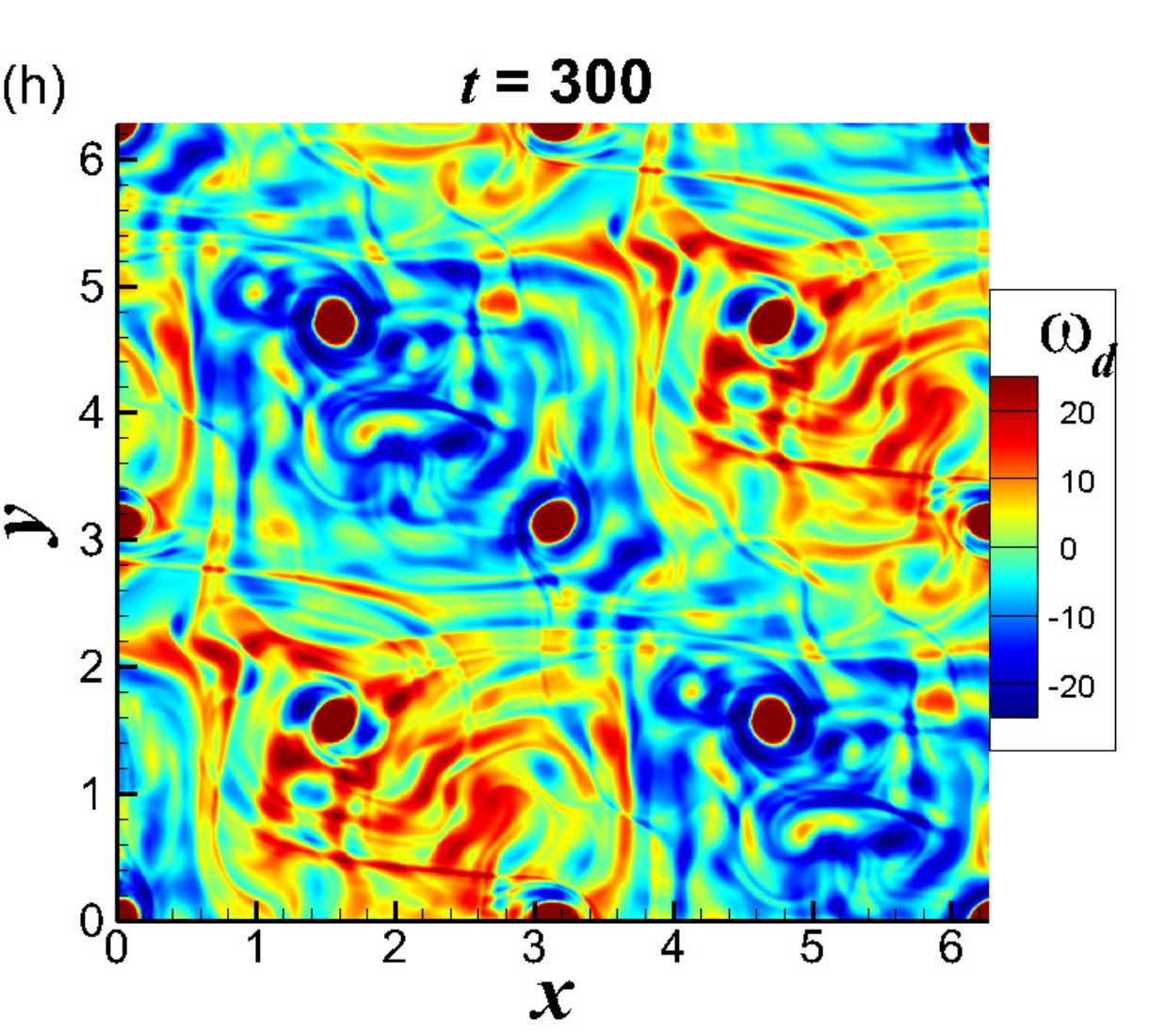}
        \end{tabular}
    \caption{Vorticity fields of the evolution $\delta_{1}(x,y,t)$, corresponding to the first disturbance $10^{-20}\sin(x+y)$ in the initial condition (\ref{initial_condition-11}) of 2D turbulent Kolmogorov flow governed by (\ref{eq_psi}) and (\ref{boundary_condition}) for $n_K=16$ and $Re=2000$ given by CNS, at the different times: \\(a) $t=0$, (b) $t=8$, (c) $t=15$, (d) $t=25$, (e) $t=30$, (f) $t=35$, (g) $t=100$, (h) $t=300$.}     \label{Vor_Evolutions-1-delta}
    \end{center}
\end{figure}

As shown in Fig.~\ref{Vor_Evolutions-1}, the vorticity field $\omega(x,y,t)$ of Flow CNS, subject to the initial condition (\ref{initial_condition}), is compared with that of Flow CNS$'$, subject to the initial condition (\ref{initial_condition-11}).   Note that Flow CNS retains the spatial symmetry (\ref{symmetry-omega:A}) of vorticity throughout the {\em whole} time interval $t\in[0,300]$.  
Obviously, the term $10^{-20} \, \sin(x+y)$ in the initial condition (\ref{initial_condition-11}) can be regarded as a micro-level disturbance added to the initial condition (\ref{initial_condition}), since it is 20 orders of magnitude smaller. Certainly, it takes some time for this tiny disturbance $10^{-20} \, \sin(x+y)$ to be enlarged to a macro-level~$O(1)$.    
Indeed, Flow CNS$'$ appears the same as Flow CNS from the beginning, for example at $t=30$ as shown in Fig.~\ref{Vor_Evolutions-1}(a) and (b), when $\delta_1(x,y,t)$ corresponding to the tiny disturbance $10^{-20} \sin(x+y)$ of the initial condition (\ref{initial_condition-11}) has {\em not} been increased to macro-level, as shown in Fig.~\ref{Vor_Evolutions-1-delta}(a)-(e), so that both Flow CNS and Flow CNS$'$ agree well and retain the {\em same} spatial symmetry (\ref{symmetry-omega:A}) of vorticity.
It is found that the vorticity of Flow CNS$'$ deviates from the spatial symmetry (\ref{symmetry-omega:A}) obviously at $t\approx35$ and thereafter loses the spatial symmetry (\ref{symmetry-omega:A}) but retains the spatial symmetry  (\ref{symmetry_psi:B}) throughout the time interval  $t\in[35,300]$ instead, as shown in Fig.~\ref{Vor_Evolutions-1}(c)-(h). 
Note that the term $\sin(x+y)$ of the initial condition (\ref{initial_condition-11}) implies the spatial symmetry of translation but no spatial symmetry of rotation since $\sin(x+y) = \sin(\pi+x +\pi + y)$ but $\sin(x+y) \neq \sin(2\pi-x + 2\pi-y)$. 
The explanation is provided by considering Fig.~\ref{Vor_Evolutions-1-delta}: $\delta_1(x,y,t)$ corresponding to the disturbance $10^{-20} \, \sin(x+y)$ in the initial condition (\ref{initial_condition-11}), which increases from a micro-level, step by step, to a macro-level at $t \approx 35$, remains the same spatial symmetry (\ref{symmetry_psi:B}) throughout the whole time interval $t\in[0,300]$ so that it destroys the spatial symmetry (\ref{symmetry-omega:A}) and triggers the transition of the spatial symmetry from (\ref{symmetry-omega:A}) to (\ref{symmetry_psi:B}) at $t\approx 35$ when it reaches a macro-level $O(1)$.
This provides us with rigorous evidence that a very small disturbance $10^{-20} \, \sin(x+y)$ to the initial condition (\ref{initial_condition-11}) indeed increases to the same order of magnitude as the exact solution of the NS equations at $t \approx 35$, which destroys the spatial symmetry (\ref{symmetry-omega:A}) and triggers the transition of the spatial symmetry from (\ref{symmetry-omega:A}) to (\ref{symmetry_psi:B}).     

\begin{figure}
    \begin{center}
        \begin{tabular}{cc}
             \includegraphics[width=2.0in]{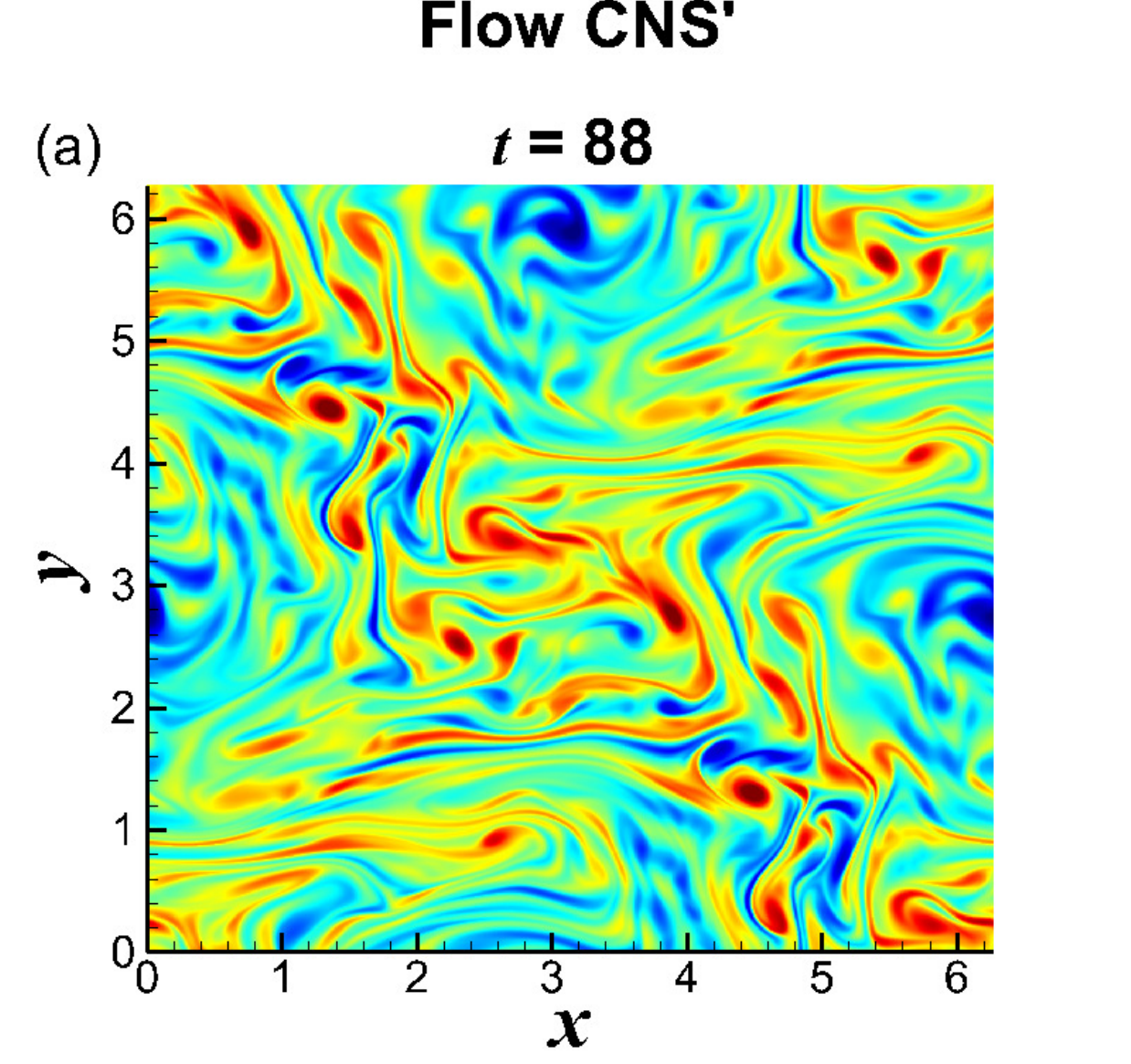}
             \includegraphics[width=2.0in]{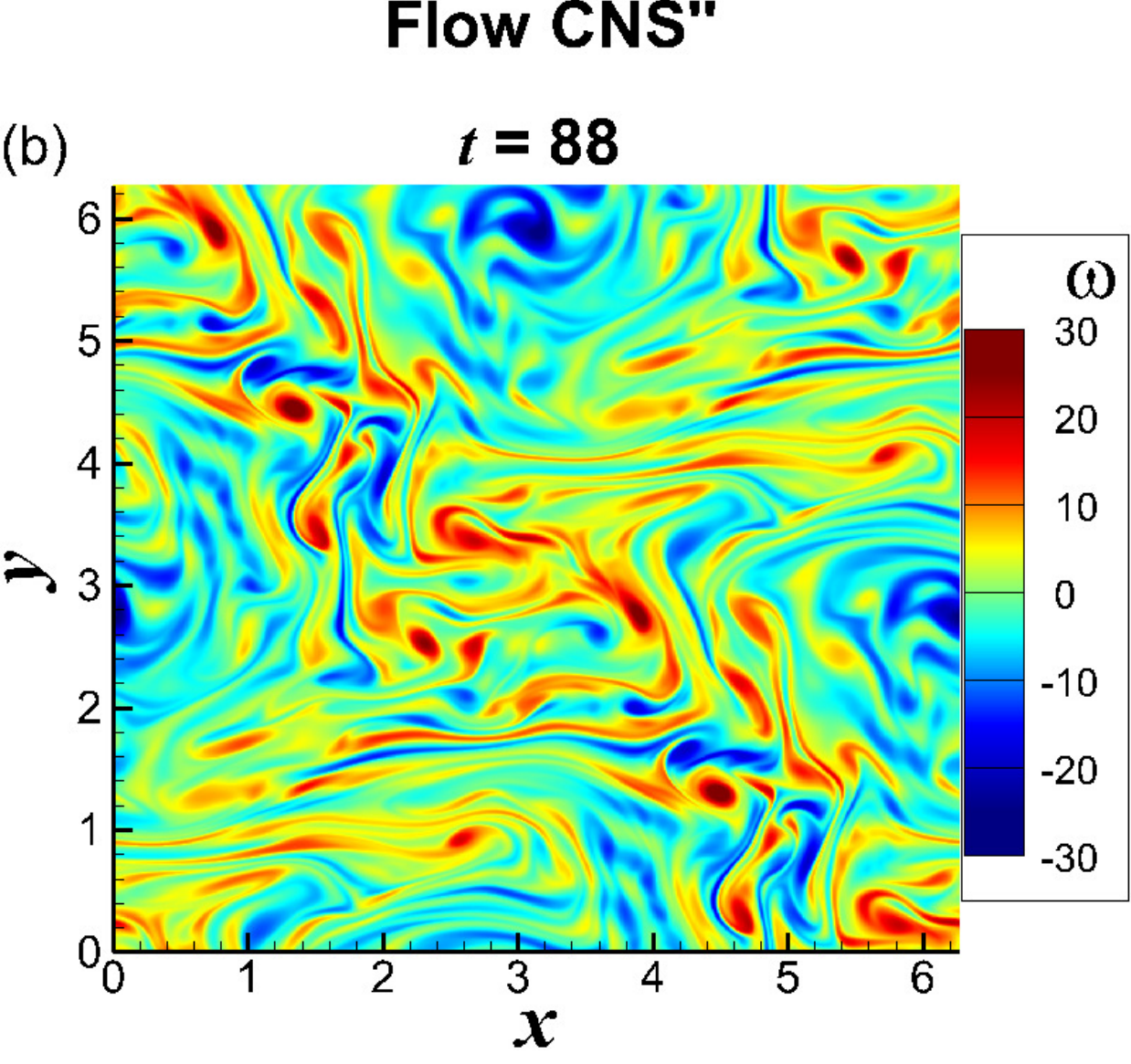}    \\
             \includegraphics[width=2.0in]{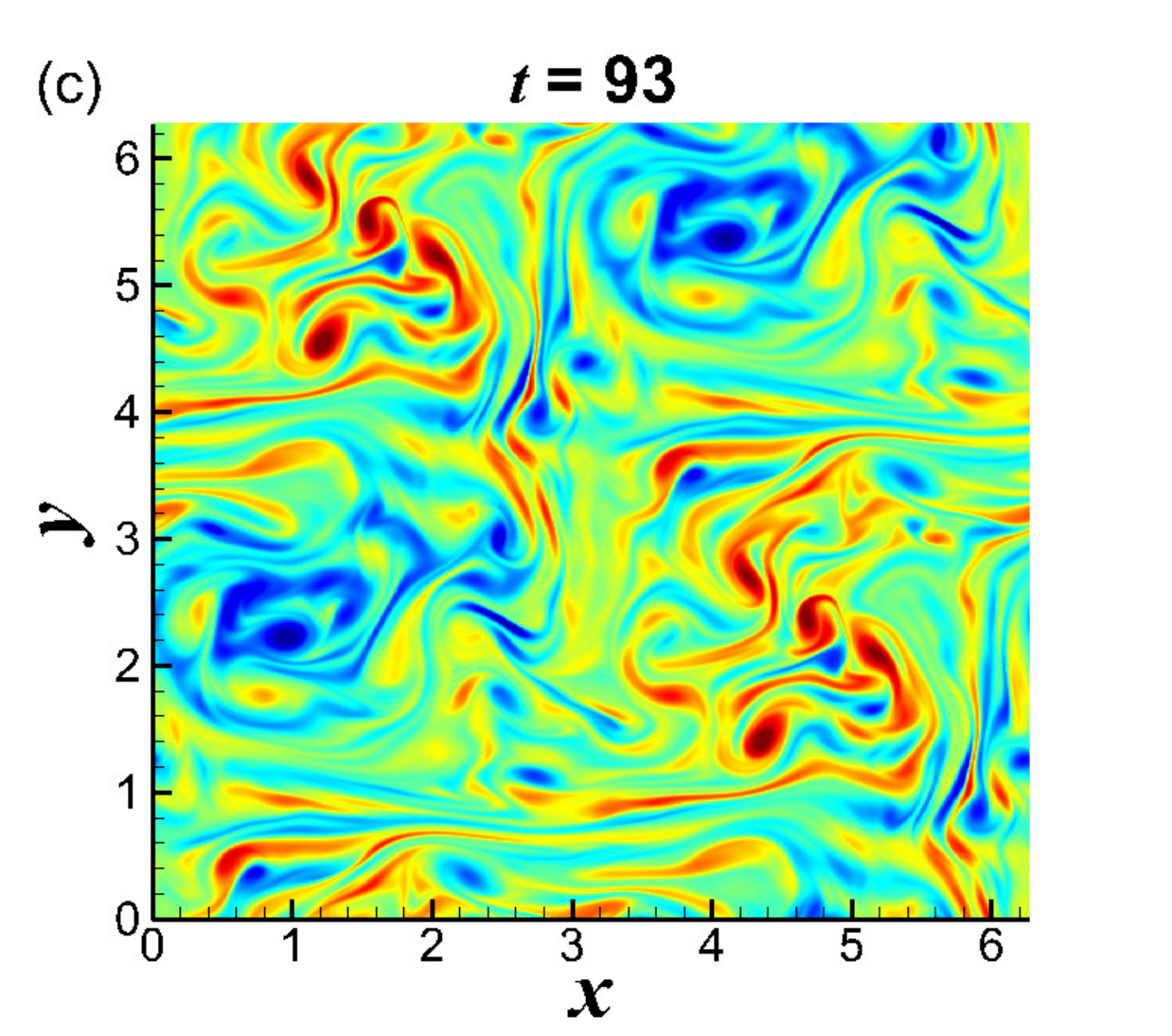}
             \includegraphics[width=2.0in]{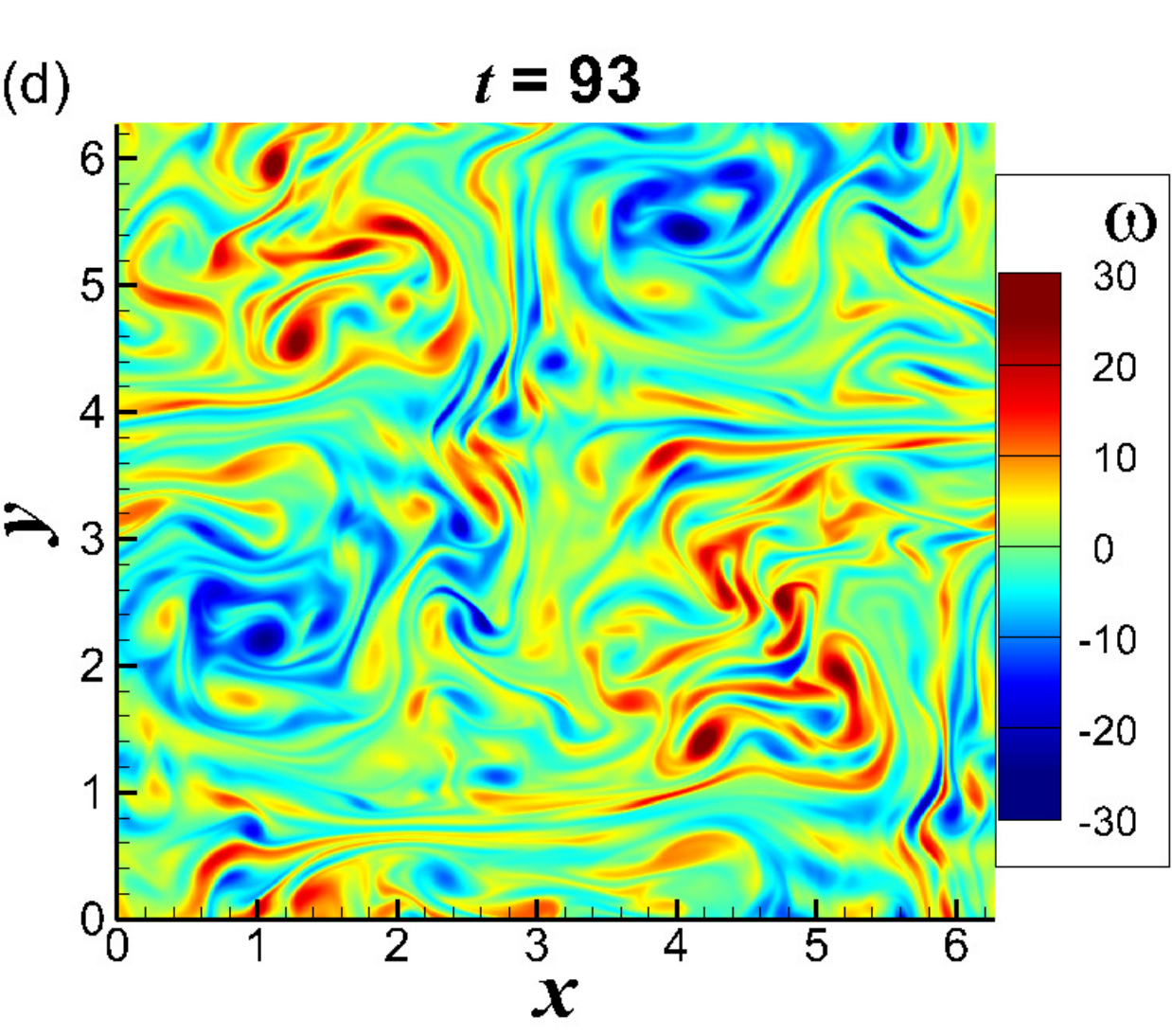}    \\
             \includegraphics[width=2.0in]{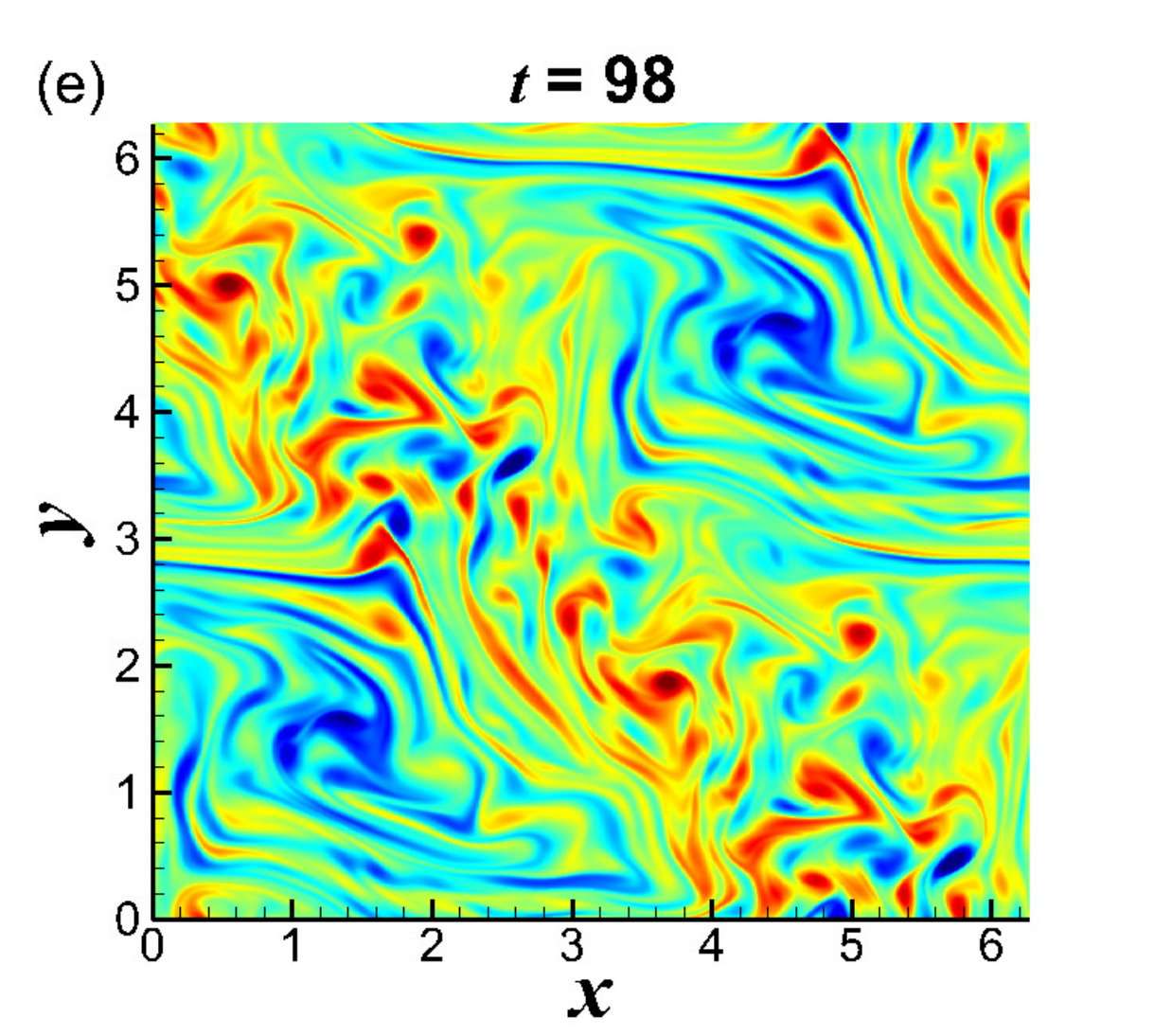}
             \includegraphics[width=2.0in]{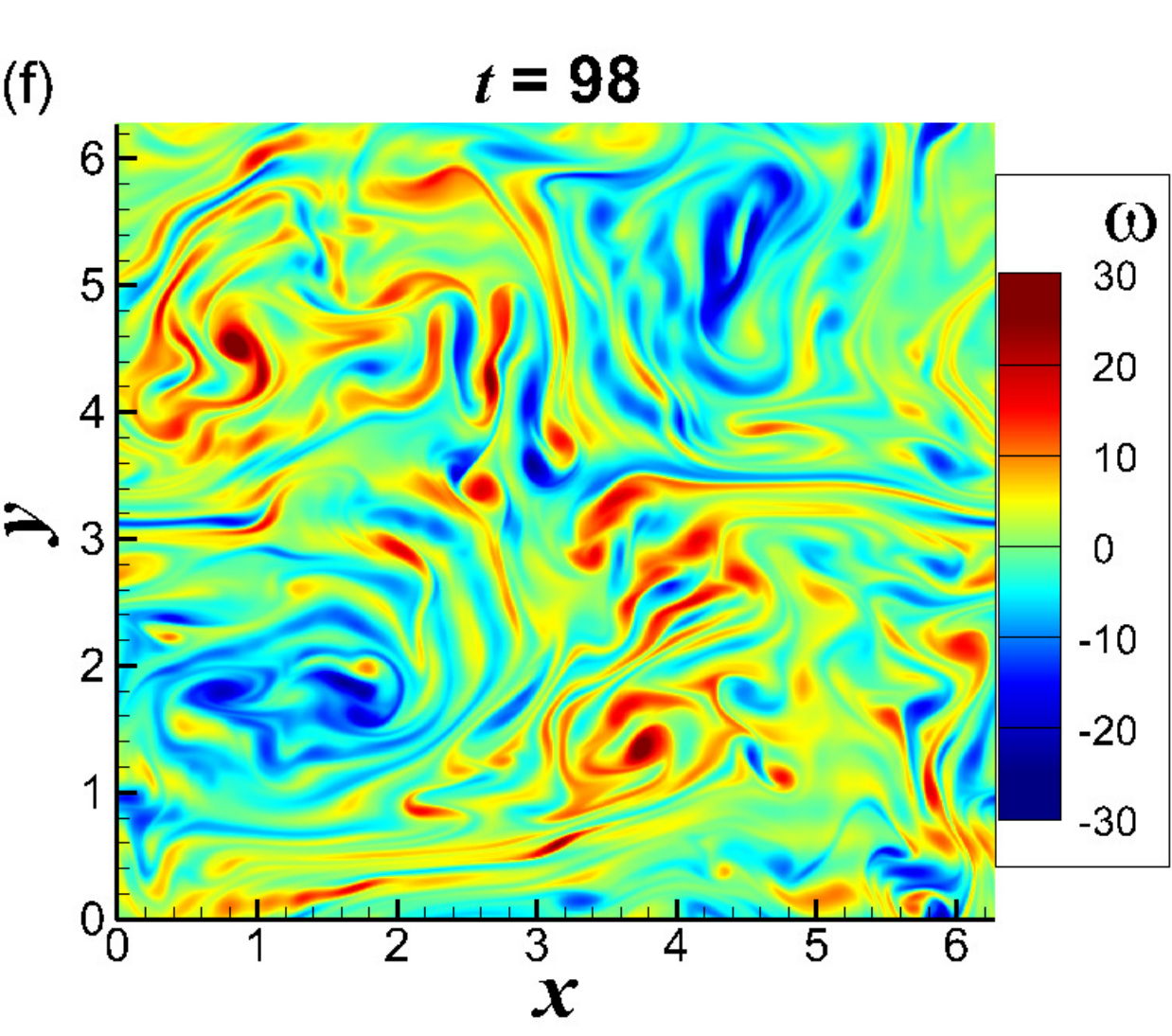}    \\
             \includegraphics[width=2.0in]{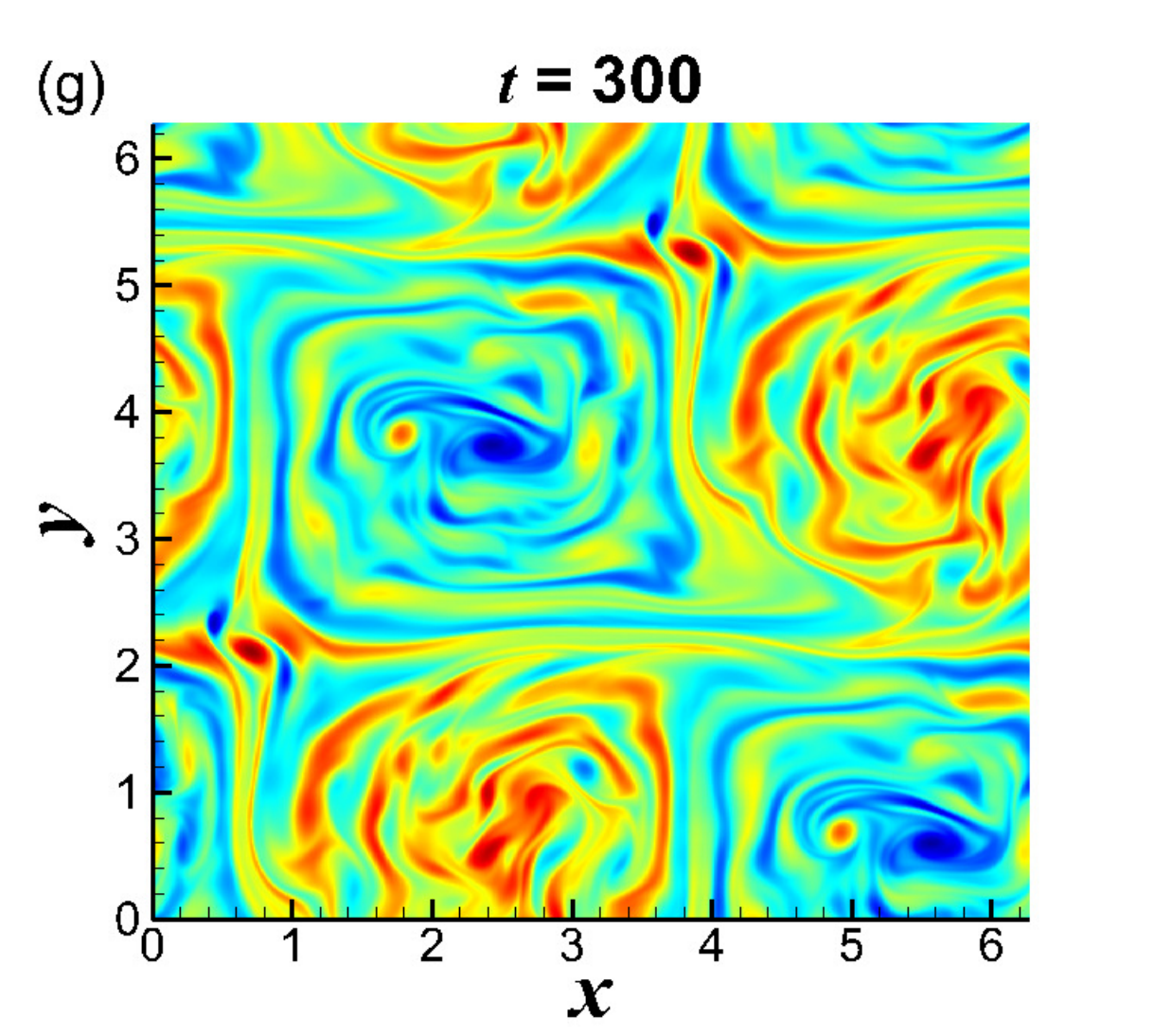}
             \includegraphics[width=2.0in]{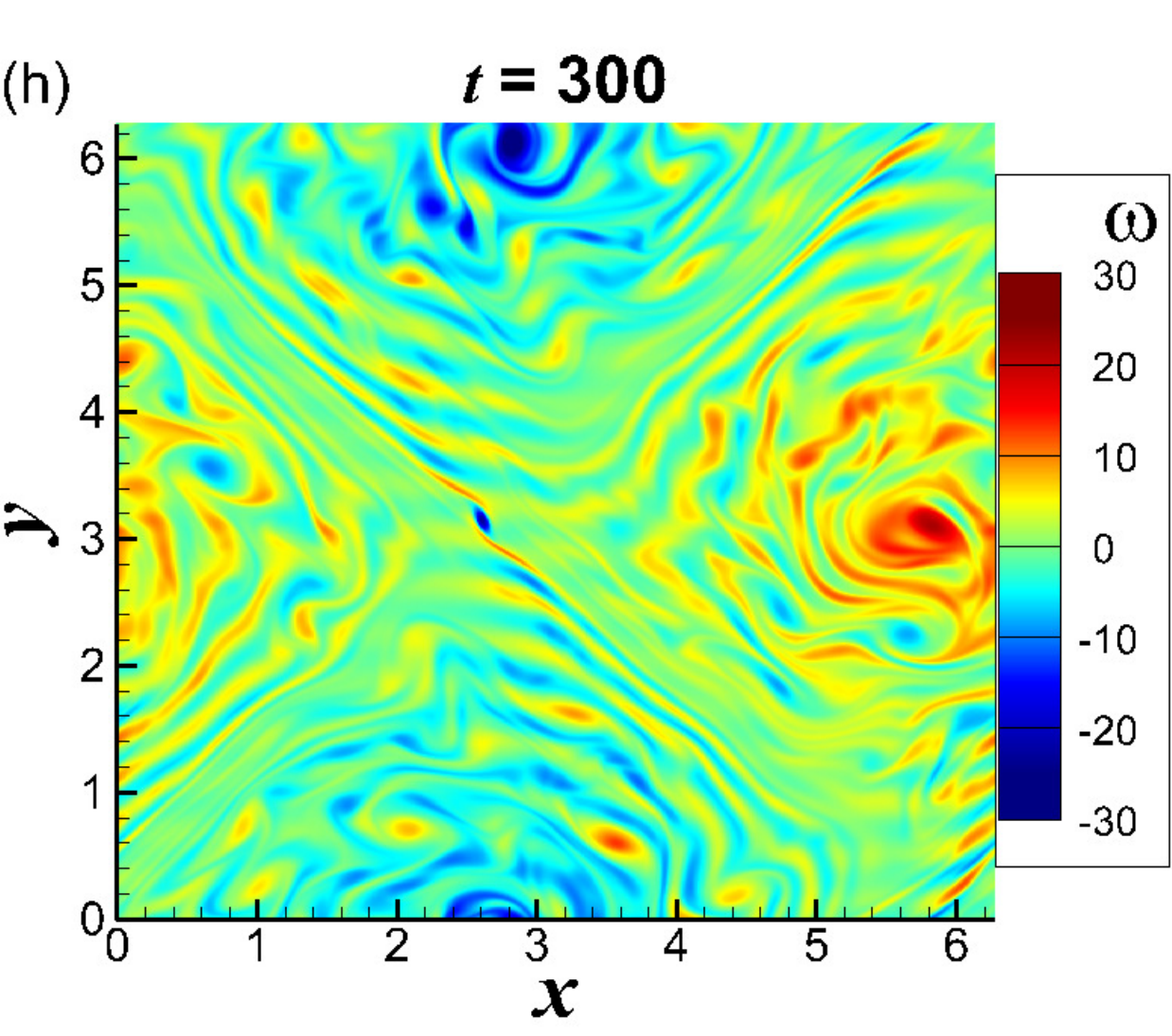}    \\
        \end{tabular}
    \caption{Vorticity fields $\omega(x,y)$ of the 2D turbulent Kolmogorov flow governed by (\ref{eq_psi}) and (\ref{boundary_condition}) for $n_K=16$ and $Re=2000$ given by CNS, subject to either the initial conditions (\ref{initial_condition-11})  (left, marked by Flow CNS$'$) or (\ref{initial_condition-22}) (right, marked by Flow CNS$''$), at different times: (a)-(b) $t=88$, (c)-(d) $t=93$, (e)-(f) $t=98$, and (g)-(h) $t=300$. See the supplementary Movie~2 for the whole evolution process, which can be downloaded via GitHub (\url{https://github.com/sjtu-liao/2D-Kolmogorov-turbulence}).
}     \label{Vor_Evolutions-2}
    \end{center}
\end{figure}

\begin{figure}
    \begin{center}
        \begin{tabular}{cc}
             \includegraphics[width=2.0in]{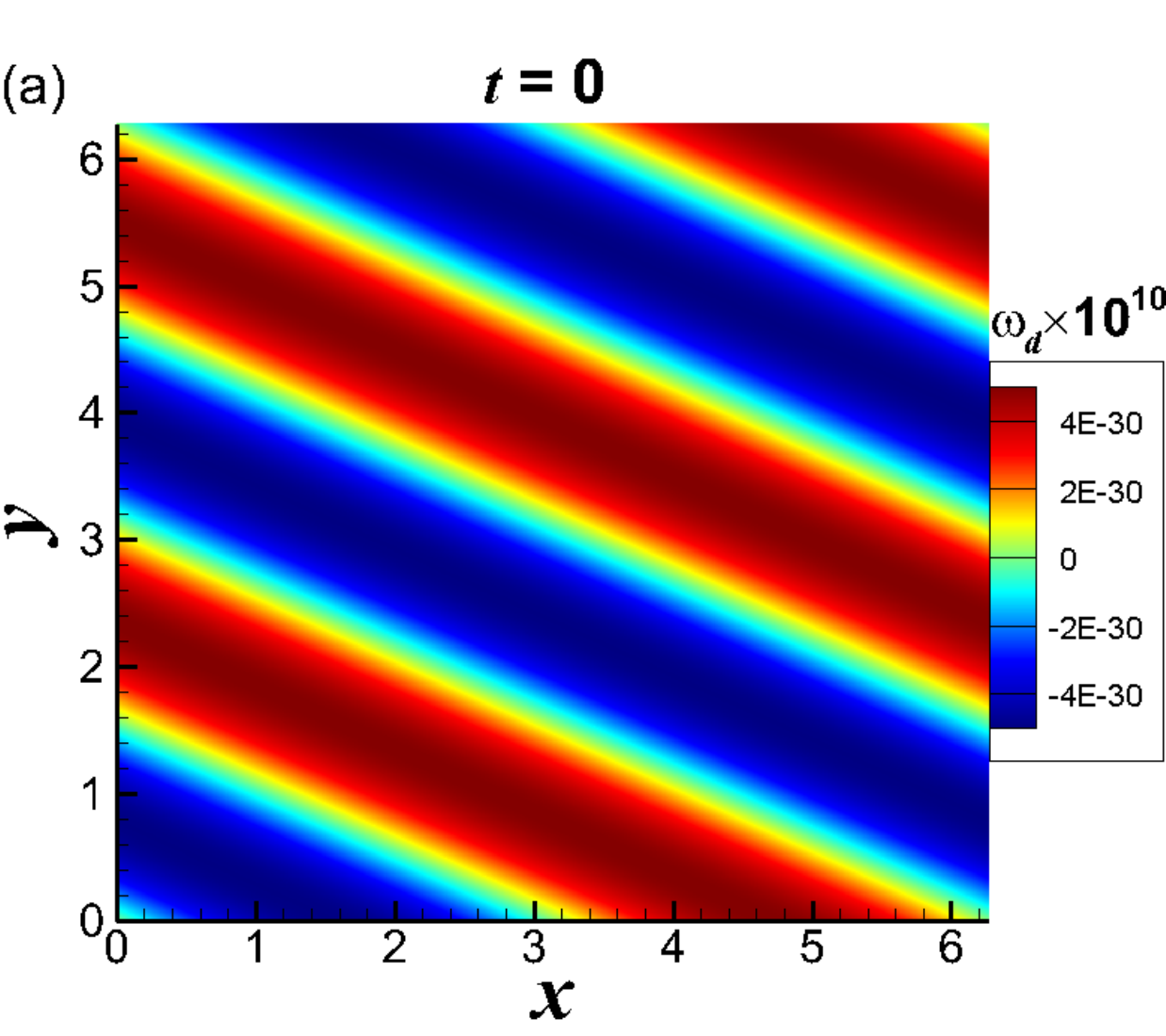}
             \includegraphics[width=2.0in]{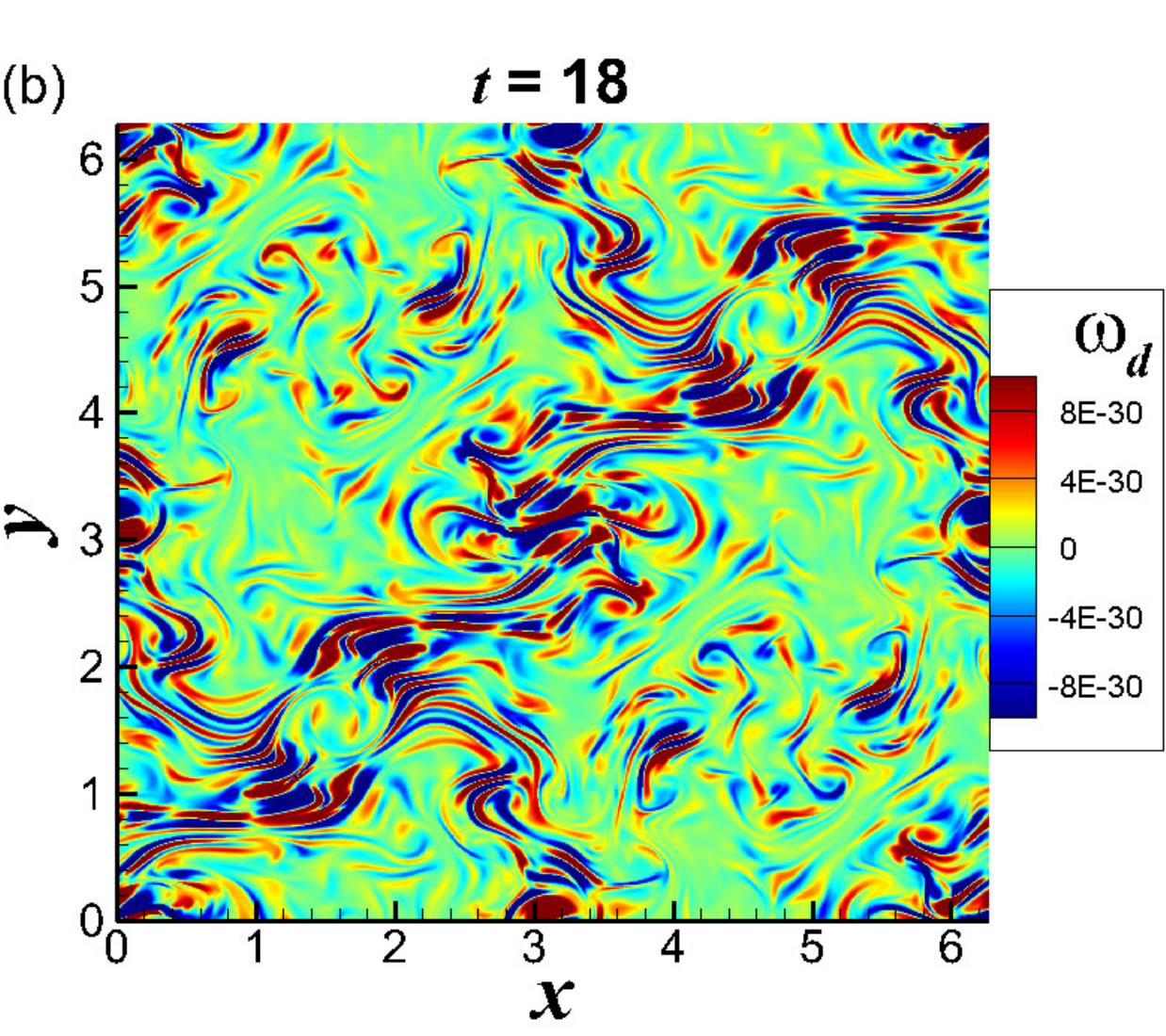}    \\
             \includegraphics[width=2.0in]{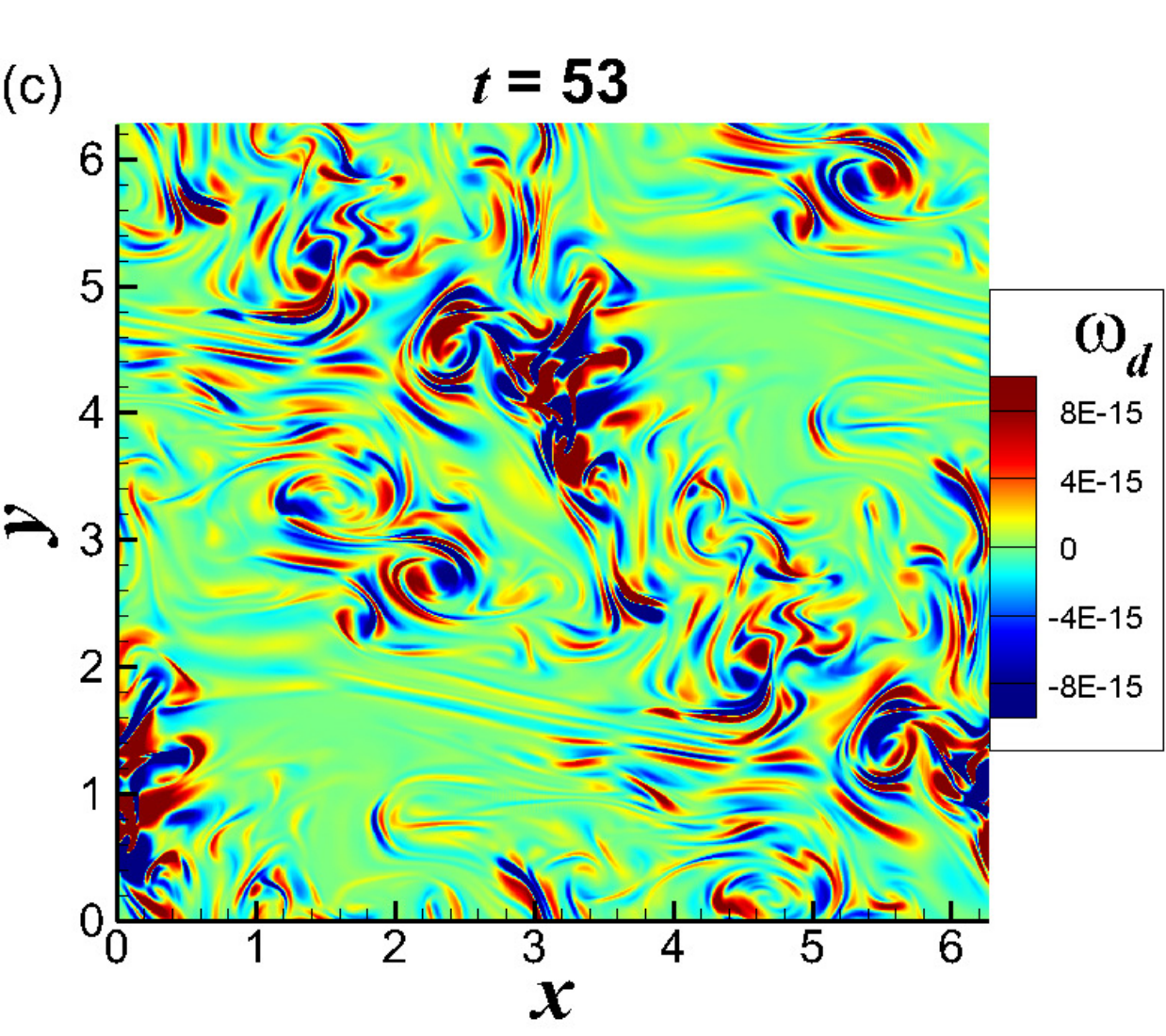}
             \includegraphics[width=2.0in]{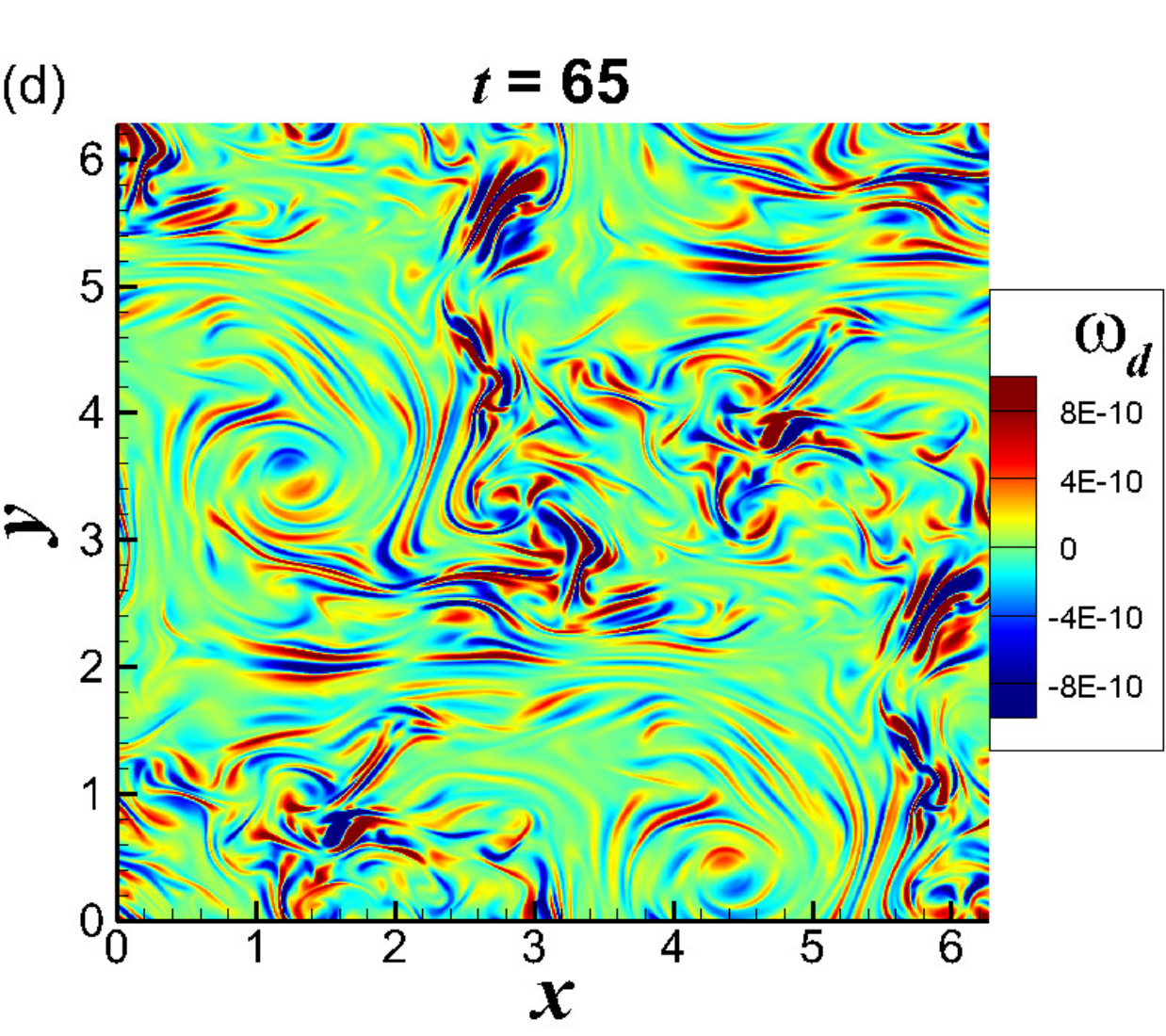}    \\
             \includegraphics[width=2.0in]{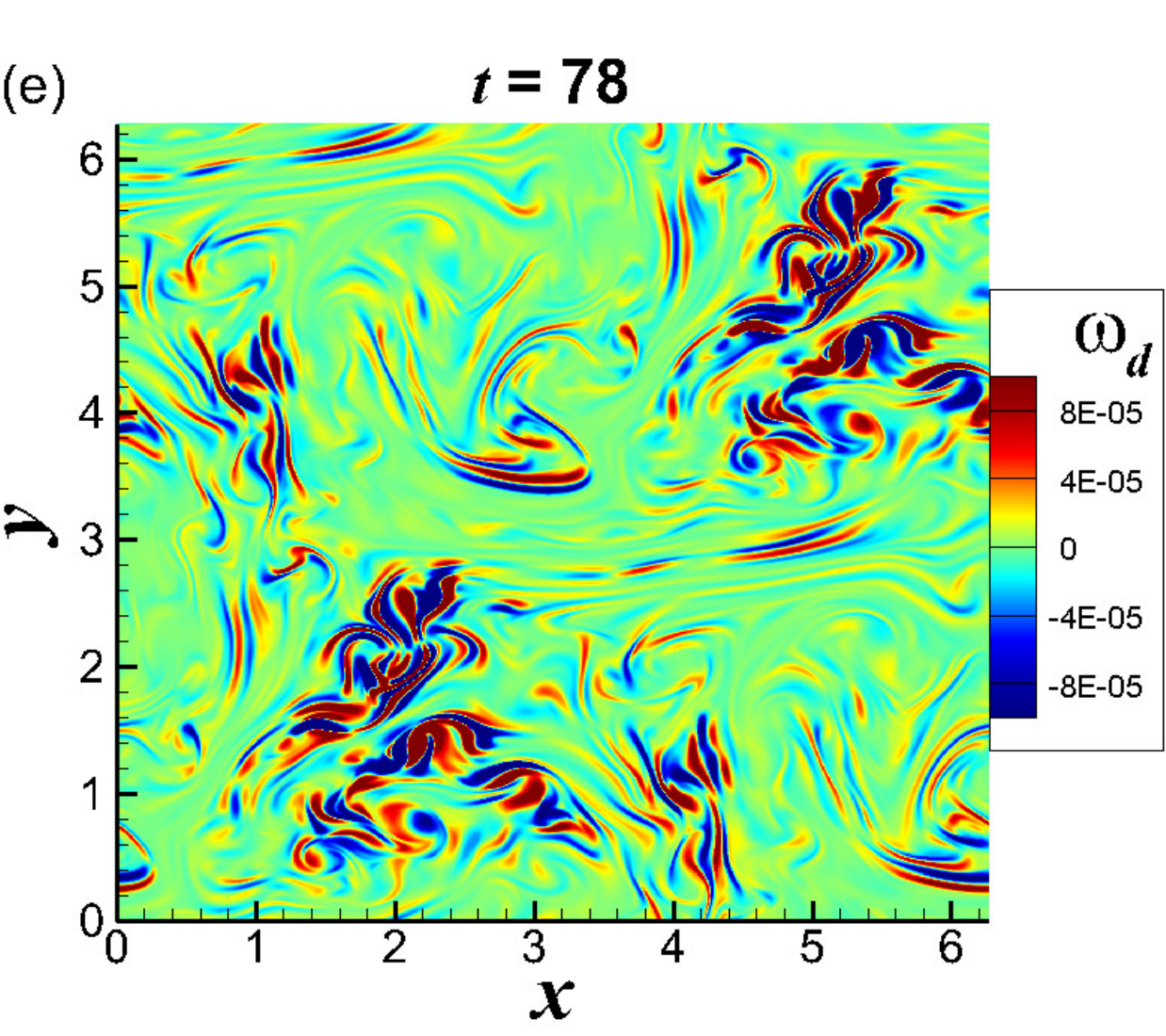}
             \includegraphics[width=2.0in]{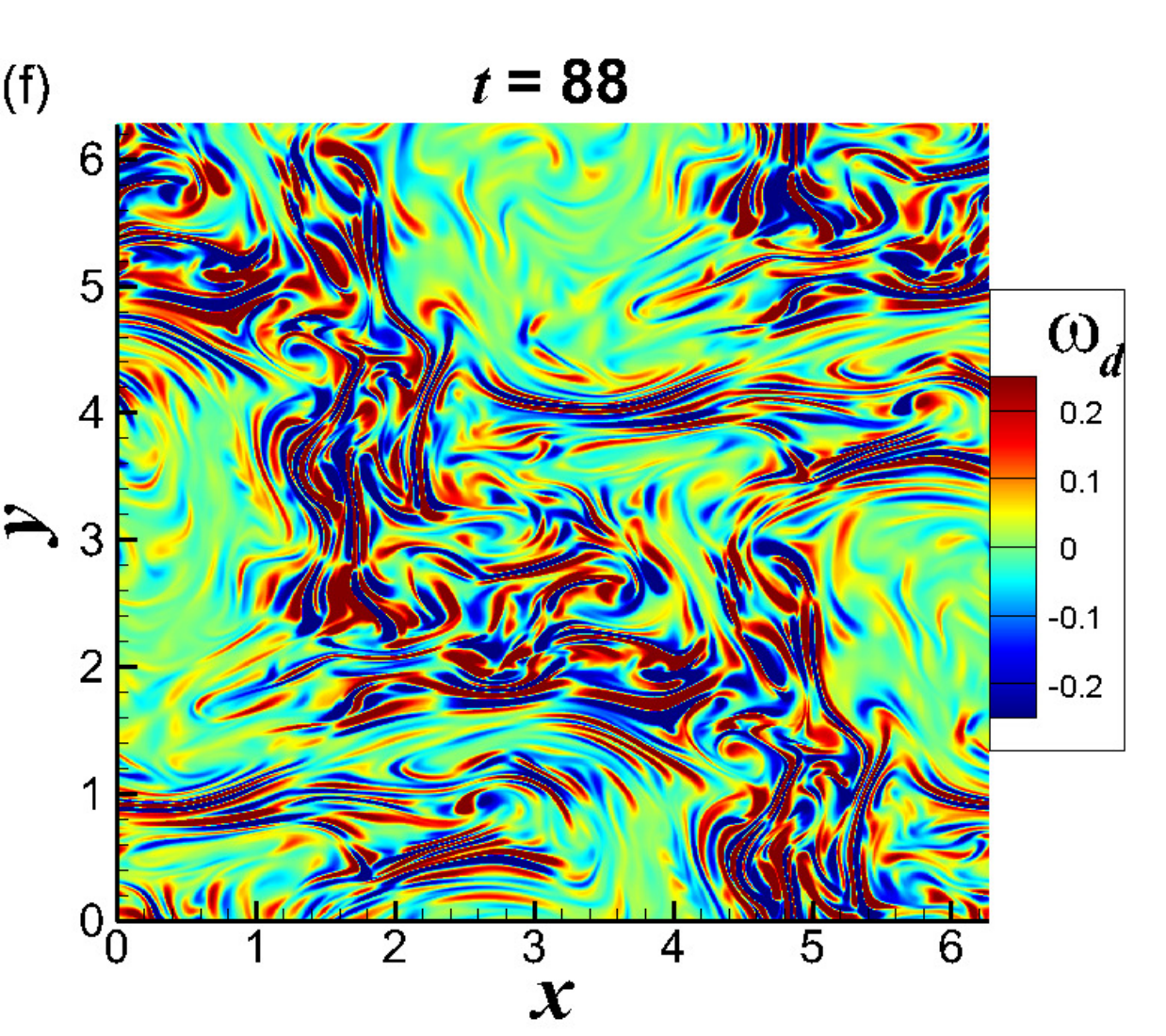}    \\
             \includegraphics[width=2.0in]{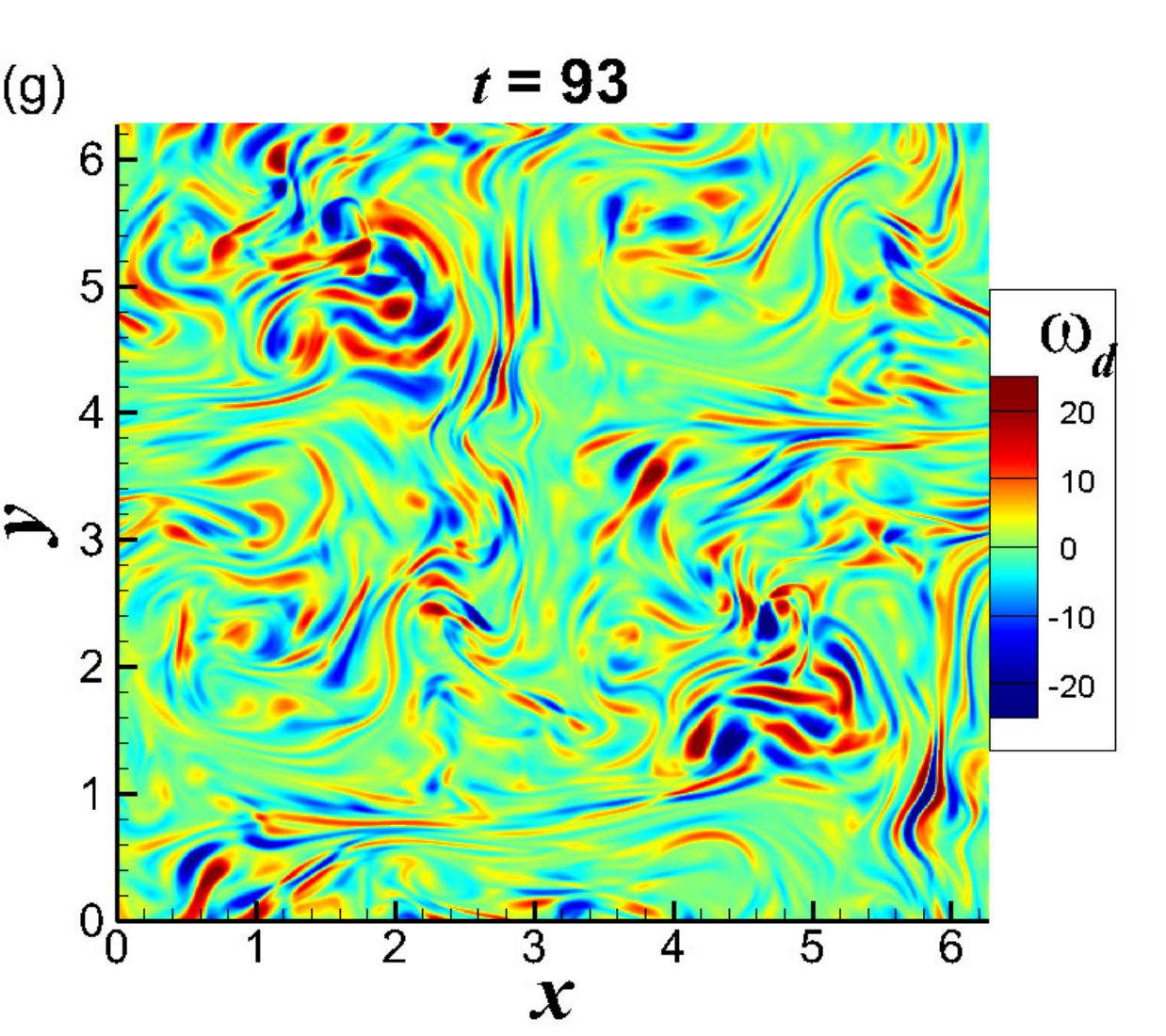}
             \includegraphics[width=2.0in]{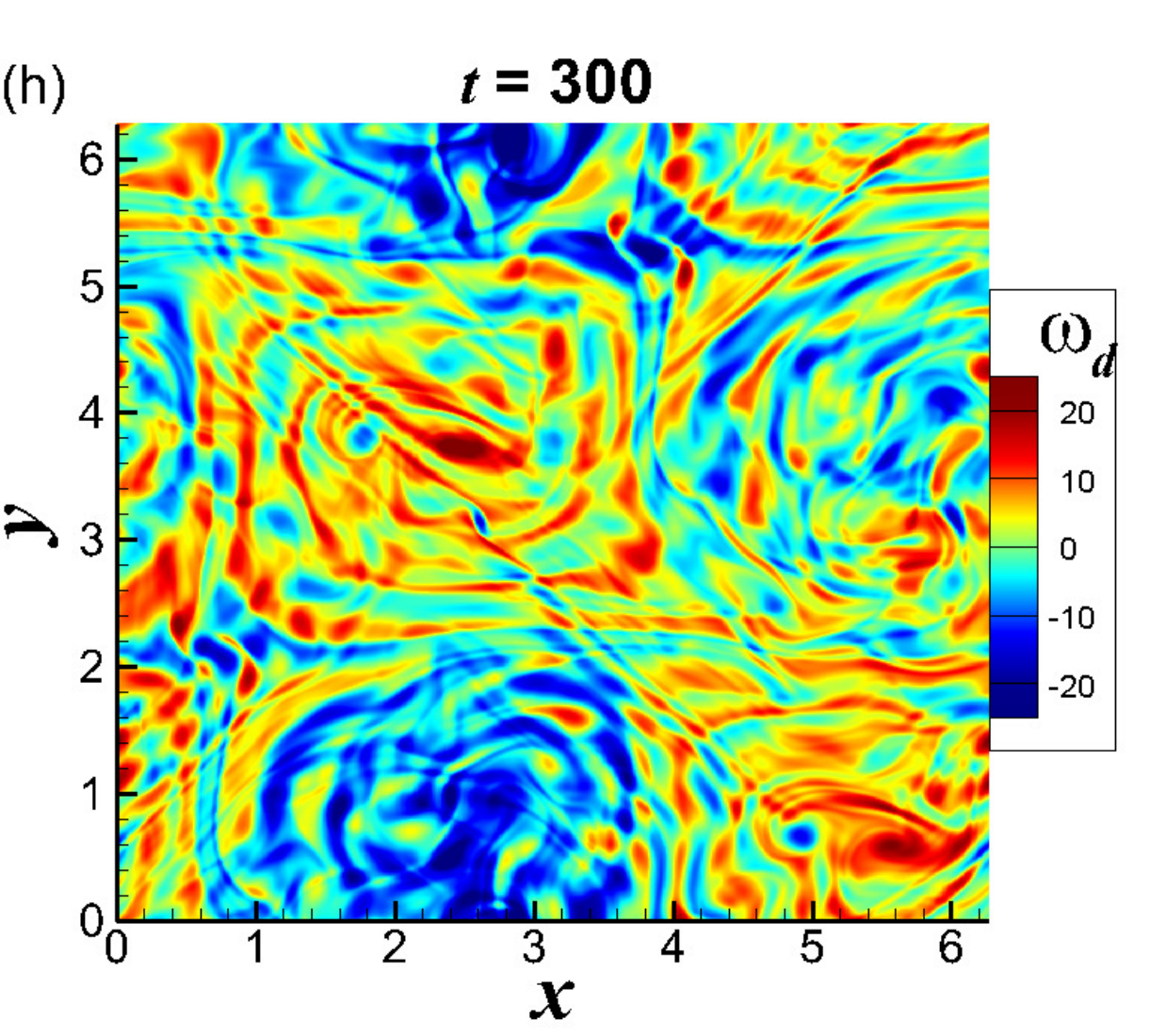}    \\
        \end{tabular}
    \caption{Vorticity fields of the evolution $\delta_{2}(x,y,t)$, corresponding to the second disturbance $10^{-40}\sin(x+2y)$ in the initial condition (\ref{initial_condition-22}) of 2D turbulent Kolmogorov flow governed by (\ref{eq_psi}) and (\ref{boundary_condition}) for $n_K=16$ and $Re=2000$ given by CNS, at the different times: (a) $t=0$, (b) $t=18$, (c) $t=53$, (d) $t=65$, (e) $t=78$, (f) $t=88$, \\(g) $t=93$, (h) $t=300$.}     \label{Vor_Evolutions-2-delta}
    \end{center}
\end{figure}

\begin{figure}
    \begin{center}
                \begin{tabular}{cc}
             \subfigure[]{\includegraphics[width=2.2in]{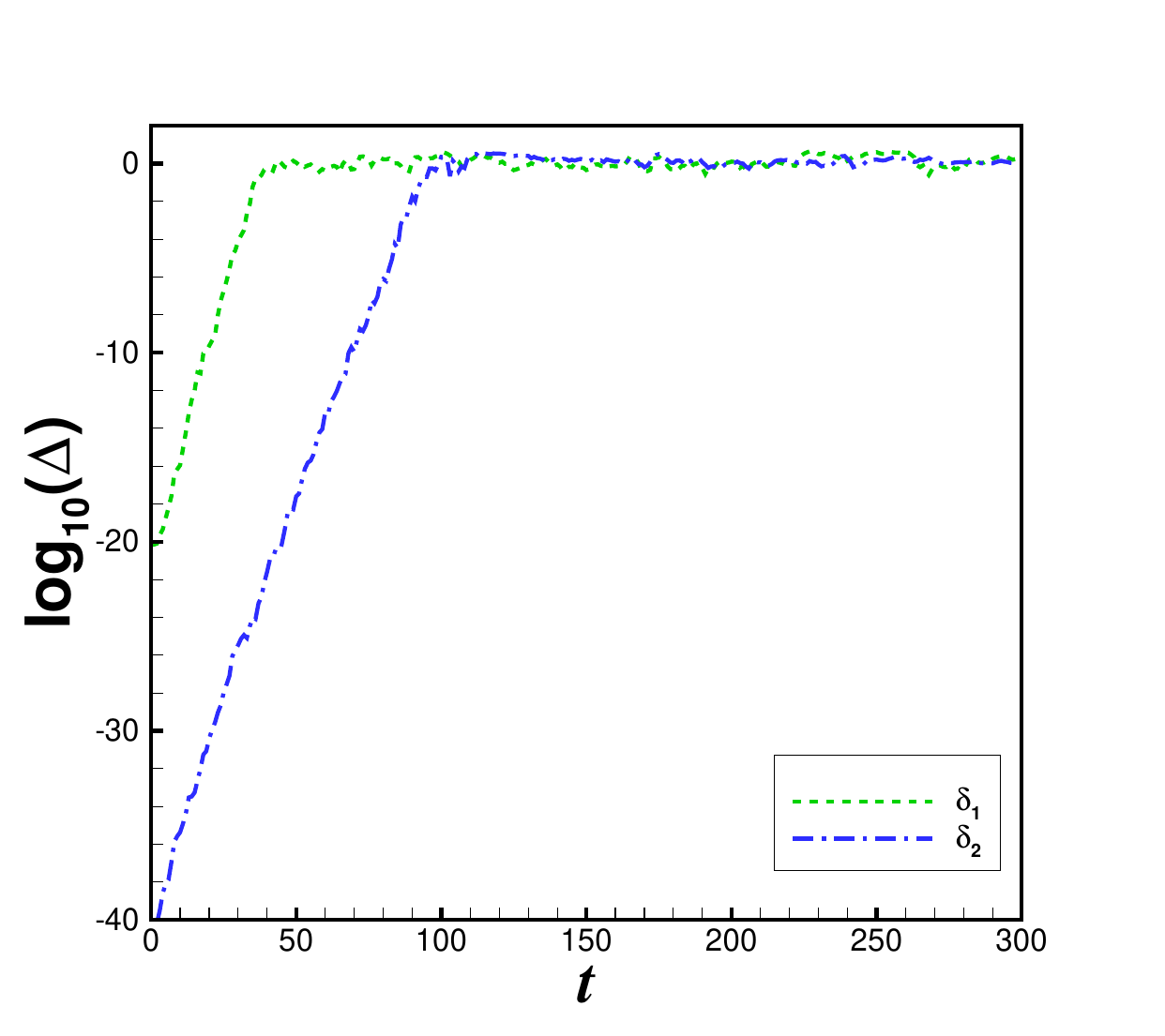}}
             \subfigure[]{\includegraphics[width=2.2in]{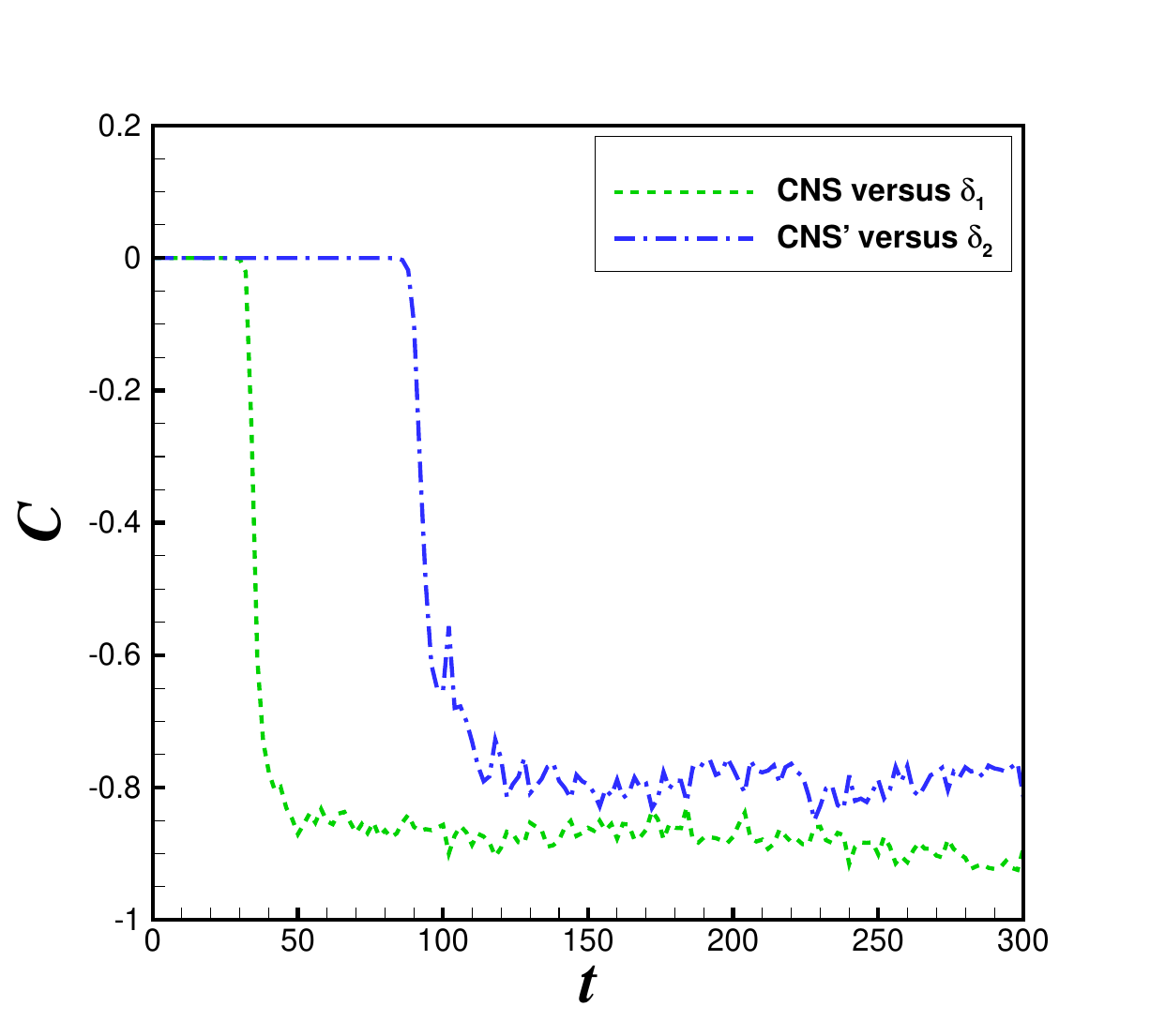}}
        \end{tabular}
    \caption{Time histories of (a) $\Delta = \sqrt{\langle \delta_{i}^{2} \rangle_A}$ with $i=1,2$, where $\delta_{1}(x,y,t)$ denotes the evolution of the first disturbance (green dashed line) and $\delta_{2}(x,y,t)$ is the evolution of the second disturbance   (blue dash-dotted line) of the initial condition (\ref{initial_condition-22}), and (b) the normalized correlation coefficient $C(t)$ of vorticity of Flow CNS versus $\delta_{1}(x,y,t)$ (green dashed line) as well as that of Flow CNS$'$ versus $\delta_{2}(x,y,t)$ (blue dash-dotted line), for the 2D turbulent Kolmogorov flow governed by Eqs.~(\ref{eq_psi}) and (\ref{boundary_condition}) in the case of $n_K=16$ and $Re=2000$ subject to the initial condition (\ref{initial_condition-22}).}     \label{delta}
    \end{center}
\end{figure}

Fig.~\ref{Vor_Evolutions-2} compares the vorticity field $\omega(x,y,t)$ of Flow CNS$'$, subject to the initial condition (\ref{initial_condition-11}), with that of Flow CNS$''$, subject to the initial condition (\ref{initial_condition-22}). Traditionally, compared with the first disturbance $10^{-20} \sin(x+y)$ in the initial condition (\ref{initial_condition-22}), the second disturbance $10^{-40} \sin(x+2y)$ could be neglected completely, since it is 20 orders of magnitude smaller.
However, this traditional viewpoint is in fact {\em wrong}: Flow CNS$''$ looks like the same as Flow CNS from the beginning until $t\approx 35$ when it loses the spatial symmetry (\ref{symmetry-omega:A}) but has the new spatial symmetry (\ref{symmetry_psi:B}) instead, indicating that  $\delta_1(x,y,t)$ corresponding to the first disturbance $10^{-20} \, \sin(x+y)$ increases to macro-level $O(1)$ and thus triggers the transition of the spatial symmetry from (\ref{symmetry-omega:A}) to (\ref{symmetry_psi:B}). More importantly, it is found that Flow CNS$''$ appears the same as Flow CNS$'$ from the beginning, for example at $t\approx 88$ as shown in Fig.~\ref{Vor_Evolutions-2}(a) and (b), until $t \approx  93$ when it loses the spatial symmetry (\ref{symmetry_psi:B}), as shown in Fig.~\ref{Vor_Evolutions-2}(c) and (d). Note that Flow CNS$''$ completely loses spatial symmetry after $t \geq 98$, as shown in Fig.~\ref{Vor_Evolutions-2}(e)-(h), clearly indicating that the evolution $\delta_2(x,y,t)$, corresponding to the second micro-level disturbance $10^{-40} \sin(x+2y)$,  must increase to a macro-level $O(1)$ and finally destroys the spatial symmetry. 
Note that the term $\sin(x+2y)$ has {\em no} spatial symmetry in rotation and translation, since $\sin(x+2y) \neq \sin(2\pi-x+4\pi-2y)$ and $\sin(x+2y) \neq \sin(\pi+x+2\pi +2y)$. So, the reason is very clear from Fig.~\ref{Vor_Evolutions-2-delta}:  $\delta_2(x,y,t)$, corresponding to the second disturbance $10^{-40} \, \sin(x+2 y)$ in the initial condition (\ref{initial_condition-22}), increases from a micro-level, step by step, to a macro-level at $t \approx 93$, which has {\em no} spatial symmetry throughout the whole time interval $t\in[0,300]$ so that it destroys the  spatial symmetry (\ref{symmetry_psi:B}) at $t\approx 93$ when it reaches a macro-level $O(1)$.    

Let us focus on the initial condition (\ref{initial_condition-22}): the second disturbance $10^{-40} \sin(x+2y)$ is 20 orders of magnitude smaller than the first disturbance $10^{-20} \sin(x+y)$. From the traditional viewpoint, the second disturbance should be negligible compared to the first one. However, on the contrary, {\em both} the first disturbance $10^{-20} \sin(x+y)$ and the second disturbance $10^{-40} \sin(x+2y)$ to the initial condition  (\ref{initial_condition-22}) enlarge {\em separately}, one by one in an inverse cascade, to macro-level $O(1)$: first the former triggers the transformation of the spatial symmetry from (\ref{symmetry-omega:A}) to (\ref{symmetry_psi:B}) at $t\approx 35$ and then the latter totally destroys the spatial symmetry of the 2D turbulent Kolmogorov flow at $t\approx 93$. This is very clear from Fig.~\ref{Vor_Evolutions-1-delta} for the evolution $\delta_1(x,y,t)$ of the first disturbance $10^{-20} \sin(x+y)$ and Fig.~\ref{Vor_Evolutions-2-delta} for the evolution $\delta_2(x,y,t)$ of the second disturbance $10^{-40} \sin(x+2 y)$ of the initial condition~(\ref{initial_condition-22}).  

Fig.~\ref{Vor_Evolutions-1-delta} shows that the vorticity field caused by $\delta_1(x,y,t)$ corresponding to the first disturbance $10^{-20}\sin(x+y)$ increases from a micro-order $O(10^{-20})$ of magnitude  
at $t=0$, step by step, to the order $O(10^{-15})$ at $t=8$, $O(10^{-10})$ at $t=15$, $O(10^{-5})$ at $t=25$, $O(10^{-2})$ at $t=30$,  until to a macro-order $O(10)$ at $t=35$.
Fig.~\ref{Vor_Evolutions-2-delta} shows that the vorticity field caused by $\delta_2(x,y,t)$ corresponding to the second disturbance $10^{-40}\sin(x+2y)$ increases from a micro-order $O(10^{-40})$ of magnitude  
at $t=0$, step by step, to the order $O(10^{-30})$ at $t=18$, $O(10^{-15})$ at $t=53$, $O(10^{-10})$ at $t=65$, $O(10^{-5})$ at $t=78$, $O(10^{-1})$ at $t=88$, until to a macro-order $O(10)$ at $t=93$. 
All of these at {\em different} orders of magnitudes often coexist with macroscopic flow field.   

Write $\Delta = \sqrt{\langle \delta_{i}^{2} \rangle_A}$ with $i=1,2$, where $\delta_{1}(x,y,t)$ and $\delta_{2}(x,y,t)$ denote the evolutions of the first disturbance $10^{-20}\sin(x+y)$ and the second disturbance $10^{-40}\sin(x+2y)$ to the initial condition~(\ref{initial_condition-22}), respectively, and $\langle \; \rangle_A$ is an operator of statistics defined in Appendix. As shown in Fig.~\ref{delta}(a), $\sqrt{\langle \delta_{1}^{2} \rangle_A}$ exponentially expands until $t\approx 35$ when it reaches a macro-level. i.e. $\sqrt{\langle \delta_{1}^{2} \rangle_A} \sim O(1)$. Similarly, $\sqrt{\langle \delta_{2}^{2} \rangle_A}$ exponentially expands until $t\approx 93$ when it is at a macro-level. i.e. $\sqrt{\langle \delta_{2}^{2} \rangle_A} \sim O(1)$. As mentioned before, Flow CNS$'$ is equal to Flow CNS plus $\delta_{1}(x,y,t)$, and Flow CNS$''$ is equal to Flow CNS$'$ plus $\delta_{2}(x,y,t)$.  
As shown in Fig.~\ref{delta}(b), the normalized correlation coefficient of vorticity of Flow CNS versus $\delta_{1}$ is very small from the beginning to $t\approx 32$, indicating that there is {\em no} correlation between them because $\delta_{1}(x,y,t)$ is negligible compared to Flow CNS, until $t\approx 35$ when their correlation suddenly becomes strong, indicating that $\delta_{1}(x,y,t)$ is at the same order of magnitude as Flow CNS and thus is {\em not} negligible thereafter. Similarly, the normalized correlation coefficient of vorticity of Flow CNS$'$ versus $\delta_{2}(x,y,t)$ is very small from the beginning to $t\approx 90$, indicating there is {\em no} correlation between them because $\delta_{2}(x,y,t)$ is negligible compared to Flow CNS$'$, until $t\approx 93$ when their correlation suddenly becomes strong, indicating that $\delta_{2}(x,y,t)$ is at the same order of magnitude as Flow CNS$'$ and thus is {\em not} negligible thereafter.       

The foregoing provide rigorous evidence that {\em all} disturbances at {\em different} orders of magnitudes to the initial condition of the NS equations increase {\em separately}, say, one by one like an inverse cascade, to macro-level, with each capable of completely altering the characteristics  (such as the vorticity spatial  symmetry) of the turbulent flow considered in this paper.  
Based on this very interesting phenomenon revealed by our clean numerical experiments mentioned above, we propose a new concept, which we call the ``noise-expansion cascade'', that closely connects the randomness of micro-level noises/disturbances to the macro-level disorder of turbulence.    

\subsection{Influence of noise-expansion cascade on statistics}

\begin{figure}
    \begin{center}
        \begin{tabular}{cc}
             \subfigure[]{\includegraphics[width=2.2in]{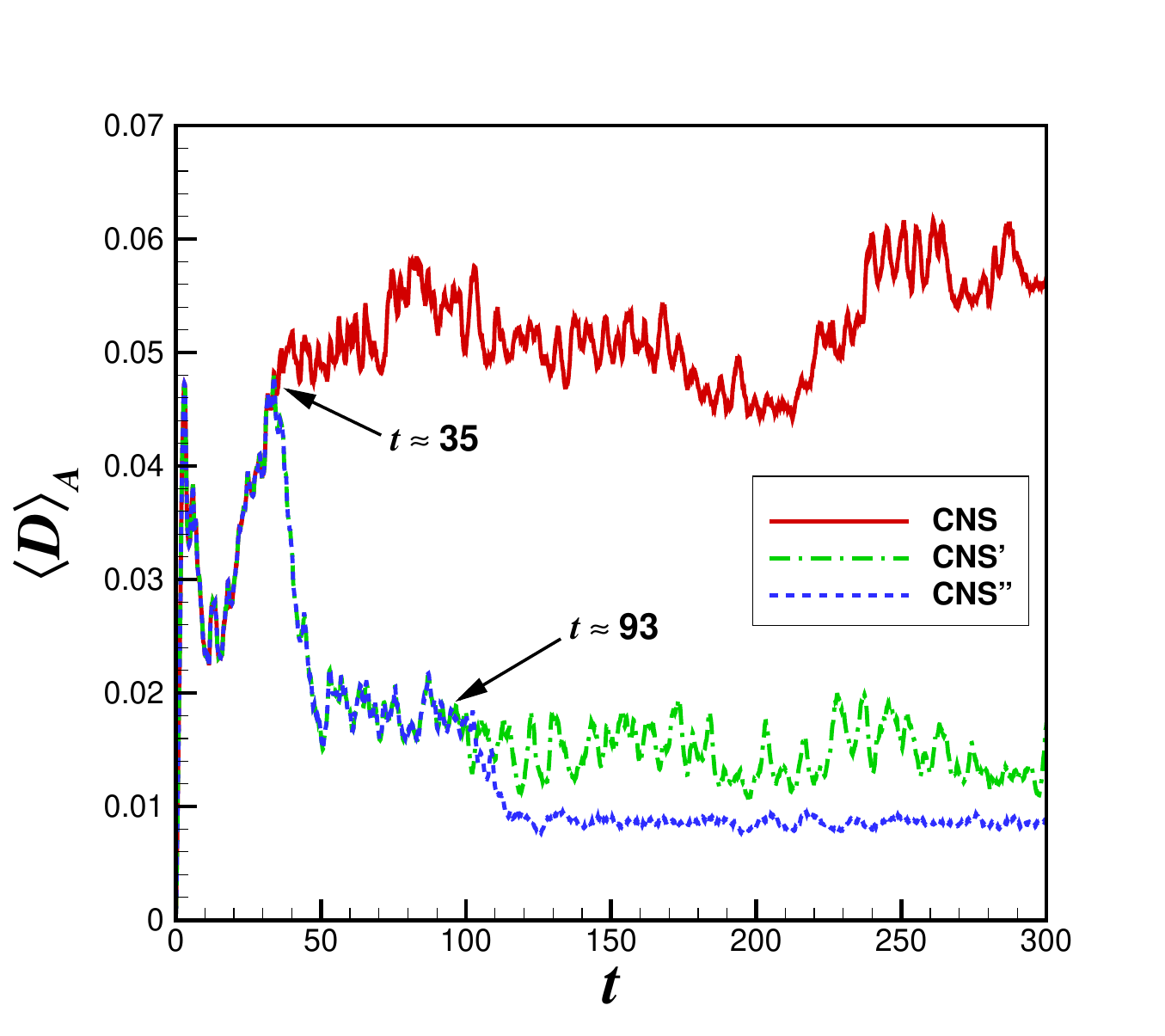}}
             \subfigure[]{\includegraphics[width=2.2in]{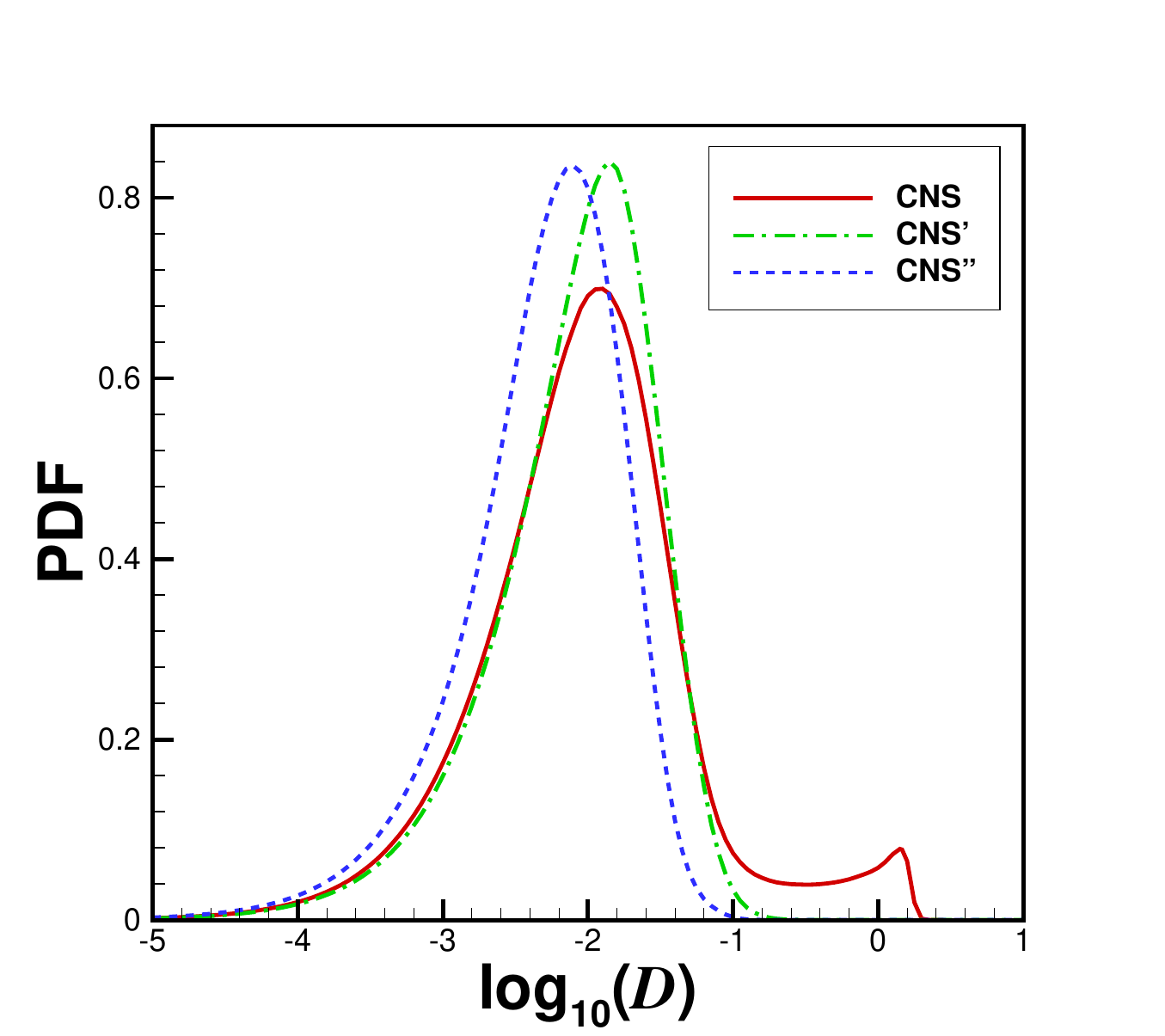}}
        \end{tabular}
    \caption{Comparisons of (a) time histories of the spatially averaged kinetic energy dissipation rate $\langle D\rangle_A$ and (b) the probability density function (PDF) of the kinetic energy dissipation $D(x,y,t)$ of the 2D turbulent Kolmogorov flow, governed by Eqs.~(\ref{eq_psi}) and (\ref{boundary_condition}) for $n_K=16$ and $Re=2000$, given by Flow CNS subject to the initial condition (\ref{initial_condition}) (solid line in red), Flow CNS$'$ subject to the initial condition (\ref{initial_condition-11}) (dash-dot line in green) and Flow CNS$''$ subject to the initial condition (\ref{initial_condition-22}) (dashed line in blue).}     \label{ED-PDF_P}
    \end{center}
\end{figure}

\begin{figure}
    \begin{center}
        \begin{tabular}{cc}
             \subfigure[]{\includegraphics[width=2.2in]{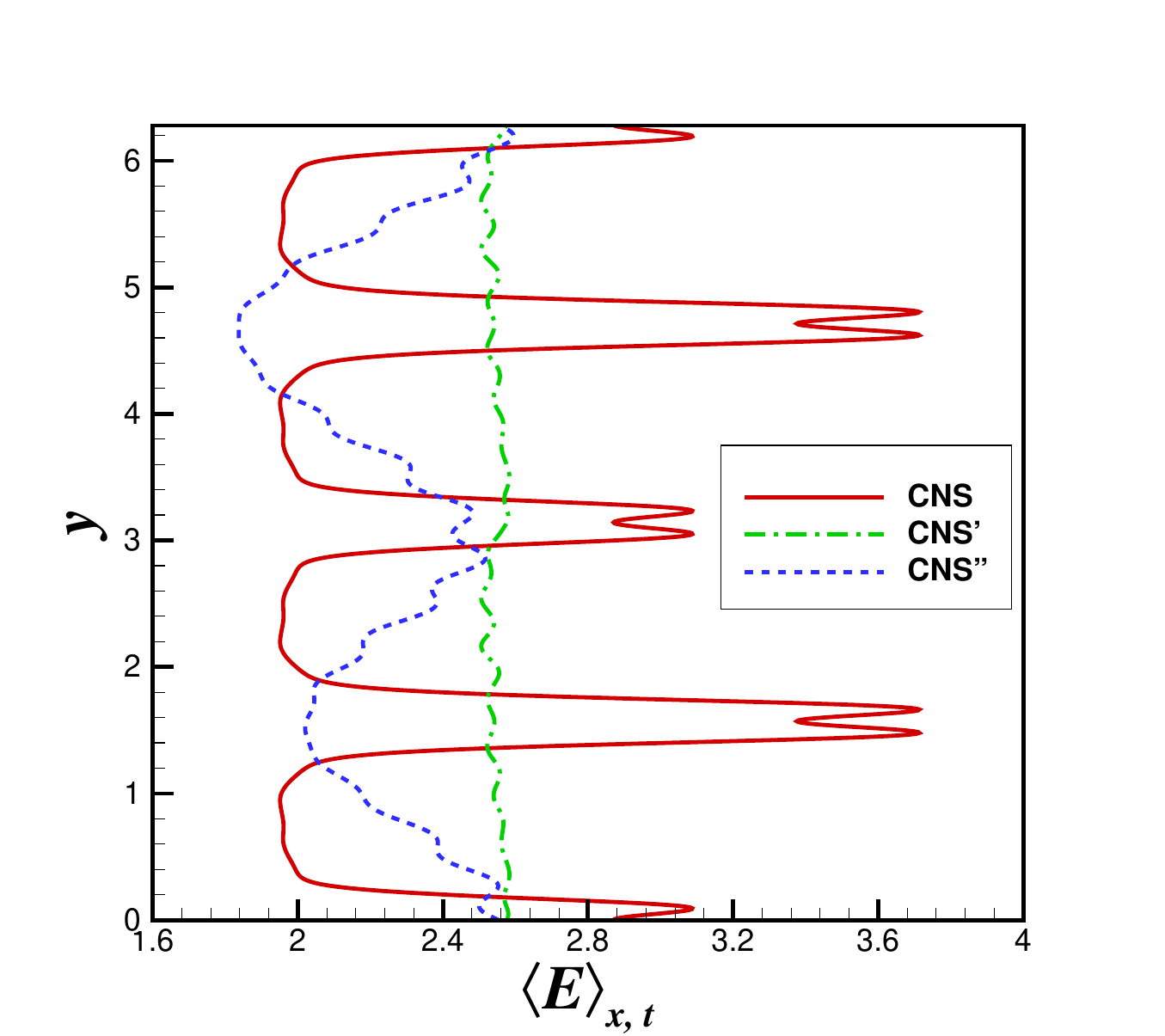}}
             \subfigure[]{\includegraphics[width=2.2in]{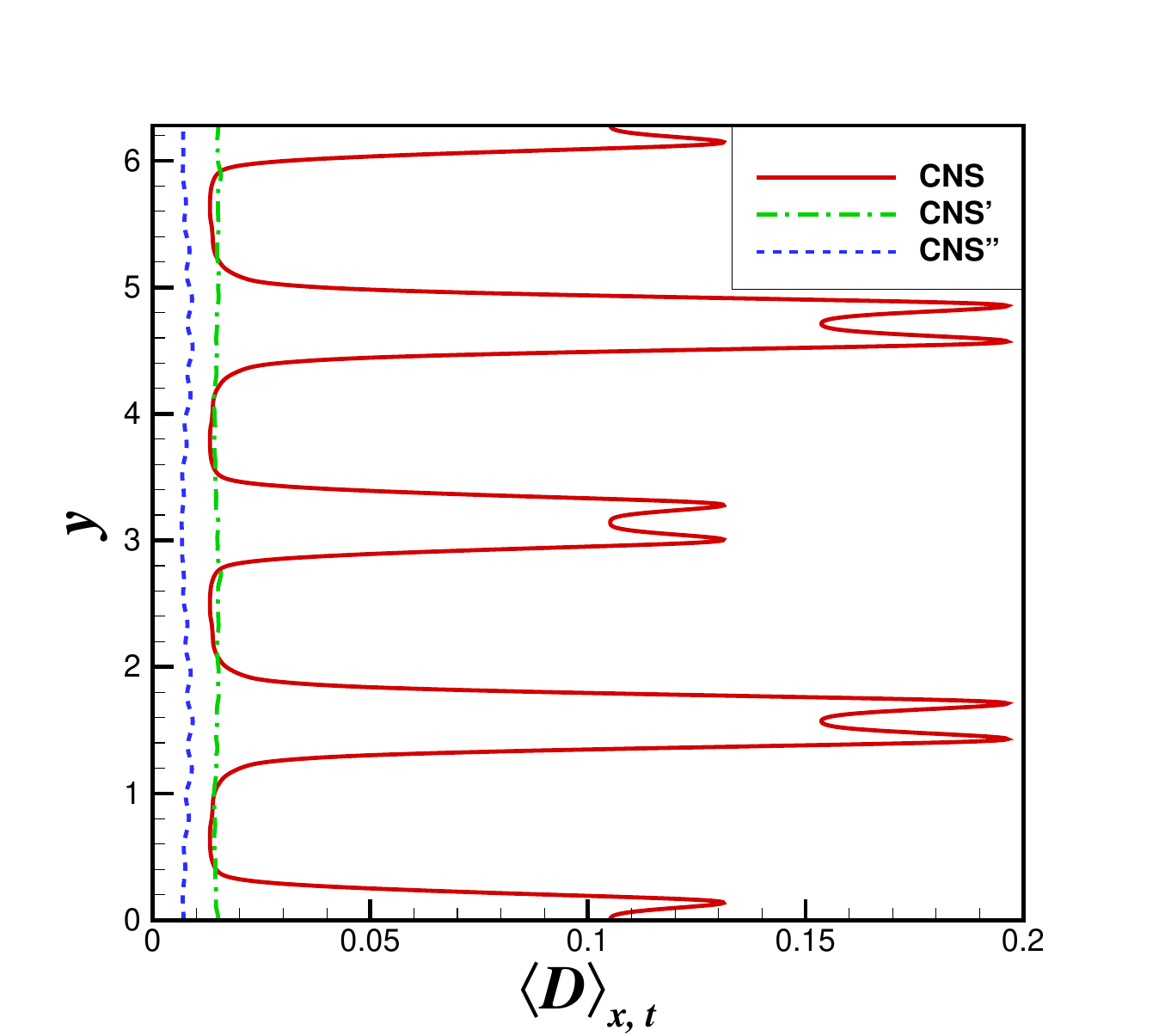}}
        \end{tabular}
    \caption{Comparisons of (a) the spatio-temporal averaged kinetic energy $\langle E\rangle_{x,t}(y)$ and (b) the spatio-temporal averaged kinetic energy dissipation rate $\langle D\rangle_{x,t}(y)$ of the 2D turbulent Kolmogorov flow, governed by Eqs.~(1) and (2) for $n_K=16$ and $Re=2000$, given by Flow CNS subject to the initial condition (3) (solid line in red), Flow CNS$'$ subject to the initial condition (\ref{initial_condition-11}) (dash-dot line in green) and Flow CNS$''$ subject to the initial condition (\ref{initial_condition-22}) (dashed line in blue).}     \label{ED_y_P}
    \end{center}
\end{figure}

Fig.~\ref{ED-PDF_P}(a) compares time histories of the spatially averaged kinetic energy dissipation rate $\langle D\rangle_A$ of Flow CNS, Flow CNS$'$ and Flow CNS$''$, where $\langle \; \rangle_A$ is an operator of statistics defined in the Appendix.    
The distinct deviation in $\langle D\rangle_A$ between Flow CNS and Flow CNS$'$ appears at $t\approx35$ when the evolution $\delta_1(x,y,t)$ of the first disturbance $10^{-20}\sin(x+y)$ increases to a macro-level $O(1)$ that finally destroys the spatial symmetry (\ref{symmetry-omega:A}) and triggers the transformation of the spatial symmetry from (\ref{symmetry-omega:A}) to (\ref{symmetry_psi:B}).
Besides, the distinct deviation in $\langle D\rangle_A$ of Flow CNS$'$ versus Flow CNS$''$ appears at $t \approx 93$ when the evolution $\delta_2(x,y,t)$ of the second disturbance $10^{-40}\sin(x+2 y)$ increases to a macro-level $O(1)$ that finally destroys all spatial symmetry. As shown in Fig.~\ref{ED-PDF_P}(a), when $t > 110$, the spatially averaged kinetic energy dissipation rate $\langle D\rangle_A$ of Flow CNS is much larger than those of Flow CNS$'$ and Flow CNS$''$. This leads to the obvious deviation between their probability density functions (PDFs), as shown in Fig.~\ref{ED-PDF_P}(b), respectively.        
Fig.~\ref{ED_y_P} compares the spatio-temporal averaged kinetic energy $\langle E\rangle_{x,t}(y)$ and the spatio-temporal averaged kinetic energy dissipation rate $\langle D\rangle_{x,t}(y)$ of Flow CNS, Flow CNS$'$ and Flow CNS$''$, where $\langle \; \rangle_{x,t}$ is an operator of statistics defined in the Appendix. Note that these statistics also exhibit obvious deviations.

\begin{figure}
    \begin{center}
        \begin{tabular}{cc}
             \subfigure[]{\includegraphics[width=2.2in]{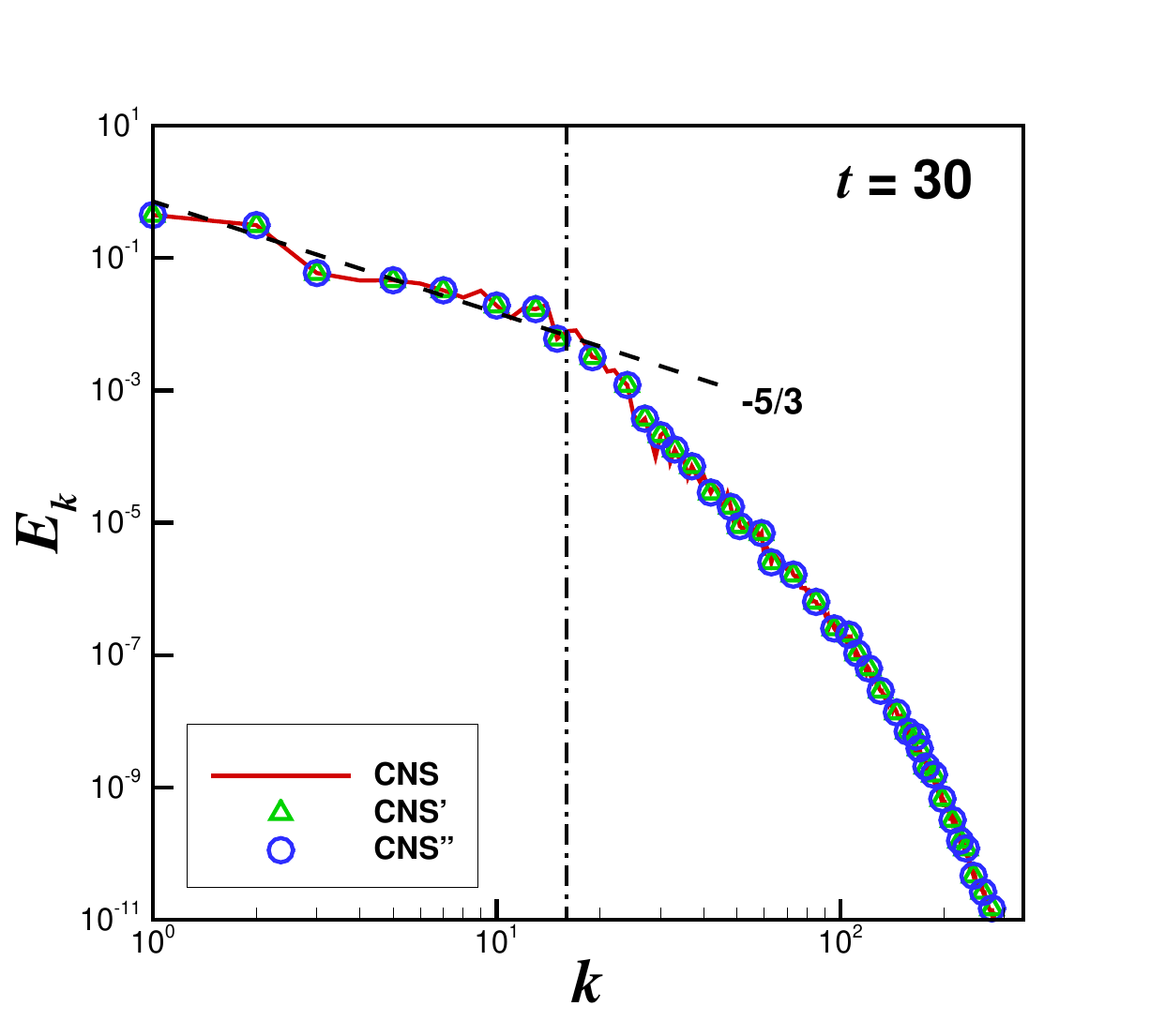}}
             \subfigure[]{\includegraphics[width=2.2in]{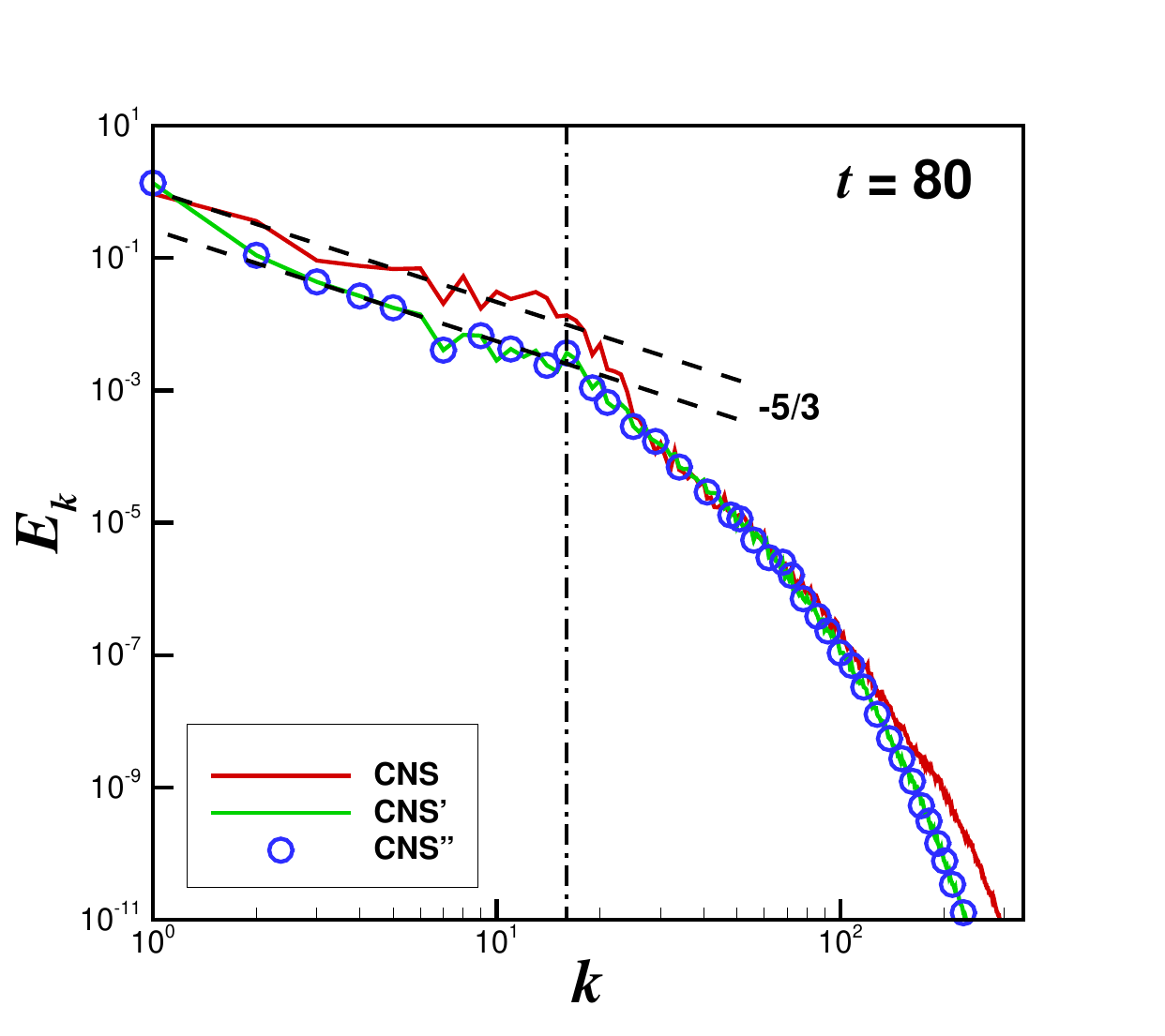}}    \\
             \subfigure[]{\includegraphics[width=2.2in]{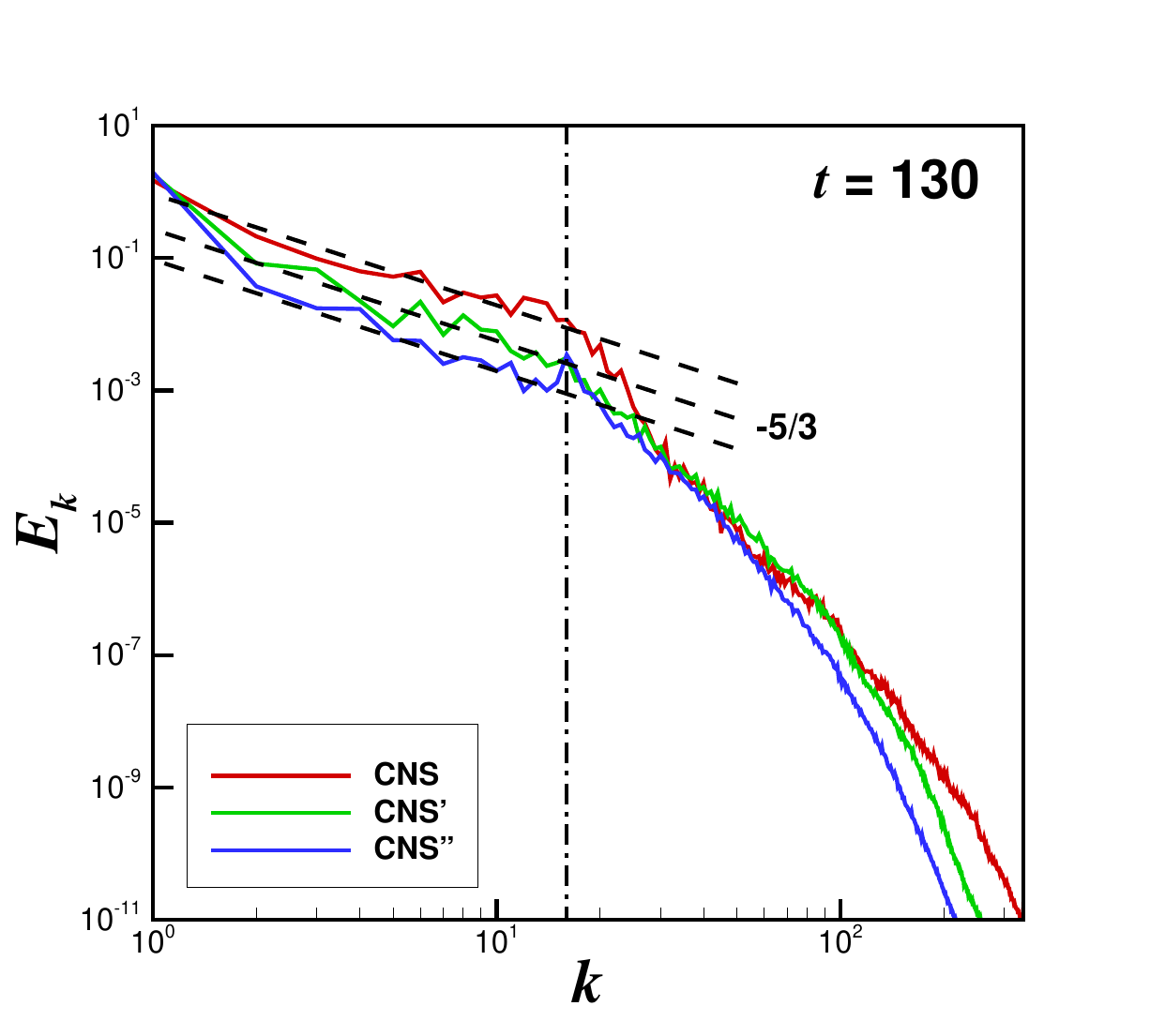}}
        \end{tabular}
    \caption{Kinetic energy spectra $E_k$ of the 2D turbulent Kolmogorov flow governed by (\ref{eq_psi}) and (\ref{boundary_condition}) for $n_K=16$ and $Re=2000$ given by Flow CNS (red line), Flow CNS$'$ (green triangle or line) and Flow CNS$''$ (blue circle or line), at the different times: \\(a) $t=30$, (b) $t=80$ and (c) $t=130$, where the black dashed line corresponds to \\$-5/3$ power law and the black dash-dot line denotes $k=n_K=16$.}     \label{E_k-T}
    \end{center}
\end{figure}

The above results can be confirmed once again by  comparing the kinetic energy spectra $E_k$ of Flow CNS, Flow CNS' and Flow CNS'',  as shown in Fig.~\ref{E_k-T}.  When $t\in[0,35]$, the deviations between the spatial-temporal trajectories of Flow CNS, Flow CNS' and Flow CNS'' are negligible, and their corresponding  kinetic energy spectrum $E_k$  agree quite well, as shown in  Fig.~\ref{E_k-T}(a) for $t=30$.   Thereafter,  the evolution $\delta_1(x,y,t)$ of the first disturbance $10^{-20}\sin(x+y)$ increases to a macro-level $O(1)$ so that the kinetic energy spectra $E_k$ of Flow CNS deviates from those of Flow CNS' and Flow CNS'' that retain the same until $t\approx 93$, as shown  in Fig.~\ref{E_k-T}~(b) for $t=80$.    When the evolution $\delta_2(x,y,t)$ of the second disturbance $10^{-40}\sin(x+2 y)$ increases to a macro-level $O(1)$ at $t \approx 93$,   all kinetic energy spectrum $E_k$ are different, as shown in Fig.~\ref{E_k-T}~(c) for $t=130$.   Note that all of them satisfy the -5/3 law, indicating that all of them are turbulent flows, no matter whether the vorticity field has spatial symmetry or not.

 It is very interesting that Flow CNS contains more kinetic energy than Flow CNS',  and Flow CNS' contains more kinetic energy than Flow CNS'', as shown in Fig.~\ref{E_k-T}.   This fact illustrates that turbulence with more spatial symmetries needs more kinetic energy to maintain.  In order words, turbulence without spatial symmetry should be optimal from the view point of energy.   This is mainly because the spatially averaged kinetic energy dissipation rate $\langle D\rangle_A$ of Flow CNS is much larger than that of Flow CNS', and the latter is larger than that of Flow CNS'', as shown in Fig.~\ref{ED-PDF_P},  indicating that turbulence with spatial symmetry indeed requires more kinetic energy to maintain.     This might be the reason why turbulent flows  have no spatial symmetry in practice.    

Thus, {\em not only} do all disturbances at different orders of magnitudes to the initial condition of the NS equations increase, one by one like an inverse cascade, to a macro-level, {\em but also} each of them is capable of completely altering the macro-level characteristics of turbulent flow, even including certain statistical properties, as illustrated above.  In other words,  micro-level noise/disturbance might have great influence on macro-level characteristics and statistics of turbulence. 

\subsection{An origin of randomness of turbulence}

The most surprising aspect of the noise-expansion cascade phenomenon is the fact that {\em all} disturbances, even if at quite different orders of magnitudes, {\em separately} enlarge to macro-level, one by one like an inverse cascade. For example, for Flow CNS$''$  subject to the initial condition (\ref{initial_condition-22}), the evolution $\delta_1(x,y,t)$ of the first disturbance $10^{-20}\sin(x+y)$ increases to macro-level at $t\approx 35$ and then triggers the transformation of the spatial symmetry of the flow field from (\ref{symmetry-omega:A}) to (\ref{symmetry_psi:B}), but the evolution $\delta_2(x,y,t)$ of the second disturbance $10^{-40}\sin(x+2 y)$ totally destroys the spatial symmetry of the turbulent flow when it increases to macro-level at $t\approx 93$. This clearly indicates that the second disturbance $10^{-40}\sin(x+2 y)$ of the initial condition (\ref{initial_condition-22}) can cause the turbulence characteristics (such as the vorticity spatial symmetry) to change greatly, even though the initial disturbance is 20 orders of magnitude smaller than the first one.
This strongly suggests that {\em all} disturbances must be considered for turbulence. This answers the second {\em open} question posed at the beginning of this paper.    

Note that internal thermal fluctuation and external environmental disturbance are  {\em unavoidable} in practice. In general, environmental disturbance is much larger than thermal fluctuation. So, traditionally, thermal fluctuation is neglected for most turbulent flows, especially those governed by the NS equations. 
However, according to the noise-expansion cascade concept proposed herein, thermal fluctuation {\em must} be considered for turbulence, even if it is many orders of magnitude smaller than environmental disturbance. 
Note that thermal fluctuation is essentially random, and this randomness can naturally transfer from the micro-level to macro-level through the noise-expansion cascade. This concept enables us to reveal an {\em origin} of randomness in turbulence, thus answering the first {\em open} question posed at the beginning of this paper. 

\section{Conclusion and discussion}

\citet{Qin2022JFM} and \citet{Qin2024JOES} recently demonstrated that the spatiotemporal trajectory given by direct numerical simulation (DNS) can rapidly become badly polluted by artificial numerical noise so that it becomes impossible for DNS to simulate accurately and rigorously the evolution and propagation of small disturbances in turbulent flow. Fortunately, clean numerical simulation (CNS) \citep{Liao2009, Liao2023book, Liao2014SCPMA, Lin2017SCPMA, Hu2020JCP, Qin2020CSF, Liao2022AAMM, Qin2023JFM,  Qin2023AAMM, Qin2024PhysicaD} produces very much smaller artificial numerical noise than DNS and so can be used to provide clean numerical experiments for turbulence.
In this paper, by means of  a ``thought experiment'' via several clean numerical experiments based on CNS for a 2D turbulent Kolmogorov flow, it is discovered, for the first time, that all disturbances at {\em different} orders of magnitudes in initial condition of NS equations quickly enlarge separately, one by one like an inverse cascade, to macro-level. More importantly, each noise/disturbance might {\em greatly} change characteristics of the turbulent flow {\em not only} in the spatial symmetry of the flow {\em but also} even in the statistics,  thus clearly indicating that  micro-level noise/disturbance might have great influence on macro-level characteristics and statistics of turbulence.  
Based on this interesting phenomenon, we propose a new concept, called the ``noise-expansion cascade'' that closely connects the randomness of micro-level noise/disturbance to the macro-level disorder of turbulence.   This reveals an {\em origin} of randomness in turbulence, thus answering the first {\em open} question posed at the beginning of the paper. 

Note that internal thermal fluctuation and external environmental disturbance are {\em unavoidable} in practice. 
So, according to the noise-expansion cascade, thermal fluctuation {\em must} be considered for turbulence, even if it is much smaller than external environmental disturbance, just as in the Landau-Lifshitz-Navier-Stokes (LLNS) equations that include the influence of thermal fluctuation \citep{LLNS1959}. This answers the second {\em open} question that we mentioned at the beginning of this paper. Note that the influence of thermal fluctuation on turbulence has been recently reported by several researchers \citep{Bandak2022PRE, JunZhang2024JFM}.

Due to the noise-expansion cascade, many micro-level random factors might exert great influence on the macro-level disorder of turbulence. Hence, the noise-expansion cascade should be viewed as a {\em bridge} connecting {\em micro-level} random fluctuation/disturbance to the {\em macro-level} disorder of turbulence. So, the concept of noise-expansion cascade provides a scientific explanation of the philosopher Heraclitus' aphorism that ``people cannot step twice into the same river''. It strongly implies that turbulence should be a {\em unity} of micro-level random noise/disturbance, their evolution and propagation,  and macro-level disorder, which are closely connected through the noise-expansion cascade. This is why turbulence is so challenging to understand.

It should be emphasized that the noise-expansion cascade as a new concept  {\em cannot} be discovered by numerical experiments based on DNS (since its artificial numerical noises increase exponentially) or  physical experiments in laboratory (since it is is impossible in practice to have such accurate micro-level disturbances): this illustrates the great potential of clean numerical experiments based on CNS.  Note that the two micro-level disturbances $10^{-20} \sin(x+y)$ and $10^{-40} \sin(x+2y)$ hardly exist in practice: this is the reason why the ``noise-expansion cascade'' as a new fundamental concept of NS turbulence can be discovered   {\em only} by such a ``{\em thought experiment}'' based on CNS.

The noise-expansion cascade exhibits obvious difference from the ``energy cascade'' which is a fundamental concept in turbulence theory. There are two types of energy cascade: the  direct energy cascade whereby energy transfers from large scale to small scale; and the inverse energy cascade whereby energy transfers from small scale to large scale. However, unlike the energy cascade, the noise-expansion cascade has only one direction: from micro-scale to large scale. Moreover, the energy cascade describes {\em spatial} transfer of energy, whereas the noise-expansion cascade describes the {\em temporal} evolution and propagation of initial noise/disturbance. 
Recently, it was reported that the average uncertainty energy of  Navier-Stokes turbulence grows exponentially \citep{aurell1996growth, boffetta2001predictability, boffetta2010evidence, boffetta2012two, Lin2017SCPMA, Vassilicos2023JFM, JunZhang2024JFM}.  This phenomenon, sometimes called ``inverse error cascade'', is observed using different initial conditions at the {\em same} order of magnitude.   However,   the noise-expansion cascade focuses on the influence of initial conditions with  {\em different} orders of magnitudes.    
This further underlines the novelty of the noise-expansion cascade as a new fundamental concept of NS turbulence.    

At each time step, DNS algorithm unavoidably contains artificial numerical noise. Thus, due to the noise-expansion cascade, the artificial numerical noise of DNS algorithm at each time step will increase consistently to reach the macro-level: this is exactly why the spatiotemporal trajectory of DNS departs quickly from the true solution, and why the noise-expansion cascade {\em cannot} be revealed by DNS.

Actually, CNS requires substantial computer resources and is time-consuming to run at present, just like DNS when \citet{Orszag1970} proposed it. For the 2D turbulent Kolmogorov flow under consideration, the parallel computing of the CNS takes 285 hours (i.e. about 12 days) using 4096 CPUs (Intel's CPU: Xeon Gold 6348, 2.60GHz) of the ``Tian-He New Generation Supercomputer'' at National Supercomputer Center in Tianjin, China.  The related code of CNS and some movies can be downloaded via GitHub (\url{https://github.com/sjtu-liao/2D-Kolmogorov-turbulence}).  
By contrast, DNS of the same case requires only 13 hours using 1024 CPUs on the same computing platform. Even so, CNS undoubtedly provides us with a new way to investigate turbulence through clean numerical experiments, i.e. with negligible artificial numerical noise, as illustrated in this paper.     

We would like to emphasize that DNS is a {\em milestone} in fluid mechanics, since it opened a new era of numerical experiment and has greatly promoted the progress of turbulence in theories, physical experiments and applications. In essence, CNS can be regarded as a {\em general} form of DNS with rigorously {\em negligible} numerical noise in a {\em finite} but long {\em enough} time interval $t\in[0,T_c]$, where $T_c$ is the so-called ``critical predictable time''.  In other words,   the critical predictable time $T_c$ of CNS is much longer than that of DNS.   Thus, CNS offers a powerful method by which to accurately investigate the influence of {\em micro-scale} physical/artificial disturbances on macro-scale characteristics and statistics of turbulent flow.

Note that, like DNS,  CNS has ``the ability to perform fundamental studies of {\em clean} flows {\em unaffected} by numerical, modelling and measurement errors'' and ``the complete control of the initial and boundary conditions, and each term in the governing equations, also leads to profound advantages over laboratory and field studies'' \citep{Coleman2010DNS}.   It is found that our conclusions are qualitatively same as those mentioned above  even if we use the different 1st disturbance  such as $10^{-10}\sin(x+y)$ and the different 2nd disturbance such as $ 10^{-20}\sin(x+2 y)$.   We emphasize  that it is impossible to accurately have such kinds of micro-level disturbances at quite different orders of magnitude by physical experiments in a laboratory, thus we in fact did a {\em thought experiment} of the NS turbulence by means of CNS in this paper.  Note that it is this {\em thought experiment} that reveals the new fundamental concept  ``noise-expansion cascade''.   
Here  we emphasize once again that the  ``noise-expansion cascade'' not only reveals an origin of randomness of turbulence but also highly indicates that micro-level noises/disturbances might have great influence on macro-level characteristics and statistics of turbulence.    
In other words,  macro-level characteristics and statistics  of turbulence should be determined {\em not only} by its  spatial domain of flow and its initial/boundary conditions {\em but also} by its micro-level noises/disturbances.  
Hopefully,  the ``noise-expansion cascade'' as a fundamental property of the NS equations could greatly deepen our understandings about turbulence, and besides is helpful for attacking the fourth  millennium problem posed by \citet{MillenniumProblem}.      

There are many interesting problems worthy of further investigation in future. For example, it might be possible that {\em artificial} numerical noise could be regarded as a kind of {\em physical} external disturbance, so long as we could prove that artificial numerical noise could have {\em same} influences as thermal fluctuation and/or environmental disturbance to turbulent flow, which unfortunately is still an {\em open} question until now. This kind of investigations might reveal the essence of artificial numerical noises to turbulent flow.    
Besides, thermal fluctuation is random and {\em discontinuous}, but it is still an {\em open} question whether or not such  {\em discontinuous} random disturbance might become {\em continuous} when it increases to macro-level due to noise-expansion cascade.  Especially,  there exist two types of chaotic system, i.e. normal-chaos and {\em ultra-chaos}  \citep{Liao2022AAMM}: unlike normal-chaos, statistics of ultra-chaos are unstable (or sensitive) to small disturbances \citep{qin_liao_2023,  Yang2023IJBC, yang2023CSF, Zhang2023PhysicaD, Zhang2024AMS, Yang2024IJBC}.  In other words, statistical non-reproducibility is an inherent property of an ultra-chaos so that an ultra-chaos is at a higher-level of disorder than a normal-chaos.  If a turbulent flow belongs to an ultra-chaos, say, its statistics are sensitive (i.e. unstable) to small noise/disturbance,  then due to the ``noise-expansion cascade'',  its statistics are unstable/sensitive to physical environmental disturbances or artificial numerical noises so that statistical reproducibility of experimental/numerical results does {\em not} exist.  However, it is an {\em open} question how to solve (or understand) such kind of ultra-chaotic turbulence without  reproducibility of statistics.           

\backsection[Acknowledgements]{Thanks to the anonymous reviewers for their valuable suggestions and constructive comments.  The calculations were performed on ``Tian-He New Generation Supercomputer'', National Supercomputer Center in Tianjing, China.  }

\backsection[Funding]{This work is supported by State Key Laboratory of Ocean Engineering, Shanghai 200240, China.}

\backsection[Declaration of Interests]{The authors report no conflict of interest.}

\backsection[Data availability statement]{The data that support the findings of this study are available on request from the corresponding author.  Besides, the related code of CNS and some movies can be downloaded via GitHub (\url{https://github.com/sjtu-liao/2D-Kolmogorov-turbulence}).}

\backsection[Author contributions]{Shijun conceived/designed the numerical experiments, proposed the new concept  ``noise-expansion cascade'',  and wrote the manuscript.  Shijie performed the numerical simulation and plotted the figures. }

\backsection[Author ORCID]{Shijun Liao, https://orcid.org/0000-0002-2372-9502; Shijie Qin, https://orcid.org/0000-0002-0809-1766}

\appendix

\section{Some definitions and measures}    \label{Key_measures}

For simplicity, the definitions of some statistical operators are briefly described below.
The operator of spatial average is defined by
\begin{align}
& \langle\,f \,\rangle_A=\frac{1}{4\pi^2}\int^{2\pi}_0\int^{2\pi}_0 f \,  \mathrm{d}x \mathrm{d}y,       \label{average_A}
\end{align}
and the operator of spatiotemporal average (along the $x$ direction) is defined by
\begin{align}
& \langle\,f \,\rangle_{x,t}=\frac{1}{2\pi (T_2-T_1)}\int^{2\pi}_0\int^{T_2}_{T_1} f\, \mathrm{d}x \mathrm{d}t,       \label{average_xt}
\end{align}
where $T_1=100$ (for Flow CNS and Flow CNS$'$) or $120$ (for Flow CNS$''$) and $T_2=300$ are used in this paper to calculate statistics.  

The kinetic energy is given by 
\begin{align}
& E(x,y,t) = \frac{1}{2}\Big[u^2(x,y,t)+v^2(x,y,t)\Big]    \label{kinetic_energy}
\end{align}
and the kinetic energy dissipation rate is defined by 
\begin{align}
& D(x,y,t)=\frac{1}{2Re}\sum_{i,j=1,2}\Big[ \partial_iu_j(x,y,t)+\partial_ju_i(x,y,t) \Big]^2,    \label{dissipation_rate}
\end{align}
where $u_1(x,y,t)=u(x,y,t)$, $u_2(x,y,t)=v(x,y,t)$, $\partial_1=\partial /\partial x$, and $\partial_2=\partial /\partial y$.

The stream function can be expanded as the Fourier series
\begin{align}
& \psi(x,y,t)\approx\sum^{\lfloor N/3 \rfloor}_{\,m=-\lfloor N/3 \rfloor}\sum^{\lfloor N/3 \rfloor}_{\,n=-\lfloor N/3 \rfloor}\Psi_{m,n}(t) \exp(\mathbf{i}\,mx)\exp(\mathbf{i}\,ny),       \label{Fourier}
\end{align}
where $m$, $n$ are integers, $\lfloor\,\,\rfloor$ stands for a floor function, $\mathbf{i}=\sqrt{-1}$ denotes the imaginary unit, and for dealiasing $\Psi_{m,n}=0$ is imposed for wavenumbers outside the above domain $\sum$. Note that, for the real number $\psi$, $\Psi_{-m,-n}=\Psi^*_{m,n}$ must be satisfied, where $\Psi^*_{m,n}$ is the conjugate of $\Psi_{m,n}$.
Therefore, kinetic energy spectrum is defined as
\begin{align}
& E_k(t)=\sum_{k-1/2 \leq \sqrt{m^2+n^2} < k+1/2}\frac{1}{2}\,(m^2+n^2)\mid \Psi_{m,n}(t) \mid^2,       \label{kinetic_energy_spectrum}
\end{align}
where the wave number $k$ is a non-negative integer.

\bibliographystyle{jfm}
\bibliography{Kolmogorov3D}

\begin{thebibliography}{48}
\expandafter\ifx\csname natexlab\endcsname\relax\def\natexlab#1{#1}\fi
\def\au#1{#1} \def\ed#1{#1} \def\yr#1{#1}\def\at#1{#1}\def\jt#1{\textit{#1}}
  \def\bt#1{#1}\def\bvol#1{\textbf{#1}} \def\vol#1{#1} \def\pg#1{#1}
  \def\publ#1{#1}\def\arxiv#1{#1}\def\org#1{#1}\def\st#1{\textit{#1}}

\bibitem[Alexakis {\em et~al.\/}(2024)Alexakis, Marino, Mininni, van Kan,
  Foldes \& Feraco]{FeracoScience2024}
{\sc \au{Alexakis, Alexandros}, \au{Marino, Raffaele}, \au{Mininni, Pablo~D.},
  \au{van Kan, Adrian}, \au{Foldes, Raffaello} \& \au{Feraco, Fabio}} \yr{2024}
   \at{Large-scale self-organization in dry turbulent atmospheres}.
  \jt{Science}  \bvol{383},  \pg{1005--1009}.

\bibitem[Aurell {\em et~al.\/}(1996)Aurell, Boffetta, Crisanti, Paladin \&
  Vulpiani]{aurell1996growth}
{\sc \au{Aurell, Erik}, \au{Boffetta, Guido}, \au{Crisanti, Andrea},
  \au{Paladin, Giovanni} \& \au{Vulpiani, Angelo}} \yr{1996}  \at{Growth of
  noninfinitesimal perturbations in turbulence}.  \jt{Phys. Rev. Lett.}
  \bvol{77}~(7),  \pg{1262}.

\bibitem[Bandak {\em et~al.\/}(2022)Bandak, Goldenfeld, Mailybaev \&
  Eyink]{Bandak2022PRE}
{\sc \au{Bandak, Dmytro}, \au{Goldenfeld, Nigel}, \au{Mailybaev, Alexei~A.} \&
  \au{Eyink, Gregory}} \yr{2022}  \at{Dissipation-range fluid turbulence and
  thermal noise}.  \jt{Phys. Rev. E}  \bvol{105},  \pg{065113}.

\bibitem[Berera \& Ho(2018)]{berera2018chaotic}
{\sc \au{Berera, Arjun} \& \au{Ho, Richard D. J.~G.}} \yr{2018}  \at{Chaotic
  properties of a turbulent isotropic fluid}.  \jt{Phys. Rev. Lett.}
  \bvol{120}~(2),  \pg{024101}.

\bibitem[Boffetta \& Ecke(2012{\natexlab{{\em a\/}}})]{boffetta2012two}
{\sc \au{Boffetta, Guido} \& \au{Ecke, Robert~E}} \yr{2012{\natexlab{{\em
  a\/}}}}  \at{Two-dimensional turbulence}.  \jt{Annu. Rev. Fluid Mech.}
  \bvol{44},  \pg{427--451}.

\bibitem[Boffetta \& Ecke(2012{\natexlab{{\em b\/}}})]{Boffetta2012ARFM}
{\sc \au{Boffetta, Guido} \& \au{Ecke, Robert~E}} \yr{2012{\natexlab{{\em
  b\/}}}}  \at{Two-dimensional turbulence}.  \jt{Annu. Rev. Fluid Mech.}
  \bvol{44},  \pg{427--51}.

\bibitem[Boffetta \& Musacchio(2001)]{boffetta2001predictability}
{\sc \au{Boffetta, Guido} \& \au{Musacchio, S}} \yr{2001}  \at{Predictability
  of the inverse energy cascade in {2D} turbulence}.  \jt{Phys. Fluids}
  \bvol{13}~(4),  \pg{1060--1062}.

\bibitem[Boffetta \& Musacchio(2010)]{boffetta2010evidence}
{\sc \au{Boffetta, Guido} \& \au{Musacchio, Stefano}} \yr{2010}  \at{Evidence
  for the double cascade scenario in two-dimensional turbulence}.  \jt{Phys.
  Rev. E}  \bvol{82}~(1),  \pg{016307}.

\bibitem[Boffetta \& Musacchio(2017)]{boffetta2017chaos}
{\sc \au{Boffetta, Guido} \& \au{Musacchio, Stefano}} \yr{2017}  \at{Chaos and
  predictability of homogeneous-isotropic turbulence}.  \jt{Phys. Rev. Lett.}
  \bvol{119}~(5),  \pg{054102}.

\bibitem[Chandler \& Kerswell(2013)]{chandler2013invariant}
{\sc \au{Chandler, Gary~J} \& \au{Kerswell, Rich~R}} \yr{2013}  \at{Invariant
  recurrent solutions embedded in a turbulent two-dimensional {Kolmogorov}
  flow}.  \jt{J. Fluid Mech.}  \bvol{722},  \pg{554--595}.

\bibitem[{Clay Mathematics Institute of Cambridge,
  Massachusetts}(2000)]{MillenniumProblem}
{\sc \au{{Clay Mathematics Institute of Cambridge, Massachusetts}}} \yr{2000}
  {The Millennium Prize Problems}.
  \url{https://www.claymath.org/millennium-problems/}.

\bibitem[Coleman \& Sandberg(2010)]{Coleman2010DNS}
{\sc \au{Coleman, Gary~N.} \& \au{Sandberg, Richard~D.}} \yr{2010}  \bt{A
  primer on direct numerical simulation of turbulence -- methods, procedures
  and guidelines}. {\em Tech. Rep.\/} AFM-09/01a.  \org{Aerodynamics \& Flight
  Mechanics Research Group, University of Southampton}, UK.

\bibitem[Crane({2017})]{NewScientist2017}
{\sc \au{Crane, L.}} \yr{{2017}}  \at{Infamous three-body problem has over a
  thousand new solutions}.  \jt{New Scientist} .

\bibitem[Davidson(2004)]{Davidson2004}
{\sc \au{Davidson, P.~A.}} \yr{2004} {\em Turbulence: an introduction for
  scientists and engineers\/}.  \publ{Oxford: Oxford University Press}.

\bibitem[Deissler(1986)]{Deissler1986PoF}
{\sc \au{Deissler, R.~G.}} \yr{1986}  \at{Is {N}avier-{S}tokes turbulence
  chaotic?}  \jt{Phys. Fluids}  \bvol{29},  \pg{1453 -- 1457}.

\bibitem[Ge {\em et~al.\/}(2023)Ge, Rolland \& Vassilicos]{Vassilicos2023JFM}
{\sc \au{Ge, Jin}, \au{Rolland, Joran} \& \au{Vassilicos, John~Christos}}
  \yr{2023}  \at{The production of uncertainty in three-dimensional
  {N}avier-{S}tokes turbulence}.  \jt{J. Fluid Mech.}  \bvol{977},  \pg{A17}.

\bibitem[Hu \& Liao(2020)]{Hu2020JCP}
{\sc \au{Hu, Tianli} \& \au{Liao, Shijun}} \yr{2020}  \at{On the risks of using
  double precision in numerical simulations of spatio-temporal chaos}.  \jt{J.
  Comput. Phys.}  \bvol{418},  \pg{109629}.

\bibitem[Landau \& Lifshitz(1959)]{LLNS1959}
{\sc \au{Landau, L.~D.} \& \au{Lifshitz, E.~M.}} \yr{1959} {\em Course of
  Theoretical Physics: Fluid Mechanics (Vol. 6)\/}.  \publ{Reading:
  Addision-Wesley}.

\bibitem[Li {\em et~al.\/}(2018)Li, Jing \& Liao]{Li2018PASJ}
{\sc \au{Li, Xiaoming}, \au{Jing, Yipeng} \& \au{Liao, Shijun}} \yr{2018}
  \at{Over a thousand new periodic orbits of a planar three-body system with
  unequal masses}.  \jt{Publ. Astron. Soc. Jpn.}  \bvol{70}~(4),  \pg{64}.

\bibitem[Li \& Liao(2017)]{Li2017SCPMA}
{\sc \au{Li, Xiaoming} \& \au{Liao, Shijun}} \yr{2017}  \at{More than six
  hundred new families of {Newtonian} periodic planar collisionless three-body
  orbits}.  \jt{Sci. China-Phys. Mech. Astron.}  \bvol{60}~(12),  \pg{129511}.

\bibitem[Liao(2009)]{Liao2009}
{\sc \au{Liao, Shijun}} \yr{2009}  \at{On the reliability of computed chaotic
  solutions of non-linear differential equations}.  \jt{Tellus Ser. A-Dyn.
  Meteorol. Oceanol.}  \bvol{61}~(4),  \pg{550--564}.

\bibitem[Liao(2023)]{Liao2023book}
{\sc \au{Liao, Shijun}} \yr{2023} {\em Clean Numerical Simulation\/}.
  \publ{Chapman and Hall/CRC}.

\bibitem[Liao {\em et~al.\/}(2022)Liao, Li \& Yang]{Liao2022NA}
{\sc \au{Liao, Shijun}, \au{Li, Xiaoming} \& \au{Yang, Yu}} \yr{2022}
  \at{Three-body problem -- from {Newton} to supercomputer plus machine
  learning}.  \jt{New Astronomy}  \bvol{96},  \pg{101850}.

\bibitem[Liao \& Qin(2022)]{Liao2022AAMM}
{\sc \au{Liao, Shijun} \& \au{Qin, Shijie}} \yr{2022}  \at{Ultra-chaos: an
  insurmountable objective obstacle of reproducibility and replication}.
  \jt{Adv. Appl. Math. Mech.}  \bvol{14}~(4),  \pg{799--815}.

\bibitem[Liao \& Wang(2014)]{Liao2014SCPMA}
{\sc \au{Liao, Shijun} \& \au{Wang, Pengfei}} \yr{2014}  \at{On the
  mathematically reliable long-term simulation of chaotic solutions of {Lorenz}
  equation in the interval [0, 10000]}.  \jt{Sci. China - Phys. Mech. Astron.}
  \bvol{57}~(2),  \pg{330 -- 335}.

\bibitem[Lin {\em et~al.\/}(2017)Lin, Wang \& Liao]{Lin2017SCPMA}
{\sc \au{Lin, Zhiliang}, \au{Wang, Lipo} \& \au{Liao, Shijun}} \yr{2017}
  \at{On the origin of intrinsic randomness of {Rayleigh-B{\'e}nard}
  turbulence}.  \jt{Sci. China-Phys. Mech. Astron.}  \bvol{60}~(1),
  \pg{1--13}.

\bibitem[Ma {\em et~al.\/}(2024)Ma, Yang, Chen, Feng, Cui \&
  Zhang]{JunZhang2024JFM}
{\sc \au{Ma, Qihan}, \au{Yang, Chunxin}, \au{Chen, Song}, \au{Feng, Kaikai},
  \au{Cui, Ziqi} \& \au{Zhang, Jun}} \yr{2024}  \at{Effect of thermal
  fluctuations on spectra and predictability in compressible decaying isotropic
  turbulence}.  \jt{J. Fluid Mech.}  \bvol{987},  \pg{A29}.

\bibitem[Nelkin(1992)]{Nelkin1992Science}
{\sc \au{Nelkin, Mark}} \yr{1992}  \at{In what sense is turbulence an unsolved
  problem?}  \jt{Science}  \bvol{255},  \pg{566--570}.

\bibitem[Obukhov(1983)]{obukhov1983kolmogorov}
{\sc \au{Obukhov, A.~M.}} \yr{1983}  \at{{Kolmogorov} flow and laboratory
  simulation of it}.  \jt{Russian Math. Surveys}  \bvol{38}~(4),
  \pg{113--126}.

\bibitem[Orszag(1970)]{Orszag1970}
{\sc \au{Orszag, S.~A.}} \yr{1970}  \at{Analytical theories of turbulence}.
  \jt{J. Fluid Mech.}  \bvol{41}~(2),  \pg{363 -- 386}.

\bibitem[Oyanarte(1990)]{oyanarte1990mp}
{\sc \au{Oyanarte, P.}} \yr{1990}  \at{{MP}-a multiple precision package}.
  \jt{Comput. Phys. Commun.}  \bvol{59}~(2),  \pg{345--358}.

\bibitem[Pope(2001)]{pope2001turbulent}
{\sc \au{Pope, S.~B.}} \yr{2001} {\em Turbulent Flows\/}.  \publ{IOP
  Publishing}.

\bibitem[Qin \& Liao(2020)]{Qin2020CSF}
{\sc \au{Qin, Shijie} \& \au{Liao, Shijun}} \yr{2020}  \at{Influence of
  numerical noises on computer-generated simulation of spatio-temporal chaos}.
  \jt{Chaos Solitons Fractals}  \bvol{136},  \pg{109790}.

\bibitem[Qin \& Liao(2022)]{Qin2022JFM}
{\sc \au{Qin, Shijie} \& \au{Liao, Shijun}} \yr{2022}  \at{Large-scale
  influence of numerical noises as artificial stochastic disturbances on a
  sustained turbulence}.  \jt{J. Fluid Mech.}  \bvol{948},  \pg{A7}.

\bibitem[Qin \& Liao(2023{\natexlab{{\em a\/}}})]{Qin2023JFM}
{\sc \au{Qin, Shijie} \& \au{Liao, Shijun}} \yr{2023{\natexlab{{\em a\/}}}}
  \at{A kind of {Lagrangian} chaotic property of the
  {Arnold-Beltrami-Childress} flow}.  \jt{J. Fluid Mech.}  \bvol{960},
  \pg{A15}.

\bibitem[Qin \& Liao(2023{\natexlab{{\em b\/}}})]{qin_liao_2023}
{\sc \au{Qin, Shijie} \& \au{Liao, Shijun}} \yr{2023{\natexlab{{\em b\/}}}}
  \at{A kind of lagrangian chaotic property of the {Arnold-Beltrami-Childress}
  flow}.  \jt{Journal of Fluid Mechanics}  \bvol{960},  \pg{A15}.

\bibitem[Qin \& Liao(2023{\natexlab{{\em c\/}}})]{Qin2023AAMM}
{\sc \au{Qin, Shijie} \& \au{Liao, Shijun}} \yr{2023{\natexlab{{\em c\/}}}}
  \at{A self-adaptive algorithm of the clean numerical simulation {(CNS)} for
  chaos}.  \jt{Adv. Appl. Math. Mech.}  \bvol{15}~(5),  \pg{1191--1215}.

\bibitem[Qin \& Liao(2024)]{Qin2024PhysicaD}
{\sc \au{Qin, Shijie} \& \au{Liao, Shijun}} \yr{2024}  \at{Influences of
  artificial numerical noise on statistics and qualitative properties of
  chaotic system}.  \jt{Physica D}  \bvol{470},  \pg{134355}.

\bibitem[Qin {\em et~al.\/}(2024)Qin, Yang, Huang, Mei, Wang \&
  Liao]{Qin2024JOES}
{\sc \au{Qin, Shijie}, \au{Yang, Yu}, \au{Huang, Yongxiang}, \au{Mei, Xinyu},
  \au{Wang, Lipo} \& \au{Liao, Shijun}} \yr{2024}  \at{Is a direct numerical
  simulation {(DNS)} of {Navier-Stokes} equations with small enough grid
  spacing and time-step definitely reliable/correct?}  \jt{Journal of Ocean
  Engineering and Science}  \bvol{9},  \pg{293 -- 310}.

\bibitem[Rogallo(1981)]{Rogallo1981NASA}
{\sc \au{Rogallo, Robert~S.}} \yr{1981}  \bt{Numerical experiments in
  homogeneous turbulence}. {\em Tech. Rep.\/} NASA-TM-81315.  \org{NASA}, USA.

\bibitem[She {\em et~al.\/}(1990)She, Jackson \& Orszag]{She1990Nature}
{\sc \au{She, Zhen-Su}, \au{Jackson, Eric} \& \au{Orszag, Steven~A.}} \yr{1990}
   \at{Intermittent vortex structures in homogeneous isotropic turbulence}.
  \jt{Nature}  \bvol{344},  \pg{226 -- 228}.

\bibitem[Whyte({2018})]{NewScientist2018}
{\sc \au{Whyte, C.}} \yr{{2018}}  \at{Watch the weird new solutions to the
  baffling three-body problem}.  \jt{New Scientist} .

\bibitem[Wu {\em et~al.\/}(2021)Wu, Schmitt, Calzavarini \&
  Wang]{wu2021quadratic}
{\sc \au{Wu, Wenwei}, \au{Schmitt, F.~G.}, \au{Calzavarini, Enrico} \&
  \au{Wang, Lipo}} \yr{2021}  \at{A quadratic {Reynolds} stress development for
  the turbulent {Kolmogorov} flow}.  \jt{Phys. Fluids}  \bvol{33},
  \pg{125129}.

\bibitem[Yang \& Liao(2024)]{Yang2024IJBC}
{\sc \au{Yang, Yu} \& \au{Liao, Shijun}} \yr{2024}  \at{Ultra-chaos: a great
  challenge for machine learning and {AI}}.  \jt{International Journal of
  Bifurcation and Chaos}  \pg{p. 2450202}.

\bibitem[Yang {\em et~al.\/}(2023{\natexlab{{\em a\/}}})Yang, Qin \&
  Li]{Yang2023IJBC}
{\sc \au{Yang, Yu}, \au{Qin, Shijie} \& \au{Li, Shijun}}
  \yr{2023{\natexlab{{\em a\/}}}}  \at{Ultra-chaos in the motion of walking
  droplet}.  \jt{International Journal of Bifurcation and Chaos}
  \bvol{33}~(16),  \pg{2350191}.

\bibitem[Yang {\em et~al.\/}(2023{\natexlab{{\em b\/}}})Yang, Qin \&
  Liao]{yang2023CSF}
{\sc \au{Yang, Yu}, \au{Qin, Shijie} \& \au{Liao, Shijun}}
  \yr{2023{\natexlab{{\em b\/}}}}  \at{Ultra-chaos of a mobile robot: A higher
  disorder than normal-chaos}.  \jt{Chaos, Solitons \& Fractals}  \bvol{167},
  \pg{113037}.

\bibitem[Zhang \& Liao(2023)]{Zhang2023PhysicaD}
{\sc \au{Zhang, Bo} \& \au{Liao, Shijun}} \yr{2023}  \at{Ultra-chaos in a
  meandering jet flow}.  \jt{Physica D}  \bvol{455},  \pg{133886}.

\bibitem[Zhang {\em et~al.\/}(2024)Zhang, Yang \& Liao]{Zhang2024AMS}
{\sc \au{Zhang, Bo}, \au{Yang, Yu} \& \au{Liao, Shijun}} \yr{2024}
  \at{Ultra-chaos of square thin plate in low earth orbit}.  \jt{Acta Mech.
  Sin.}  \bvol{40},  \pg{523428}.

\end{thebibliography}

\end{sloppypar}
\end{document}